\documentclass[prl,reprint, twocolumn,showpacs,preprintnumbers,amsmath,amssymb,nofootinbib,floatfix]{revtex4-1} 
\usepackage{graphicx}

\usepackage{mathrsfs}
\usepackage{hyperref}

\usepackage{slashed}

\usepackage{color}

\usepackage{gensymb}

\usepackage{amsmath}
\allowdisplaybreaks[4]

\usepackage{bm}

\usepackage{multirow}
\usepackage{booktabs}
\usepackage{placeins}

\def \matrix #1 {\left(\begin{array}{cc} #1 \end{array}\right)}

\def\II{\hbox{{1}\kern-.25em\hbox{l}}}

\newcommand{\blue}[1]{\textcolor[rgb]{0.00,0.00,1.00}{#1}}

\begin{document}

\title{Enhanced Three-Particle Contribution to Electroweak Penguin $B$-Meson Decays}

\author{Yong-Kang Huang}
\email{huangyongkang@mail.nankai.edu.cn}

\author{Yu-Ming Wang}
\email{corresponding author: wangyuming@nankai.edu.cn}

\author{Xue-Chen Zhao}
\email{corresponding author: zxc@mail.nankai.edu.cn}

\affiliation{\vspace{0.2 cm}
School of Physics, Nankai University, \\
Weijin Road 94, Tianjin 300071, P.R. China \\}

\date{\today}

\begin{abstract}
\noindent

We compute for the first time the subleading twist correction to  the exclusive rare $b \to \left \{s, d \right \}  \ell^{+} \ell^{-}$ decays from the three-particle $B$-meson  distribution amplitude at next-to-leading order accuracy
by employing the soft-collinear effective theory.
This constitutes the last missing ingredient for  the complete factorization analysis of the hadronic matrix element
of the weak effective Hamiltonian at leading power in the heavy quark expansion.
Incorporating further a variety of the next-to-next-to-leading-order QCD  corrections  to  the  spectator-scattering amplitude
and the weak annihilation topology, we then present the improved  field-theoretic  predictions for
phenomenologically interesting  observables in  the  electroweak penguin
$B \to \left \{K, \pi \right \} \ell^{+} \ell^{-}$ decays.
\\[0.4em]

\end{abstract}


\maketitle

%
\section{Introduction}
%

It is generally accepted that the exclusive  flavour-changing neutral current (FCNC)  decays of bottom hadrons
are of utmost importance for providing decisive tests of the quark-flavour mixing mechanism in the Standard Model (SM) framework
and for probing sensitively the tantalising  New Physics (NP) signatures above the electro-weak scale
at the ongoing collider  experiments.
The persistent deviations of experimental measurements for numerous  $b \to  s  \ell^{+} \ell^{-}$
decay observables  from  their  theoretical  predictions have therefore triggered  tremendous  efforts on
unravelling the ultimate mystery  of these long-standing discrepancies.
Among the prominent flavour anomalies in  the   electroweak penguin $B$-meson decays are
the  celebrated   tension  between the measured angular observable $P_5^{\prime}(B \to K^{\ast} \mu^{+} \mu^{-})$
and the current SM prediction at  large  hadronic recoil  \cite{LHCb:2020lmf,LHCb:2025mqb,CMS:2024atz}
and the systematic deviations of all  measured branching fractions for exclusive $b \to  s  \ell^{+} \ell^{-}$ decay processes
with respect to the state-of-the-art SM computations in  the entire kinematic region \cite{LHCb:2025mqb,LHCb:2014cxe,LHCb:2021zwz}.
The  complementary and encouraging  measurements for  the angular observables and  the   branching fractions
of the exclusive rare $b \to  d  \ell^{+} \ell^{-}$ decays are,  however,  in nice  agreement with
the available  SM   predictions \cite{LHCb:2015hsa,LHCb:2026xvw,LHCb:2026dfm,LHCb:2018rym}.
As a consequence, advancing  further our theoretical description of these  exclusive semileptonic FCNC decays from first principles
will  be highly beneficial for deciphering the observed  intriguing  flavour patterns in an unambiguous fashion.

The model-independent and fully  quantitative formalism for evaluating the exclusive electroweak penguin
$B$-meson decay matrix elements has been  established robustly in the heavy-quark limit
with the diagrammatic factorization method \cite{Beneke:2001at,Beneke:2004dp,Grinstein:2004vb,Beylich:2011aq}
and with the modern effective field theory technique \cite{Chay:2003kb,Becher:2005fg,Ali:2006ew,Ali:2007sj}.
In contrast with QCD factorization for the hadronic two-body decays of  bottom mesons \cite{Beneke:1999br,Beneke:2000ry,Beneke:2001ev,Beneke:2003zv,Lu:2022kos},
the weak annihilation contribution  to  the exclusive  $b \to \left \{s, d \right \} \ell^{+} \ell^{-}$ decays
turns  out to be unsuppressed  in the ${\Lambda_{\rm QCD} / m_b}$ expansion
and hence appears in the perturbative factorization formula at leading power.
The next-to-leading-order (NLO)  computation  of  such  annihilation  amplitude
is evidently  indispensable for  achieving   the desired  factorization-scale invariance
of measurable quantities at order ${\cal O}(\alpha_s)$ \cite{Huang:2024xii}.
The isospin-breaking  power corrections to the transverse amplitude of
the semileptonic $B \to V \ell^{+} \ell^{-}$ decay  arising from  the virtual photon radiation off the  spectator quark
have been successfully  determined with  the soft-collinear  factorization  prescription \cite{Feldmann:2002iw,Beneke:2004dp}.
Subsequently, the non-factorizable  and power-suppressed charming penguin contributions to
the exclusive $B \to K^{(\ast)} \ell^{+} \ell^{-}$ decay amplitudes  were estimated from QCD light-cone sum rules
with the higher-twist $B$-meson distribution amplitudes in the region of spacelike momentum transfer \cite{Khodjamirian:2010vf,Khodjamirian:2012rm,Gubernari:2020eft,Gubernari:2022hxn},
by implementing  the dedicated  power-counting scheme $m_b \sim m_c \gg \sqrt{\Lambda_{\rm QCD} \, m_b} \gg \Lambda_{\rm QCD}$ \cite{Beneke:2000ry,Gambino:2012rd} (see, however,  \cite{Benzke:2010js,Kozachuk:2018yxf,Melikhov:2022wct,Belov:2023xqk,Belov:2024vkv,Qin:2022rlk,Huang:2023jdu,Bartocci:2024bbf,Bartocci:2026lhe}
for applications of an alternative counting scheme in the FCNC $B$-meson decays).
Extrapolating the QCD-based calculation  of  non-local hadronic matrix elements towards the physical kinematic region
can be further carried out by invoking a dispersion relation for the considered correlation function
in the variable of the dilepton invariant mass $q^2$, at the price of introducing an additional model dependence
of the particular ans\"{a}tz for the hadronic  spectral density above  the open charm threshold \cite{Khodjamirian:2023wol}.
Despite these encouraging   progresses on addressing the next-to-leading power corrections,
the truly  complete proof of QCD factorization for the exclusive $b \to \left \{s, d \right \} \ell^{+} \ell^{-}$
transition matrix elements cannot be  eventually  accomplished even at  leading-power accuracy
without  an in-depth  and  meticulous  analysis of  the dynamical  contribution from the non-leading Fock components of the $B$-meson
and the energetic light  meson (see e.g.,  \cite{Beneke:2000ry,Bosch:2003fc,Beneke:2003pa,Hill:2002vw,Ball:2003bf,Lange:2003pk,Wang:2015vgv}
for  discussions in distinct contexts).

In order to better explore the factorization structure of the exclusive electroweak penguin $B$-meson decay,
we will concentrate on  constructing  the  soft-collinear factorization theorem  for
$B \to P \, \ell^{+} \ell^{-}$  (with $P=\pi, \, K$) in the low-$q^2$ region with an emphasis on
the  non-valence Fock state contribution   in this Letter.
In particular, we will report on a novel observation of the leading-power contribution to
the $B \to P \, \ell^{+} \ell^{-}$ decay amplitude from  the  non-minimal partonic configuration of the bottom meson,
which can be  expressed in terms of the perturbatively calculable short-distance matching coefficients
and the three-particle $B$-meson distribution amplitude on the light-cone.
Importantly,  the yielding  convolution integral in the factorized expression for this new mechanism
can be further demonstrated to be  convergent  by taking advantage of the asymptotic behaviour of
the subleading twist $B$-meson distribution amplitude  of our interest.
Apart from obtaining the full infrared structure of the semileptonic $B \to P \, \ell^{+} \ell^{-}$  amplitude
in the leading-power approximation, we will  determine   analytically  three  distinct classes of
the next-to-next-to-leading-order (NNLO) radiative  corrections to the  hard and hard-collinear  functions
entering the QCD factorization formula for  the non-local $B \to P$  matrix elements
by applying the two-step matching  strategy  in soft-collinear effective theory (SCET) \cite{Bauer:2001yt,Bauer:2002nz,Beneke:2002ph,Beneke:2002ni} (see \cite{Becher:2014oda,Beneke:2015wfa}  for an overview).
Phenomenological implications of the  thus  obtained    leading-power   three-particle   contribution
and higher-order perturbative  corrections  to the exclusive  $B \to \left \{K, \pi \right \} \ell^{+} \ell^{-}$
decay observables will be  then investigated  comprehensively by  taking into account  the latest lattice QCD constraints
on  non-perturbative shape parameters for the light-cone distribution amplitudes of the light pseudoscalar mesons.

\section{General Analysis}

The low-energy effective weak Hamiltonian in the SM  for the semileptonic FCNC  $b \to D  \, \ell^{+} \ell^{-}$
(with $D=d, \, s$)  transition  takes the form \cite{Beneke:2001at,Beneke:2004dp}
\begin{widetext}
\begin{eqnarray}
{\cal H}_{\rm eff} &=&  - {G_F \over \sqrt{2}} \,  \left [   V_{u b} V_{u D}^{\ast} \,
\underbrace{\sum_{i=1}^{2} \, C_i \, \left ( {\cal Q}_i^{c} - {\cal Q}_i^{u} \right )}
+ \,  V_{t b} V_{t D}^{\ast} \, \underbrace{\left (  C_1  \, {\cal Q}_1^{c} + C_2  \, {\cal Q}_2^{c}
+ \sum_{i=3}^{10} \, C_i \, {\cal Q}_i  \right )} \right  ] + {\rm h.c.} \,,
\label{effective weak Hamiltonian}
\\
&& \hspace{3.5 cm} \equiv  {\cal H}_{\rm eff}^{(u)} \hspace{4.0 cm} \equiv  {\cal H}_{\rm eff}^{(t)}
\nonumber
\end{eqnarray}
\end{widetext}
where we employ further  the effective operator basis as proposed in \cite{Chetyrkin:1996vx,Chetyrkin:1997gb}
ensuring the consistent use of  anticommuting $\gamma_5$ matrices in  multi-loop computations.
The two  matrix elements of the semileptonic operators ${\cal Q}_{9, \, 10}$ can be readily expressed through
the heavy-to-light $B$-meson  decay form factor  and the (axial)-vector leptonic tensors.
When combined with  the electromagnetic  interactions  of  quarks and leptons,
both the  four-quark  operators  and the magnetic dipole operators can  contribute to
the  $B \to P \, \ell^{+} \ell^{-}$ transition amplitude.
It then becomes apparent that we are  required  to introduce
the  generalized  $B \to P \, \gamma^{\ast}$  form factors  defined by
\begin{eqnarray}
&& \left \langle P(p^{\prime}) \,  \gamma^{\ast}(q, \mu) \left |
{\cal H}_{\rm eff}^{(t, \, u)} \right | \bar B(p) \right \rangle
= - { g_{\rm em} \over 4  \pi^2} \, { m_{b, \, \rm {PS}} \over m_B} \,  T_{P}^{(t, \, u)}(q^2) \,
\nonumber \\
&& \hspace{2.0 cm}
\times \, \left [ q^2 \left ( p^{\mu} + p^{\prime \, \mu} \right )  - (m_B^2 - m_P^2) \, q^{\mu} \right ] \,,
\label{definition of non-local form factors}
\end{eqnarray}
where $m_{b, \, \rm {PS}}$  refers to the bottom-quark mass in the potential-subtracted renormalization scheme \cite{Beneke:1998rk}
and the factorizable  contribution from the magnetic penguin operator ${\cal Q}_{7}$
can be  represented by the customary  tensor form factor for the $B \to P$  transition.
The light flavour state  $| P \rangle$ denotes $| \pi^{-} \rangle$ ($| K^{-} \rangle$) for the  $B^{-}$-meson decay
and $-\sqrt{2} \, | \pi^{0} \rangle$ ($| \bar K^{0} \rangle$) for the $\bar B^{0}$-meson decay.
These non-local  corrections  to the  exclusive semileptonic  $B \to \left \{K, \pi \right \} \ell^{+} \ell^{-}$ decay amplitudes
can be conveniently  included  in the process- and $q^2$-dependent  ``effective"  Wilson coefficients
\begin{eqnarray}
{\cal C}_{9, \, P}^{(p)}(q^2) = C_9 \, \delta^{p  t} + {2  \,  m_{b, \, \rm {PS}} \over m_B} \,
\frac{T_{P}^{(p)}(q^2)}{f_{BP}^{+}(q^2)} \,,
\label{definition of effective Wilson coefficients}
\end{eqnarray}
where  the superscript  $p=t, \, u$  characterizes  the dedicated  Cabibbo-Kobayashi-Maskawa (CKM) structure of
the effective weak  Hamiltonian (\ref{effective weak Hamiltonian}).
Disentangling the multi-scale strong interaction  dynamics of our  problem   with the effective field theory technique
enables us to  derive  systematically  the  soft-collinear factorization formula for
the  emerged   hadronic  quantities  $T_{P}^{(t, \, u)}$ at leading power  in the  ${\Lambda_{\rm QCD} / m_b}$  expansion
\begin{widetext}
\begin{eqnarray}
T_{P}^{(t, \, u)} &=& {\cal C}_{\rm eff,\, A}^{(t, \, u)} (\bar n \cdot p^{\prime},  n \cdot q) \,\, f_{BP}^{+}(q^2)
\nonumber \\
&&   - \, 4 \, \pi^2 \, {{\cal F}_B \, f_P  \over m_{b, \, \rm {PS}}} \,
  \bigg  \{ \int_0^{\infty} \, {d \omega \over \omega }  \, \int_0^{1} \, d u \, \int_0^{1} \, d \tau  \,\,
{\cal C}_{\rm eff,\, B}^{(t, \, u)} (\bar n \cdot p^{\prime},  n \cdot q, \tau)  \,
{\cal \overline J}_{\rm B}(\bar n \cdot p^{\prime}, \omega, \tau, u)
\,\,  \phi_{B, \, +}(\omega) \,\, \phi_P(u)
\nonumber \\
&&
+ \, \int_0^{\infty} \, {d \omega \over \omega }  \, \int_0^{1} \, d u  \,\,
{\cal C}_{\rm eff,\, C}^{(t, \, u)} (\bar n \cdot p^{\prime},  n \cdot q,  u) \,\,
{\cal J}_{\rm C}^{(\rm 2P)}(n \cdot q, \bar n \cdot q,  \omega) \,\,  \phi_{B, \, -}(\omega) \,\, \phi_P(u)
\nonumber \\
&&
{\color{magenta}  + \,   \int_0^{\infty} \, {d \omega_1 \over \omega_1 }  \int_0^{\infty} \, {d \omega_2 \over \omega_{2}^{2} }
\, \int_0^{1} \, d u  \,\, {\cal C}_{\rm eff,\, C}^{(t, \, u)} (\bar n \cdot p^{\prime},  n \cdot q,  u) \,\,
\, {\cal J}_{\rm C}^{(\rm 3P)}(n \cdot q, \bar n \cdot q,  \omega_1,  \omega_2)  \,\,
\phi_{B, \,  3}(\omega_1,  \omega_2) \,\, \phi_P(u) }
\nonumber \\
&& {\color{blue} + \, \int_0^{\infty} \, {d \omega \over \omega }  \, \int_0^{1} \, d u \, \int_0^{1} \, d \tau
\,\,  {\cal C}_{\rm eff,\, D}^{(t, \, u)} (\bar n \cdot p^{\prime},  n \cdot q,  \tau,  u)
\,\, {\cal J}_{\rm D}(n \cdot q, \bar n \cdot q, \omega, \tau)
\,\,  \phi_{B, \, +}(\omega) \,\, \phi_P(u) } \bigg \}  \,,
\label{master factorization formula in SCET}
\end{eqnarray}
\end{widetext}
where the  hard-collinear  photon moves nearly in the direction of the light-cone vector $\bar n$
and the light $\pi$ ($K$) meson moves in the opposite light-cone direction $n$.
The non-perturbative QCD  fluctuations encoded in the two invariant amplitudes $T_{P}^{(t, \, u)}$  are
parameterized  by   the  physical $B \to P$  transition form factor  $f_{BP}^{+}$
\cite{Beneke:2000wa,Khodjamirian:2011ub,Lu:2018cfc,Cui:2022zwm}, decay constants ${\cal F}_B$ and $f_P$,
together with light-cone distribution amplitudes of the $B$-meson
and the energetic light meson $\phi_{X}$  \cite{Grozin:1996pq,Braun:2017liq,Lepage:1980fj,Efremov:1979qk} in the heavy quark limit.
The hard and hard-collinear  coefficient functions  ${\cal C}_{{\rm eff},\, i}^{(t, \, u)}$ and ${\cal J}_{k}$ arise from
the two different short-distance  scales $m_b$ and $\sqrt{m_b \, \Lambda_{\rm QCD}}$, respectively,
such that they can be computed in perturbation theory with the implementation of  the two-step matching program
${\rm QCD} \to {\rm SCET}_{\rm I} \to {\rm SCET}_{\rm II}$ in sequence.
In comparison with the classic diagrammatic analysis \cite{Beneke:2001at,Beneke:2004dp},
the distinctive feature of the factorized expression  (\ref{master factorization formula in SCET})
consists in  the very appearance of the higher Fock state contribution from
the three-particle $B$-meson distribution amplitude $\phi_{B, \, 3}$  
as well as  the additional leading-twist contribution
${\cal C}_{\rm eff,\, D}^{(t, \, u)} \star {\cal J}_{\rm D} \star \phi_{B, \, +} \star  \phi_P$
dependent on  the electric charge of the spectator quark in the bottom meson,
the latter of which has actually been determined  at the  ${\cal O}(\alpha_s)$ accuracy
in the QCD factorization framework \cite{Huang:2024xii}.

\section{Short-distance matching coefficients}

We are now in a position to compute the short-distance matching coefficients
by integrating out the hard and hard-collinear field modes of the QCD correlation functions
appearing in (\ref{definition of non-local form factors}).
To achieve this goal, we first establish  the  power counting scheme for the four-momenta
of the light meson and the off-shell photon
\begin{eqnarray}
p^{\prime}_{\mu} & \equiv   (n \cdot p^{\prime}, \,\, \bar n \cdot p^{\prime},  \,\,  p^{\prime}_{\perp \, \mu}) \,\,
& \sim (\lambda^4, \,\,  1, \,\,  \lambda^2) \,\, m_b  \,,
\nonumber \\
q_{\mu}  & \equiv   (n \cdot q, \,\, \bar n \cdot q,  \,\,  q_{\perp \, \mu}) \,\,
& \sim (1, \,\,  \lambda^2, \,\,  \lambda) \,\, m_b  \,,
\label{power counting scheme}
\end{eqnarray}
where the familiar  light-cone decomposition for an arbitrary vector quantity
$P_{\mu} \equiv  {n \cdot P \over 2}  \, \bar n_{\mu} + {\bar n \cdot P \over 2}  \,  n_{\mu} + P_{\perp \, \mu}$
is  consistently employed   and the expansion parameter $\lambda$ scales as $\sqrt{\Lambda_{\rm QCD} / m_b}$.
In addition to the two relevant  momentum modes defined above, we further need to introduce  the momentum scalings
$P_{{s}, \, \mu} \sim  (\lambda^2, \,\,  \lambda^2, \,\,  \lambda^2) \,\, m_b$
for the  soft  parton  of  the  $B$-meson,
$P_{{h}, \, \mu} \sim  (1, \,\,  1, \,\,  1) \,\, m_b$ for the external heavy $b$-quark in full QCD,
and $P_{\overline{{hc}}, \, \mu} \sim  (\lambda^2, \,\,  1, \,\,  \lambda) \,\, m_b$
for the internal anti-hard-collinear propagator due to  the gluon exchange
between the soft  spectator  quark and the  anti-collinear  quark.
A systematic decoupling of the strong interaction dynamics at the hard scale $m_b$
amounts to performing  the ${\rm QCD} \to {\rm SCET}_{\rm I}$ matching
for the effective weak  Hamiltonians ${\cal H}_{\rm eff}^{(t, \, u)} $
in the presence of  electromagnetic interactions
\begin{eqnarray}
{\cal H}_{\rm eff}^{(t, \, u)}
& \rightarrow &  \int d \hat{r} \, d \hat{s}   \,\,  \widetilde{{\cal C}}_{\rm eff,\, A}^{(t, \, u)}(\hat{r}, \hat{s})  \,\,
{\cal O}_{\rm  A}(r, s)
\nonumber \\
&& +  \int d \hat{r} \, d \hat{t}  \,  d \hat{s}   \,\,
\widetilde{{\cal C}}_{\rm eff, \, B}^{(t, \, u)}(\hat{r}, \, \hat{t}, \, \hat{s})  \,\,
{\cal O}_{\rm B}(r, t, s)
\nonumber \\
&&  + \,  \int d \hat{a} \, d \hat{t}   \,\,  \widetilde{{\cal C}}_{\rm eff, \, C}^{(t, \, u)}(\hat{a}, \, \hat{t})  \,\,
{\cal O}_{\rm C}(a, t)
\nonumber \\
&&  + \,  \int d \hat{a} \, d \hat{s}   \, d \hat{t}   \,\,
\widetilde{{\cal C}}_{\rm eff, \, D}^{(t, \, u)}(\hat{a}, \, \hat{s},  \, \hat{t})  \,\,
{\cal O}_{\rm D}(a, s, t) \,,
\label{QCD to SCET-I matching}
\end{eqnarray}
where we have defined the   dimensionless and boost-invariant  convolution variables
$\{ \hat r, \, \hat t \}  \equiv \{ r,  \, t\}  \, {m_B  \over  n \cdot v}$
and $\{ \hat a, \, \hat s \}  \equiv \{ a,  \, s\}  \, {m_B  \over  \bar n \cdot v}$.
The complete ${\rm SCET}_{\rm I} $ operator basis  relevant to our analysis   can be constructed as follows
\begin{eqnarray}
{\cal O}_{\rm  A}  &=&  m_b \,
\bigg  [ \widetilde{C}_{f_{+}}^{(\rm A0)}  \, (\bar \chi W_{\bar{c}})(r \, \bar n) \,  h_v(0)
 + \, {1 \over m_b} \,  \int d \hat t  \,
 \nonumber \\
&&  \widetilde{C}_{f_{+}}^{(\rm B1)}(\hat t) \,
(\bar \chi W_{\bar c})(r \, \bar n)  \,  ( W_{\bar c}^{\dagger} \, i \slashed {D}_{\perp \bar c} W_{\bar c} )(t \bar n) \,  h_v(0) \bigg ] \nonumber \\
&&  i \, g_{\rm em} \, n_{\alpha}  \, \bar n_{\beta} \, F_{c}^{\alpha \beta}(s n) \,,
\nonumber \\
{\cal O}_{\rm  B} &=&   \left [  (\bar \chi W_{\bar c})(r \, \bar n)  \,
( W_{\bar c}^{\dagger} \, i \slashed {D}_{\perp \bar c} W_{\bar c} )(t \bar n) \,  h_v(0)  \right ]
\nonumber \\
&& i \, g_{\rm em} \, n_{\alpha}  \, \bar n_{\beta} \, F_{c}^{\alpha \beta}(s n)  \,,
\nonumber  \\
{\cal O}_{\rm  C} &=&  \left [  (\bar \chi W_{\bar c})(t \, \bar n)  \,  { \slashed {\bar n} \over 2}  \,
(1 - \gamma_5)  \,  (W_{\bar c}^{\dagger} \, \chi)(0)  \right ]
\nonumber \\
&& \left [ (\bar \xi W_c)( a n) \, \slashed{n} (1 - \gamma_5)  \, h_v(0) \right ],
\nonumber \\
{\cal O}_{\rm  D} &=&  {1 \over m_b} \, \left [  (\bar \chi W_{\bar c})(t \, \bar n)  \,  { \slashed {\bar n} \over 2}  \,
(1 - \gamma_5)  \,  (W_{\bar c}^{\dagger} \, \chi)(0) \right ]
\nonumber \\
&&   \hspace{-0.80 cm}  \left [ (\bar \xi W_c)( a n) \, {\slashed{n} \over 2 }   \,
( W_{c}^{\dagger} \, i \slashed {D}_{\perp c} W_{c} )(s n) \, (1 + \gamma_5) \, h_v(0) \right ],
\end{eqnarray}
where the square bracket in the ${\rm A}$-type effective operator has been  adjusted to ensure  that
its matrix element coincides with the full QCD form factor  $f_{BP}^{+}$ (apart from  an overall factor of $\bar n \cdot p^{\prime}$)
and the elementary building block  $F_{c}^{\alpha \beta}$  stands for  the hard-collinear  photon field-strength tensor.
In accordance with  the  established  SCET conventions for exclusive heavy-meson  decays \cite{Beneke:2004rc,Beneke:2005gs,Beneke:2005vv},
the  hard-collinear and anti-hard-collinear quark  fields moving into the  light-ray  directions of
$\bar n$ and $n$ are labelled by $\xi$ and $\chi$, which evidently fulfill the  equations-of-motion  constraints
$\slashed{\bar n}  \, \xi=0$  and $\slashed{n}  \, \chi=0$.
Moreover,  the two light-cone Wilson lines  $W_{c}$ and $W_{\bar c}$    have  been introduced  to
maintain  the (anti)-hard-collinear gauge invariance for  the effective operators \cite{Beneke:2002ph,Beneke:2003pa}.
The momentum-space coefficient functions ${\cal C}_{{\rm eff},\, i}^{(t, \, u)}$
in the factorization formula (\ref{master factorization formula in SCET}) can  be obtained from the Fourier transformation
of the position-space coefficient functions $\widetilde{{\cal C}}_{{\rm eff},\, i}^{(t, \, u)}$
\begin{eqnarray}
&& {\cal C}_{\rm eff,\, A}^{(t, \, u)} (\bar n \cdot p^{\prime},  n \cdot q)
=  \int d \hat{r} \,  d \hat{s}  \,\, e^{ i \, (\bar n \cdot \hat{p}^{\prime}  \, \hat{r} + n \cdot \hat{q} \, \hat{s}) }  \,\,
\widetilde{{\cal C}}_{\rm eff,\, A}^{(t, \, u)}(\hat{r}, \hat{s}),
\nonumber  \\
&& {\cal C}_{\rm eff,\, B}^{(t, \, u)} (\bar n \cdot p^{\prime},  n \cdot q, \tau)
=  \int d \hat{r}  \,  d \hat{t}   \,   d \hat{s}
\,\, e^{ i \, \left [ \bar n \cdot \hat{p}^{\prime}  \, \left ( \bar \tau \,  \hat{r} +  \tau \, \hat{t} \right )
+ n \cdot \hat{q} \, \hat{s} \right ] }
\nonumber \\
&& \hspace{3.5 cm}   \widetilde{{\cal C}}_{\rm eff,\, B}^{(t, \, u)}(\hat{r}, \hat{t} , \hat{s}),
\nonumber \\
&& {\cal C}_{\rm eff,\, C}^{(t, \, u)} (\bar n \cdot p^{\prime},  n \cdot q, u)
=  \int d \hat{a} \,  d \hat{t}  \, e^{ i \, (n \cdot \hat{q} \, \hat{a} + u \, \bar n \cdot \hat{p}^{\prime} \, \hat{t}) }  \,
\widetilde{{\cal C}}_{\rm eff,\, C}^{(t, \, u)}(\hat{a}, \hat{t}),
\nonumber \\
&& {\cal C}_{\rm eff,\, D}^{(t, \, u)} (\bar n \cdot p^{\prime},  n \cdot q, \tau, u)
=  \int d \hat{a}  \,  d \hat{s}   \,   d \hat{t}
\,\, e^{ i \, \left [ n \cdot \hat{q}  \, \left ( \bar \tau \,  \hat{a} +  \tau \, \hat{s} \right )
+ u \, \bar n \cdot \hat{p}^{\prime} \, \hat{t} \right ] }
\nonumber \\
&& \hspace{3.5 cm}  \widetilde{{\cal C}}_{\rm eff,\, D}^{(t, \, u)}(\hat{a}, \hat{s} , \hat{t}),
\end{eqnarray}
where  $\{ \bar n \cdot \hat{p}^{\prime}, \,  n \cdot  \hat{q} \} \equiv
\{ \bar n \cdot p^{\prime} \, n \cdot v, \,\,  n \cdot q \, \bar n \cdot v   \} / m_B$,
and  the ``bar notation"  $\bar \tau \equiv 1 - \tau$ is further  employed  for the convolution variable $\tau$
which corresponds to  the momentum fraction carried by the transverse (anti)-hard-collinear gluon.
These hard-scattering kernels can be expanded perturbatively  in terms of the strong coupling $\alpha_s$
(similarly for any other QCD quantity)
\begin{eqnarray}
{\cal C}_{{\rm eff},\, i}^{(t, \, u)} = \sum_{\ell =0}^{\infty}  \, \left ( {\alpha_s \over 4 \, \pi} \right )^{\ell} \,\,
{\cal C}_{{\rm eff},\, i}^{{(\ell)}, \, (t, \, u)} \,.
\end{eqnarray}
The leading-order (LO) and NLO  expressions  for the hard matching coefficients ${\cal C}_{{\rm eff},\, i}^{(t, \, u)}$
(with an exception of the ${\rm D}$-type correction at ${\cal O}(\alpha_s)$)
can be inferred from  \cite{Beneke:2001at,Beneke:2004dp,Huang:2024xii}.
We then proceed to improve on  the   previous factorization analysis by  extracting analytically
two  new  coefficient  functions  ${\cal C}_{{\rm eff},\, {\rm C}}^{{(2)}, \, (t, \, u)}$
and ${\cal C}_{{\rm eff},\, {\rm D}}^{{(1)}, \,  (t, \, u)}$ in Supplemental Material,
which constitute   the  necessary  ingredients for   continually pinning down the  theoretical  uncertainties
of  phenomenological observables in the exclusive semileptonic  $B \to \left \{K, \pi \right \} \, \ell^{+} \ell^{-}$   decays.

Subsequently,  we implement   the second-step   ${\rm SCET}_{\rm I} \to {\rm SCET}_{\rm II}$ matching procedure
for the effective operators ${\cal O}_{i}$ by integrating out  the (anti)-hard-collinear  fluctuations
from  the dynamical  scale $\sqrt{m_b \, \Lambda_{\rm QCD}}$.
Bearing in mind that the hard-collinear and  anti-hard-collinear fields are
already decoupled at the hard scale \cite{Chay:2003ju,Bauer:2004tj},
the  matrix elements of ${\cal O}_{\rm C}$ and ${\cal O}_{\rm D}$ fall apart into two individual sectors each.
Consequently, we are led to the  four  non-local  ${\rm SCET}_{\rm I}$  matrix elements  below
\begin{widetext}
\begin{eqnarray}
&&  \bar n \cdot \hat{p}^{\prime}  \, \int { d \hat{t} \over  2 \pi}  \,
e^{- i \, \tau \, \bar n \cdot \hat{p}^{\prime} \, \hat{t}}  \,
\left \langle P(p^{\prime})   \left |   (\bar \chi W_{\bar c})(0)  \,
( W_{\bar c}^{\dagger} \, i \slashed {D}_{\perp \bar c} W_{\bar c} )(t \bar n) \,  h_v(0) \right | \bar B(p) \right  \rangle
= m_b \,\,  \bar n \cdot p^{\prime} \,\,  \Xi_P(\tau, \bar n \cdot p^{\prime}),
\nonumber \\
&&   \bar n \cdot \hat{p}^{\prime}  \,
\int  { d \hat{t} \over  2 \pi}  \,  e^{- i \, u \, \bar n \cdot \hat{p}^{\prime} \, \hat{t}}  \,
\left \langle P(p^{\prime})   \left |  (\bar \chi W_{\bar c})(t \, \bar n)  \,
{ \slashed {\bar n} \over 2}  \, (1 - \gamma_5)  \,  (W_{\bar c}^{\dagger} \, \chi)(0)   \right | 0 \right  \rangle
= i \, f_P \, {\bar n \cdot p^{\prime} \over 2} \, \phi_P(u),
\nonumber \\
&&  \int d^4 x \, e^{i q \cdot x}  \, \left \langle 0  \left |
{\rm T}  \left \{ j_{\rm em}^{\mu}(x), \,  (\bar \xi W_c)(0) \,  \gamma_5  \, h_v(0)  \right \}
\right | \bar B(p) \right  \rangle
=  \left [ {\cal T}_{\gamma^{\ast}}^{(\rm 2P)}(n \cdot q)
+ {\color{magenta}  {\cal T}_{\gamma^{\ast}}^{(\rm 3P)}(n \cdot q) }   \right ] \, \bar n^{\mu},
\nonumber \\
&&  n \cdot \hat{q}  \int { d \hat{s} \over  2 \pi}
e^{- i \,  \tau \,  n \cdot \hat{q} \, \hat{s}  }  \, \int d^4 x \, e^{i q \cdot x}  \,
\left \langle 0  \left |  {\rm T}  \left \{j_{\rm em}^{\mu}(x), \,
(\bar \xi W_c)(0) \,  \gamma_5   \, ( W_{c}^{\dagger} i \slashed {D}_{\perp c} W_{c} )(s n)
\,  h_v(0) \right \} \right | \bar B(p) \right  \rangle
= {\color{blue} \Xi_{\gamma^{\ast}}(\tau, n \cdot q) } \,\,  \bar n^{\mu}.
\hspace{1.0  cm}
\end{eqnarray}
\end{widetext}
The ${\rm SCET_I}$ representation for the electromagnetic current of our interest  can be written as
\begin{eqnarray}
j_{\rm em}^{\mu}(x) &=&
\sum_{f} Q_f \, \left [ \bar \xi(x) \, { \slashed{n} \over 2}  \,   \xi(x) \right ]  \, \bar n^{\mu}
\nonumber \\
&& + \, \sum_{f} Q_f \, \left [  \bar  q_s(x_{-}) \,  {\slashed{n}  \over 2 } \,  (W_{c}^{\dagger} \xi)(x)  \right ]
\, \bar n^{\mu}
\nonumber \\
&\equiv&  \sum_{f} Q_f \,  \left [  j_{\rm em}^{(0)}(x)  + j_{\rm em}^{(2)}(x) \right ] \,\, \bar n^{\mu}  \,,
\end{eqnarray}
where the soft quark field $\bar q_s$ is  evaluated at the position of
$x_{-}^{\nu} \equiv  \left ( n \cdot x \right )  \, {\bar n^{\nu} /  2}$
due to the multipole expansion.
The soft-collinear factorization  formula  of  the  ${\rm B}$-type  effective   form factor
$\Xi_P = \left ( {m_B \over 4 \, m_b} \right ) \,  \mathbb{\overline J}_{\rm B} \, \star  \left (  {\cal F}_B \, \phi_{B, \, +} \right )
\, \star  \left (f_P  \, \phi_{P} \right )$ has been constructed for  investigating  the  spectator-scattering  corrections
to heavy-to-light $B$-meson form factors at large hadronic recoil \cite{Beneke:2005gs,Becher:2004kk,Hill:2004if}.
The   non-hadronic    radiative   form factors  ${\cal T}_{\gamma^{\ast}}^{(\rm 2P)}$  and $\Xi_{\gamma^{\ast}}$
that parameterize  the  effective   $B \to \gamma^{\ast}$  matrix elements of the two-body and three-body  ${\rm SCET_I}$  operators
have also been determined  in \cite{Huang:2024xii} at the ${\cal O}(\alpha_s)$ accuracy
\begin{eqnarray}
{\cal T}_{\gamma^{\ast}}^{(\rm 2P)}  &=&  \frac{{\cal F}_B \, m_B}{2} \, \int_0^{\infty}   d \omega  \,
\mathbb{J}_{\rm C}^{(\rm 2P)}(n \cdot q, \bar n \cdot q,  \omega) \,  \phi_{B, \, -}(\omega),
\nonumber \\
\Xi_{\gamma^{\ast}} &=&  \frac{{\cal F}_B \, m_B}{2} \, \int_0^{\infty}   d \omega  \,
\mathbb{J}_{\rm D} (n \cdot q, \bar n \cdot q,  \omega, \tau) \,  \phi_{B, \, +}(\omega).
\hspace{0.80 cm}
\end{eqnarray}
It is then straightforward to express  the  three  jet  functions  ${\cal \overline J}_{\rm B}$,
${\cal J}_{\rm C}^{(\rm 2P)}$  and ${\cal J}_{\rm D}$ entering the  QCD  factorization formula
(\ref{master factorization formula in SCET}) in terms of the short-distance   matching coefficients
$\mathbb{\overline J}_{\rm B}$,  $\mathbb{J}_{\rm C}^{(\rm 2P)}$  and  $\mathbb{J}_{\rm D}$ defined  above,
whose lengthy expressions  are summarized explicitly in Supplemental Material.
It remains important to point out that we have  taken into account the particular  NNLO  corrections to
the  ${\rm B}$-type contributions of  the invariant amplitudes $T_{P}^{(t, \, u)}$
from the  anti-hard-collinear quantum fluctuations,  by  incorporating further the  order  ${\cal O}(\alpha_s^2)$ effect
in the renormalized  coefficient function   ${\cal \overline J}_{\rm B}$.

\begin{figure}[h!]
\begin{center}
\includegraphics[width=1.0 \columnwidth]{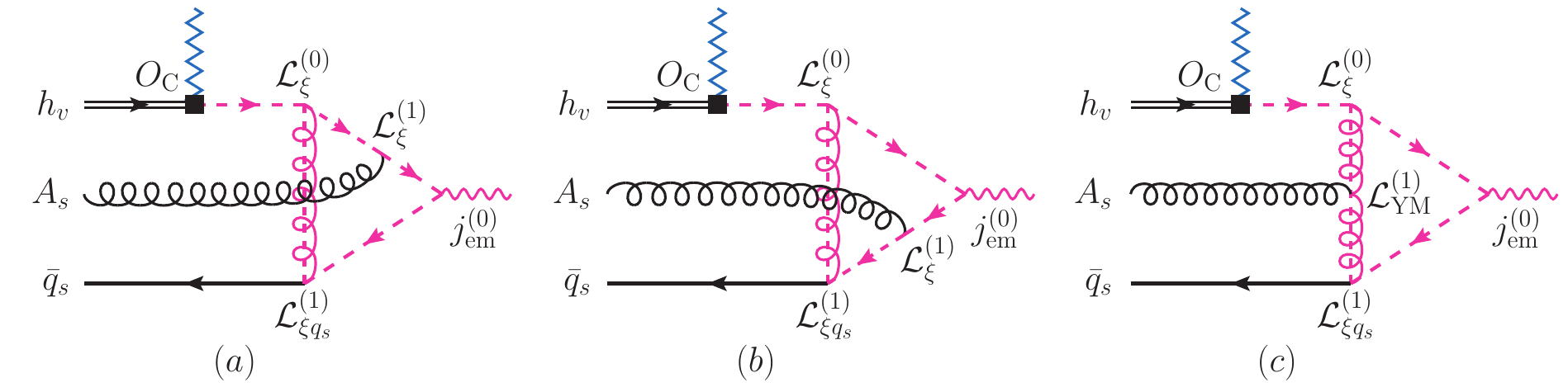}
\caption{Sample one-loop  Feynman diagrams for the  non-local  radiative form factor  ${\cal T}_{\gamma^{\ast}}^{(\rm 3P)}$
defined in  (\ref{SCET-I correlator of the 3P effect}).
The square box marks  an insertion of the effective   heavy-to-light  current $O_{\rm C}(0) = (\bar \xi W_c)(0) \,  \gamma_5  \, h_v(0)$.
The leading-power term  of the electromagnetic current in ${\rm SCET}_{\rm I}$   reads
$j_{\rm em}^{(0)}(x) =  \bar \xi(x) \, { \slashed{n} \over 2}  \,   \xi(x)$.}
\label{fig: sample SCET diagrams for the 3P effect}
\end{center}
\end{figure}

We can now turn to derive the soft-collinear  factorization formula for  the non-valence Fock state contribution
of the ${\rm C}$-type ${\rm SCET}_{\rm I}$  correlation function
from the three-particle twist-three $B$-meson distribution amplitude.
This essential new result can be achieved   by first identifying  the leading-power contribution to
the  effective  form factor  ${\cal T}_{\gamma^{\ast}}^{(\rm 3P)}$  in the heavy quark expansion
\begin{widetext}
\begin{eqnarray}
{\cal T}_{\gamma^{\ast}}^{(\rm 3P)}
&=&  \int d^4 x \, e^{i q \cdot x}  \, \int d^4 y \,  \int d^4z   \,
 \left \langle 0  \left | {\rm T}  \left \{ j_{\rm em}^{(0)}(x), \,\,  i {\cal L}_{\xi q_s}^{(1)}(y),  \,\,
i {\cal L}_{\xi}^{(1)}(z),  \,\, (\bar \xi W_c)(0) \,  \gamma_5  \, h_v(0)  \right \}
\right | \bar B(p) \right  \rangle
\nonumber \\
&& + \, \int d^4 x \, e^{i q \cdot x}  \, \int d^4 y \,  \int d^4z   \,
 \left \langle 0  \left | {\rm T}  \left \{ j_{\rm em}^{(0)}(x), \,\,  i {\cal L}_{\xi q_s}^{(1)}(y),  \,\,
i {\cal L}_{\rm YM}^{(1)}(z),  \,\, (\bar \xi W_c)(0) \,  \gamma_5  \, h_v(0)  \right \}
\right | \bar B(p) \right  \rangle,
\label{SCET-I correlator of the 3P effect}
\end{eqnarray}
\end{widetext}
where the first order power-suppressed interaction terms ${\cal L}_{\xi q_s}^{(1)}$, ${\cal L}_{\xi}^{(1)}$
and ${\cal L}_{\rm YM}^{(1)}$  in the ${\rm SCET}_{\rm I}$ Lagrangian have already been  presented in \cite{Beneke:2002ni}
(see \cite{Beneke:2018rbh}  for the manifest  expressions of their Feynman rules).
Unsurprisingly,  the  three-particle  Fock component of  the heavy-meson wave function under discussion
can  only  bring about  the leading (in the ${\Lambda_{\rm QCD} / m_b}$  expansion)  contribution to
the rare  $B \to P \, \ell^{+} \ell^{-}$ decay amplitude  starting at  one-loop order.
Three sample Feynman diagrams describing the lowest-order contribution to
the exclusive radiative  form factor ${\cal T}_{\gamma^{\ast}}^{(\rm 3P)}$
are depicted in Figure  \ref{fig: sample SCET diagrams for the 3P effect} explicitly.

Matching  the emerged  ${\rm SCET}_{\rm I}$ correlation function in  (\ref{SCET-I correlator of the 3P effect})
onto  ${\rm SCET}_{\rm II}$  then  allows us to write down  the  desired  soft-collinear  factorization formula
at leading power
\begin{eqnarray}
{\cal T}_{\gamma^{\ast}}^{(\rm 3P)}
&=& {\mathcal{F}_B \, m_B \over 2}  \, \int_0^{\infty} \, d \omega_1  \,   \int_0^{\infty} \, d \omega_2
\nonumber \\
&&  \mathbb{J}_{\rm C}^{(\rm 3P)}(n \cdot q, \bar n \cdot q,  \omega_1, \omega_2)
\,\,  \phi_{B, \, 3}( \omega_1, \omega_2)  \,.
\label{factorization formuula of the 3P effect}
\end{eqnarray}
The  resulting  jet  function  can be cast in the form
\begin{eqnarray}
\mathbb{J}_{\rm C}^{(\rm 3P)} &=& {1 \over \omega_1 \, \omega_2^2} \,\,
{\alpha_s \over 4 \, \pi}
\left  [   ( 2 \, C_F  - C_A) \,   \mathbb{K}_{\rm I}
+ C_A  \,  \mathbb{K}_{\rm II}  \right ]
 +   {\cal O}(\alpha_s^2)   \,,
\nonumber \\
\mathbb{K}_{\rm I} &=&  \left [ 2  \, \ln { {\hat \mu}^2  \over \omega  - \bar n \cdot q - i \, 0}
+ 2 \, \ln (1 + \eta)  + 5  \right ]  \,
\nonumber \\
&&  \times \, \bigg [ {\bar v}^2  \, \ln (1 + \eta) - (1 - v \, \tilde{\eta}) \, \ln (1 + \bar v \, \eta)
+ v \, \bar v \, \tilde{\eta} \bigg ]
\nonumber \\
&& -  \,   {\bar v}^2  \, \ln^2  (1 + \eta) + (1 - v \, \tilde{\eta}) \, \ln^2 (1 + \bar v \, \eta)
\nonumber \\
&& - \, \ln (1 + \eta) +  \ln  (1 +  v \, \eta) + (1 - v \, \tilde{\eta}) \,  \ln (1 + \bar v \, \eta) \,,
\nonumber \\
\mathbb{K}_{\rm II} &=&  \left [ 2  \, \ln { {\hat \mu}^2  \over \omega  - \bar n \cdot q - i \,  0}
+ 2 \, \ln (1 + \eta)  + 5  \right ]  \,
\nonumber \\
&&  \times \, \bigg [ ({\bar v}^2 - 1) \, \ln (1 + \eta) +  \ln (1 + v \, \eta) + v \, \bar v \, \tilde{\eta} \bigg ]
\nonumber \\
&& + \, (1 - {\bar v}^2) \, \ln^2  (1 + \eta) -  \ln^2 (1 +  v \, \eta)
\nonumber \\
&&  - 2 \, \bar v \, \tilde{\eta}  \, \ln (1 + v \, \eta)  \,,
\label{result of the 3P jet function}
\end{eqnarray}
where we have employed  the following conventions
\begin{eqnarray}
{\hat{\mu}}^2  & \equiv & {\mu^2  \over n \cdot q} \,,  \qquad    \omega  \equiv \omega_1 + \omega_2 \,,
\qquad  \eta  \equiv   - { \omega  \over \bar n \cdot q}   \,,
\nonumber \\
\tilde{\eta} & \equiv &  {\eta  \over 1 + \eta}  \,,   \qquad   v \equiv  { \omega_1 \over \omega_1+ \omega_2} \,,
\qquad   \bar v \equiv 1 - v   \,.
\end{eqnarray}
In addition,  we  have endeavoured to  determine   this  one-loop  jet  function with an alternative  strategy
by evaluating the appropriate    full diagrams for the three-particle correction  to the QCD matrix element
$ \left \langle \gamma^{\ast} (q, \mu)  \left | \bar q  \,  \gamma_5 \, \slashed{n}  \, b  \right | \bar B(p) \right  \rangle$
and by  merely  extracting  the hard-collinear contributions from  these diagrams with the method of regions \cite{Beneke:1997zp,Smirnov:1998vk,Smirnov:2002pj,Ma:2025emu}.
It turns out that  the thus obtained   matching coefficient  $\mathbb{J}_{\rm C}^{(\rm 3P)}$
reproduces  precisely the  analytic   result  displayed in  (\ref{result of the 3P jet function}),
providing further a strong validation of our SCET  calculation.
On the basis of the asymptotic behaviours  of the three-particle distribution amplitude  $\phi_{B, \, 3}$
at  large and small quark/gluon momenta  \cite{Braun:2017liq,Braun:2015pha},
we can readily verify that  the soft convolution  integrations over  the two soft variables
$\omega_1$ and $\omega_2$  in the factorized expression (\ref{factorization formuula of the 3P effect})
are indeed convergent at their endpoints.   We can then continue to  demonstrate  that the yielding  ${\rm C}$-type contribution
to the  non-local  $B \to P \, \ell^{+} \ell^{-}$  decay amplitude  is  truly independent of
the  factorization scale at order ${\cal O}(\alpha_s)$ with the inclusion of
the  effective form factor  ${\cal T}_{\gamma^{\ast}}^{(\rm 3P)}$,
by applying the one-loop  renormalization-group (RG)  equation for the twist-three
$B$-meson   distribution amplitude $\phi_{B, \, -}$  \cite{Descotes-Genon:2009jif}.

\section{Numerical analysis}

We are now prepared to explore phenomenological implications  of our improved  QCD   factorization formula
for the exclusive rare  $B \to P \, \ell^{+} \ell^{-}$  decay amplitude,
including  the  identified leading-power  contributions  from the non-leading Fock state of the $B$-meson
and the yet higher-order radiative corrections to three short-distance matching coefficients ${\cal \overline J}_{\rm B}$,
${\cal C}_{{\rm eff},\, {\rm C}}^{(t, \, u)}$ and ${\cal C}_{{\rm eff},\, {\rm D}}^{(t, \, u)}$.
It is then instructive to employ an attractive  nonperturbative   model for both the leading-twist
and higher-twist $B$-meson distribution amplitudes  introduced in \cite{Beneke:2018wjp}
(see \cite{Wang:2021yrr}  for the further extension to  twist-six accuracy
and \cite{Wang:2019msf,Han:2024fkr,LatticeParton:2024zko,LPC:2026ffe,LPC:2026vyv,Giusti:2023pot,Giusti:2025ibe}
for the alternative strategies of extracting these fundamental non-perturbative functions on the lattice).
The two lowest  Gegenbauer moments for  the leading-twist pion and kaon distribution amplitudes
are determined from the  modern lattice QCD simulation \cite{RQCD:2019osh} with $N_f=2+1$ flavours of
dynamical Wilson-clover fermions and with the three-loop conversion from the regularization-invariant symmetric
momentum subtraction (${\rm RI^{\prime}/SMOM}$) scheme to the  $\overline {\rm MS}$ scheme \cite{Kniehl:2020sgo,Kniehl:2020nhw}.
Moreover, the   state-of-the-art  lattice calculations for  the leptonic decay constants  of  the pion, kaon and  $B$-meson in QCD \cite{FlavourLatticeAveragingGroupFLAG:2024oxs} (see e.g., \cite{Gasser:2010wz,Carrasco:2015xwa,Cornella:2026lkp}
for additional discussions on the structure-dependent QED corrections to  these  hadronic quantities)
will be  adopted in the subsequent numerical analysis.
An all-order summation of parametrically large  logarithms of $m_b / \Lambda_{\rm QCD}$
in the  perturbative  factorization formula  (\ref{master factorization formula in SCET})
can be   analytically performed   with  the momentum-space evolution equations
for the short-distance matching coefficients \cite{Huang:2024xii}
and  for the  pertinent  light-cone distribution amplitudes of the $B$-meson \cite{Braun:2019wyx,Liu:2020ydl,Galda:2020epp,Braun:2015pha,Braun:2017liq}
and the light pseudoscalar meson \cite{Braun:2017cih,Strohmaier:2018tjo}.
We  further  prefer to  employ  the updated  numerical  predictions  for  the physical  form factors
$f_{B \pi}^{+}$ and $f_{B K}^{+}$   obtained by fitting the  Bourrely-Caprini-Lellouch (BCL)
parametrization \cite{Bourrely:2008za} against the SCET light-cone sum rule determinations \cite{Cui:2022zwm}
(see \cite{DeFazio:2005dx,DeFazio:2007hw,Gao:2019lta,Gao:2021sav,Cui:2023jiw,Wang:2017jow,Wang:2016qii}
for numerous applications in heavy quark decays)
and the existing lattice QCD data points \cite{FermilabLattice:2015mwy,Flynn:2015mha,FermilabLattice:2015cdh,Bailey:2015dka}.
The allowed numerical intervals for the remaining theory inputs
are identical to  those   summarized in  \cite{Shen:2020hfq,InputParameter:2026}.

\begin{figure}[h!]
\begin{center}
\includegraphics[width=0.95  \columnwidth]{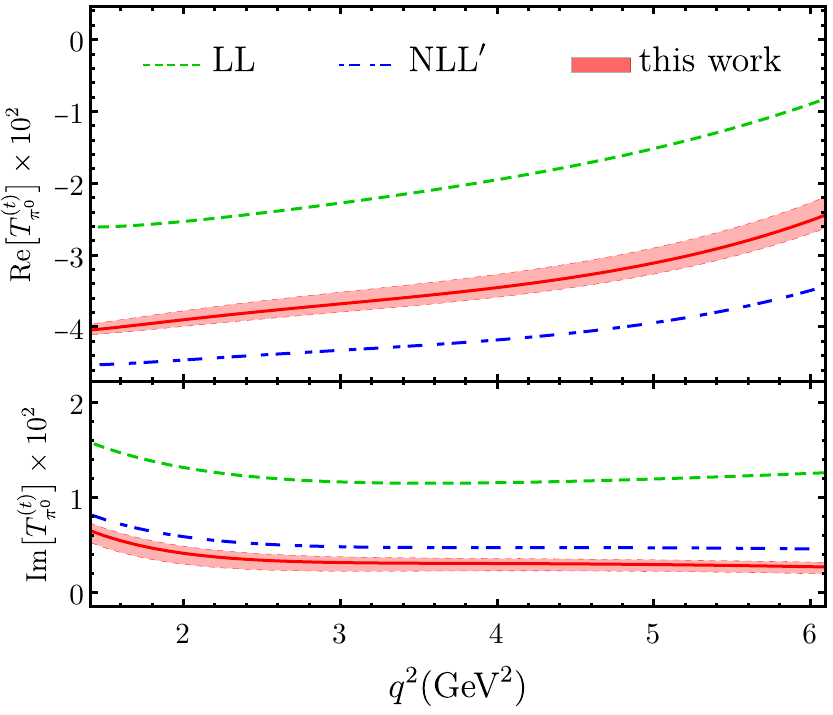}
\caption{Theory predictions of   the  invariant amplitude  $T_{\pi^{0}}^{(t)}$
for the exclusive rare  $\bar B^{0} \to \pi^{0} \, \ell^{+} \ell^{-}$ decay
from our updated  SCET  computations as described in the text (pink curves),
where the  uncertainty bands  are  obtained from varying the hard-collinear scale.
We further display the  numerical  predictions   for this  quantity
at the LL order (green curves) and at the ${\rm NLL}^{\prime}$ order (blue curves) for an exploratory comparison.}
\label{fig: numerical result for the invariant amplitude}
\end{center}
\end{figure}

In order to  facilitate an in-depth exploration  of the distinct  strong interaction  mechanisms
dictating the exclusive   radiative $B \to P \, \gamma^{\ast}$  form factors,
we explicitly  display  in Figure  \ref{fig: numerical result for the invariant amplitude}
the yielding  theory   predictions for   the   sample  invariant function  $T_{\pi^{0}}^{(t)}$
of  the  electroweak   penguin  $\bar B^{0} \to \pi^{0} \, \ell^{+} \ell^{-}$  decay
obtained with  the leading-logarithmic (LL) and   next-to-leading-logarithmic (${\rm NLL}^{\prime}$) approximations
as well as  with adding  further  the newly  determined    leading-power contributions discussed above \cite{NLLprime:2026}.
It is  then evident that  the very inclusion of the  unsuppressed  non-valence   Fock state  contribution
and the higher-order perturbative   corrections   to  the  short-distance coefficient functions
can  generate   the  noticeable  impacts on the ${\rm NLL}^{\prime}$  QCD  predictions for both the real and imaginary parts of
the decay  amplitude $T_{\pi^{0}}^{(t)}$ in the top-sector:
numerically at the level of ${\cal O}\, (30 \, \%)$ in the kinematic range  $q^2 \in  \left [1.5, \,  4.0 \right ] \, {\rm GeV^2}$.
This distinctive feature  can be attributed to the  substantial  cancellation of the  dominant first-order corrections
from  the  dedicated  ${\rm A}$-type and  ${\rm B}$-type  matrix elements
and to the  even more pronounced  suppression   of the leading  form-factor contributions
from the four-quark operators ${\cal Q}_i^{(c)}$ (with $i=1,..., 6$)  and the  magnetic dipole  operator ${\cal Q}_{7}$.
It then  becomes  straightforward to   verify  that  the  achieved  results   for the  other  three  amplitudes
$\left \{ T_{\pi^{-}}^{(t)}, \,  T_{K^{-}}^{(t)}, \, T_{\bar K^{0}}^{(t)}  \right \}$
in the top-sector  can be again  prominently affected when taking into account
the previously missing  leading-power  effects  in the heavy quark expansion.
As far as  the  four  invariant functions   in the up-sector are concerned,
the obtained  predictions  for  the  numerically  suppressed    amplitudes    ${\rm Re}   \left [T_{\pi^{0}}^{(u)}  \right ] $
and ${\rm Re}  \left  [T_{\bar K^{0}}^{(u)} \right ]$  of   the neutral $B$-meson decay channels
turn out to be  most sensitive to the   spectator-scattering contributions   derived in this work.

\begin{figure}[h!]
\begin{center}
\includegraphics[width=0.95 \columnwidth]{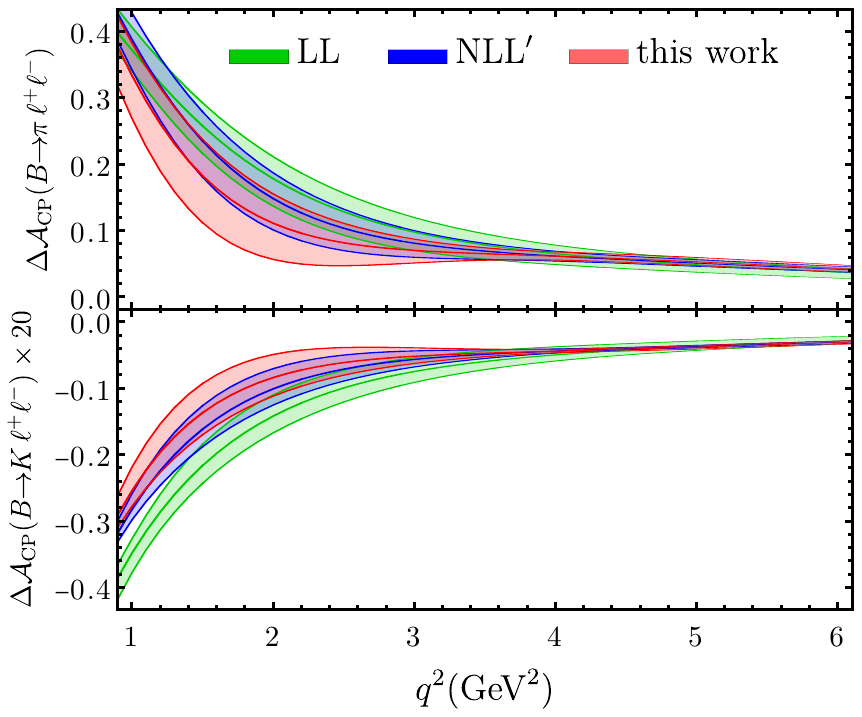}
\caption{Theory predictions of the CP asymmetry differences for   $B \to \pi  \,  \ell^{+} \ell^{-}$
(upper panel) and $B \to K  \,  \ell^{+} \ell^{-}$  (lower  panel)  obtained from the leading-power   factorization  analysis
at the LL order (green curves) and at the ${\rm NLL}^{\prime}$ order (blue curves) and from the further inclusion of
the various   spectator-scattering  corrections  discussed  in the text (pink curves).
The combined  uncertainties from adding  all separate errors in quadrature are indicated by the colour bands.}
\label{fig: numerical result for the CP asymmetry differences}
\end{center}
\end{figure}

Subsequently,   we turn to  investigate   numerical   impacts  of  these   newly  computed    non-local hadronic corrections
to the  invariant  amplitudes  on the  experimental observables of  the  electroweak penguin
$B \to P \,  \ell^{+} \ell^{-}$ decays.
It is plainly  not unexpected that   our  updated   predictions   for  the CP-averaged branching   fractions
of both the charged and neutral  bottom-meson decays   appear to be   marginally different from  those
determined in the previous  analysis   \cite{Huang:2024xii}.
As a matter of fact,  the smallness of the considered ${\cal O}(\alpha_s^2)$  corrections to   the ${\rm C}$-type and ${\rm D}$-type
matrix   elements   can be  already  anticipated   from  the  destructive interference of the  various    NNLO   terms
in  non-leptonic  two-body   $B$-meson  decay  amplitudes \cite{Beneke:2009ek,Bell:2020qus}.
We further observe that   the CP-averaged isospin-breaking effects  in  both the $B \to K  \, \ell^{+} \ell^{-}$
and  $B \to \pi  \, \ell^{+} \ell^{-}$  decay modes  turn out to be   rather stable against
the currently  obtained  higher-twist and higher-order   corrections,    thus   enabling these flavour  asymmetries
to become   theoretically clean  probes of the  intricate  spectator interactions.
In an attempt to  characterize  the  numerical  significance of these  dynamical corrections  for   CP-violating  asymmetries,
it proves to be  most  advantageous to concentrate   on   the following   quantities    defined by
the differences   of  the direct CP asymmetries between the charged and   neutral  $B$-meson   decay modes
(in  close  analogy to the  prominent observable  $\Delta {\cal A}_{\rm CP}$   for  the  non-leptonic $B \to \pi K$  decay  system \cite{Buras:2003yc,Mishima:2004um,Fleischer:2007mq,Fleischer:2018bld,Fang:2026fhl})
\begin{eqnarray}
&&  \Delta {\cal A}_{\rm CP} (B \to  P  \, \ell^{+} \ell^{-})
\nonumber \\
&&  \equiv  {\cal A}_{\rm CP} (B^{0} \to P^{0}  \,  \ell^{+} \ell^{-})
- {\cal A}_{\rm CP} (B^{+} \to  P^{+} \, \ell^{+} \ell^{-}).
\hspace{0.5 cm}
\end{eqnarray}
Inspecting the comparative predictions for  these  CP-violating   quantities   of
the exclusive  $B \to \left \{K, \pi \right \} \, \ell^{+} \ell^{-}$    decays  displayed in Figure
\ref{fig: numerical result for the CP asymmetry differences}
indicates that  the  presently   established   leading-power  spectator-scattering  mechanisms
can  profoundly    shift the  yielding   results   for both  two asymmetry observables
in the  ${\rm NLL}^{\prime}$ approximation:  numerically around  $(15 -20) \, \%$ corrections
in the large hadronic recoil region.
Interestingly,   the    three-particle  twist-three  contributions of the ${\rm C}$-type  non-local   matrix elements
yield   the most significant corrections to the  thus  determined    CP asymmetry differences
$ \Delta {\cal A}_{\rm CP} (B \to  P   \ell^{+} \ell^{-})$,  on account of the  non-trivial  strong phase
of   the obtained   hard-collinear coefficient  function $\mathbb{J}_{\rm C}^{(\rm 3P)}$  in  (\ref{result of the 3P jet function}).
We finally  present  our  updated   predictions    of  the CP-averaged  branching fractions (${\cal BR}$)
and    isospin asymmetries (${\cal A}_{\rm I}$) as well as   the CP-violating observables (${\cal A}_{\rm CP}$)
for  the  semileptonic    $B \to \left \{K, \pi \right \} \, \ell^{+} \ell^{-}$ decays
in Tables \ref{tab:B-to-Kll-Observables} and \ref{tab:B-to-pill-Observables} of the  Supplemental Material,
confronting    with  the available    experimental  measurements
from the Belle \cite{Belle:2009zue,BELLE:2019xld}, BaBar \cite{BaBar:2012mrf},  CDF \cite{CDF:2011grz,CDF:2011buy},
LHCb \cite{LHCb:2014cxe,LHCb:2014mit,LHCb:2026suh,LHCb:2015hsa,LHCb:2026huw} and  CMS  \cite{CMS:2024syx} Collaborations.

\section{Conclusions}

In summary,  we have  endeavoured   to construct  for the first time  the complete soft-collinear factorization formula
of  the  flagship  exclusive   $B \to P \,  \ell^{+} \ell^{-}$  decays at leading power,
by  identifying  an  intriguing   non-valence   Fock state  contribution
from the  three-particle twist-three    heavy-meson distribution amplitude.
Applying the modern effective field theory formalism  then   enabled  us to  extract   analytically
the NLO matching coefficients  in  the factorized  expressions for  the  non-local  $B \to P \, \gamma^{\ast}$  form factors.
Including  further  the  various  higher-order QCD  corrections to   the leading  two-particle  valence  contributions,
we  continued to  provide  our   updated   field-theoretic predictions of  the  differential  branching fractions,
the isospin asymmetries, and the  direct CP asymmetries for  the FCNC $B \to \left \{K, \pi \right \} \, \ell^{+} \ell^{-}$ decays
in the kinematic regime of  small  dilepton  invariant  mass.
In particular, we have  demonstrated   that   these  newly  computed    leading-power   contributions
of the exclusive bottom-meson decay matrix elements  can  bring    about   the  enormous  impacts  on the  predicted
CP asymmetry differences  $\Delta {\cal A}_{\rm CP} (B \to  P  \,  \ell^{+} \ell^{-})$ numerically.
Extending and developing further  our factorization analysis to the four-body leptonic $B$-meson decays,
to the  very   rare $B^{0}   \to  \bar D^{0} \, \ell^{+} \ell^{-}$   and  $B^{+}   \to  D_{s}^{+} \, \ell^{+} \ell^{-}$
decays,  and   to the  more  topical    $B \to V \,  \ell^{+} \ell^{-}$  decays
(with $V=K^{\ast}, \, \rho, \, \omega$)  will  apparently be  beneficial  for deepening our  understanding  towards
the multi-scale  strong interaction     dynamics  of   these exclusive heavy-hadron decays
and for exploiting  the enigmatic  non-standard flavour dynamics above the electroweak scale in a meticulous manner.

%
\begin{acknowledgments}
\section*{Acknowledgements}

We would like to thank  Matthew Birch  and   Xin-Qiang Li  for illuminating  discussions.
This research of Y.M.W. is supported in part by the  National Natural Science Foundation of China with Grants No. 12475097 and No. 12535006,
by the Natural Science Foundation of Tianjin with Grant No. 25JCZDJC01190,
and by the Fundamental Research Funds for the Central Universities  with Grant No. 63261180.
X.C.Z. is supported  by   the China Postdoctoral Science Foundation with
Grants  No.  GZB20250794,  No. 2026T190897 and  No.  2026M793702.

\end{acknowledgments}

\section{Data Availability}

The data that support the findings of this manuscript are openly available
in the ancillary files  attached to the arXiv preprint version of this Letter.

\appendix

\begin{widetext}

\section{SUPPLEMENTAL MATERIAL}

\subsection{Analytic Expressions for the  Short-Distance Functions}

We collect here the analytic expressions for  the short-distance matching coefficients
appearing in the soft-collinear  factorization formula (\ref{master factorization formula in SCET})
of  the two  invariant functions  $T_{P}^{(t, \, u)}$ at leading power in the  ${\Lambda_{\rm QCD} / m_b}$ expansion.
The four types of  hard functions from  the ${\rm SCET}_{\rm I}$ expansion of the effective weak  Hamiltonians
can be summarized as follows
\begin{eqnarray}
{\cal C}_{\rm eff,\, A}^{(t)} &=& \left \{  C_{7}^{\rm eff} +  {m_B \over 2 \, m_{b, \, \rm {PS}}} \, Y(q^2) \right \}
+  \left ( {\alpha_s \over 4 \, \pi} \right ) \,
\bigg  \{ C_F \, C_{7}^{\rm eff} \, \left [  4 \, \ln {m_b^2 \over \mu^2}
+ 2 \, \left (1 -   {m_b^2 \over q^2} \right ) \, \ln \left (1 - {q^2 \over m_b^2} \right )
+ 4 \, {\mu_{f} \over m_b} - 4 \, \right ]
\nonumber \\
&& -  \left (C_2 - {1 \over 6}  C_1 \right )   F_{2}^{(7)} - C_{8}^{\rm eff}  F_{8}^{(7)}
-   {m_B \over 2 \, m_{b, \, \rm {PS}}}  \left [ \left (C_2 - {1 \over 6}  C_1 \right )   F_{2}^{(9)}
+ C_1   \left  (  F_{1}^{(9)} + {1 \over 6} \, F_{2}^{(9)}  \right )
+ C_{8}^{\rm eff}   F_{8}^{(9)} \right ]  \bigg  \} + {\cal O}(\alpha_s^2),
\nonumber \\
{\cal C}_{\rm eff,\, A}^{(u)} &=&  \left \{  {m_B \over 2 \, m_{b, \, \rm {PS}}} \, \left ( {4 \over 3} \, C_1 + C_2 \right ) \,
\left [ h(q^2, m_c) - h(q^2, 0) \right ] \right \}  +  \left ( {\alpha_s \over 4 \, \pi} \right ) \,
\bigg \{ - \left (C_2 - {1 \over 6} \, C_1 \right )  \, \left (F_{2}^{(7)} + F_{2, u}^{(7)} \right )
\nonumber \\
&& -  \, {m_B \over 2 \, m_{b, \, \rm {PS}}} \, \left [ \left (C_2 - {1 \over 6} \, C_1 \right )  \,
\left (F_{2}^{(9)} + F_{2, u}^{(9)} \right )
+ C_1 \, \left ( F_{1}^{(9)} + {1 \over 6} \, F_{2}^{(9)}
+ F_{1, \, u}^{(9)} + {1 \over 6} \, F_{2, \, u}^{(9)}  \right )   \right ] \bigg  \}
+ {\cal O}(\alpha_s^2),
\nonumber \\
{\cal C}_{\rm eff,\, B}^{(t)} &=&  -{1 \over 4 \, \pi^2} \, {m_B \over 2 \, m_b} \,
\bigg \{  2 \, C_{7}^{\rm eff}  +
{1 \over 2} \, \bigg [ Q_u  \,  \left ( -{1 \over 6}  C_1 + C_2 + 6  C_6  \right ) \, \mathbb{G}(\tau, m_c)
+ Q_d  \,  \left (  C_3 - {1 \over 6}  C_4 + 16  C_5 +  {10 \over 3}   C_6  \right ) \, \mathbb{G}(\tau, m_b)
\nonumber \\
&& +  \,  Q_d  \,  \left (  C_3 - {1 \over 6}  \,  C_4 + 16 \,  C_5 -  {8 \over 3} \, C_6  \right ) \, \mathbb{G}(\tau, 0) \bigg ]  \bigg \}
+ {\cal O}(\alpha_s)  \,,
\nonumber \\
{\cal C}_{\rm eff,\, B}^{(u)} &=&  -{1 \over 8 \, \pi^2} \, {m_B \over 2 \, m_b} \,
\left \{ Q_u \,  \left (C_2 - {1 \over 6} \, C_1 \right )  \,
\left [ \mathbb{G}(\tau, m_c)  - \mathbb{G}(\tau, 0)  \right ] \right \} + {\cal O}(\alpha_s) \,,
\nonumber \\
{\cal C}_{\rm eff,\, C}^{(t)} &=&  \sum_{i=1}^{2} C_{i}  \, \widetilde{\mathbb{H}}_{i,  \, {\rm pen}}^{\rm I,  (c)}
+ \sum_{i=3}^{6} C_{i}  \, \widetilde{\mathbb{H}}_{i}^{\rm I} + C_{8} \,  \widetilde{\mathbb{H}}_{8}^{\rm I}  \,,
\nonumber \\
{\cal C}_{\rm eff,\, C}^{(u)} &=& (-1) \, \left [ \delta_{q u} \sum_{i=1}^{2} C_{i}  \,  \mathbb{H}_{i}^{\rm I,  (u)}
-  \,  \delta_{q d} \,  \delta_{P \pi} \sum_{i=1}^{2} C_{i}  \, \widetilde{\mathbb{H}}_{i}^{\rm I, (u)}  \right ]
+  \sum_{i=1}^{2} C_{i}  \, \widetilde{\mathbb{H}}_{i, \, {\rm pen}}^{\rm I, (u)}  \,,
\nonumber \\
{\cal C}_{\rm eff,\, D}^{(t)} &=&  \sum_{i=1}^{2} C_{i}  \, \widetilde{\mathbb{H}}_{i,  \, {\rm pen}}^{\rm II,  (c)}
+ \sum_{i=3}^{6} C_{i}  \, \widetilde{\mathbb{H}}_{i}^{\rm II} + C_{8} \,  \widetilde{\mathbb{H}}_{8}^{\rm II} \,,
\nonumber \\
{\cal C}_{\rm eff,\, D}^{(u)} &=&  (-1) \, \left [ \delta_{q u} \sum_{i=1}^{2} C_{i}  \,  \mathbb{H}_{i}^{\rm II,  (u)}
-  \,  \delta_{q d} \,  \delta_{P \pi} \sum_{i=1}^{2} C_{i}  \, \widetilde{\mathbb{H}}_{i}^{\rm II, (u)}  \right ]
+  \sum_{i=1}^{2} C_{i}  \, \widetilde{\mathbb{H}}_{i, \, {\rm pen}}^{\rm II, (u)} \,.
\end{eqnarray}
The  manifest definitions of the LO   coefficients  $Y(q^2)$ and  $h(q^2, m_q)$ due to the factorizable quark loop diagrams
(generated by  the four-quark operators)  entering   the ${\rm A}$-type hard functions  can be  extracted from \cite{Beneke:2001at}
and   the corresponding  NLO quantities  $F_{i}^{(7, 9)}$  and $F_{i, \,  u}^{(7, 9)}$ (with $i=1, \, 2, \, 8$)
have already been determined   analytically  in     \cite{Seidel:2004jh,Asatryan:2001zw,Asatrian:2019kbk}.
Additionally,  the  LO  primitive kernel  $\mathbb{G}(\tau, m_q)$ in  the  ${\rm B}$-type hard functions
${\cal C}_{\rm eff,\, B}^{(t, u)} $ takes the following  form
\begin{eqnarray}
\mathbb{G}(\tau, m_q) = 2 \, \left \{ I_1(m_q)
+  \frac{\tau \, m_B^2 + \bar \tau \, q^2}{\tau \, \bar n \cdot p^{\prime} \, m_B} \,
\left [ B_0 \left (\tau m_B^2 + \bar \tau q^2, \, m_q \right ) - B_0 \left (q^2, m_q \right ) \right ] \right \},
\end{eqnarray}
where the perturbative  functions $B_0$ and $I_1$  are defined in (29) and (30) of \cite{Beneke:2001at}.
The obtained  master formulae for a variety of  hard-scattering  kernels   in the ${\rm C}$-type hard functions
${\cal C}_{\rm eff,\, C}^{(t, u)} $ can be further written as
\begin{eqnarray}
&& \widetilde{\mathbb{H}}_{i,  \, {\rm pen}}^{\rm I,  (c)} =  0
+ \left (  {\alpha_s \over 4 \, \pi} \right )   \, \widetilde{\mathbb{T}}_{i,  \, {\rm pen}}^{\rm (1),  (c)}
+ {\color{magenta} \left (  {\alpha_s \over 4 \, \pi} \right )^2   \,
\left [  C_{\rm FF}^{(1)}  \,\,   \widetilde{\mathbb{T}}_{i,  \, {\rm pen}}^{\rm (1),  (c)}
+ \widetilde{\mathbb{T}}_{i,  \, {\rm pen}}^{\rm (2),  (c)}   \right ] }  + {\cal O}(\alpha_s^3)
\qquad   (i=1, \, 2)\,,
\nonumber \\
&& \widetilde{\mathbb{H}}_{i}^{\rm I}  =   \widetilde{\mathbb{T}}_{i}^{(0)}
+ \left (  {\alpha_s \over 4 \, \pi} \right )   \,  \left [  C_{\rm FF}^{(1)}  \,\,  \widetilde{\mathbb{T}}_{i}^{(0)}
+  \widetilde{\mathbb{T}}_{i}^{(1)}  \right ]
+ {\color{magenta} \left (  {\alpha_s \over 4 \, \pi} \right )^2 \,\,
\left [  \left (  C_{\rm FF}^{(2)}  +  C_{\rm  \bar q q}^{(2)} \right ) \,\,  \widetilde{\mathbb{T}}_{i}^{(0)}
+  C_{\rm FF}^{(1)}   \,   \widetilde{\mathbb{T}}_{i}^{(1)}  +  \widetilde{\mathbb{T}}_{i}^{(2)}  \right ] }
+ {\cal O}(\alpha_s^3)   \qquad   (i=3, ..., 6) \,,
\nonumber \\
&& \widetilde{\mathbb{H}}_{8}^{\rm I}  =   0
+ \left (  {\alpha_s \over 4 \, \pi} \right )   \, \widetilde{\mathbb{T}}_{8}^{\rm (1)}
+ {\color{magenta} \left (  {\alpha_s \over 4 \, \pi} \right )^2 \,\,  \left [  C_{\rm FF}^{(1)}   \,\, \widetilde{\mathbb{T}}_{8}^{\rm (1)}
+ \widetilde{\mathbb{T}}_{8}^{\rm (2)}   \right ] }  + {\cal O}(\alpha_s^3)   \,,
\nonumber \\
&& \mathbb{H}_{i}^{\rm I,  (u)}   =   \mathbb{T}_{i}^{(0)} + \left (  {\alpha_s \over 4 \, \pi} \right )   \,
\left [  C_{\rm FF}^{(1)}  \, \mathbb{T}_{i}^{(0)} + \mathbb{T}_{i}^{(1)} \right ]
+ {\color{magenta} \left (  {\alpha_s \over 4 \, \pi} \right )^2    \,
\left [  \left (  C_{\rm FF}^{(2)}  +  C_{\rm  \bar q q}^{(2)}  \right )  \, \mathbb{T}_{i}^{(0)}
+ C_{\rm FF}^{(1)}  \, \, \mathbb{T}_{i}^{(1)}  + \mathbb{T}_{i}^{(2)}    \right ] } + {\cal O}(\alpha_s^3)
\qquad   (i=1, \, 2)  \,,
\nonumber \\
&& \widetilde{\mathbb{H}}_{i}^{\rm I, (u)} =   \widetilde{\mathbb{T}}_{i}^{(0)} + \left (  {\alpha_s \over 4 \, \pi} \right )   \,
\left [  C_{\rm FF}^{(1)}  \, \widetilde{\mathbb{T}}_{i}^{(0)} + \widetilde{\mathbb{T}}_{i}^{(1)} \right ]
+ {\color{magenta} \left (  {\alpha_s \over 4 \, \pi} \right )^2    \,
\left [  \left (  C_{\rm FF}^{(2)}  +  C_{\rm  \bar q q}^{(2)}  \right )  \, \widetilde{\mathbb{T}}_{i}^{(0)}
+ C_{\rm FF}^{(1)}  \, \, \widetilde{\mathbb{T}}_{i}^{(1)}  + \widetilde{\mathbb{T}}_{i}^{(2)}    \right ]  }
+ {\cal O}(\alpha_s^3)  \qquad   (i=1, \, 2)   \,,
\nonumber \\
&& \widetilde{\mathbb{H}}_{i, \, {\rm pen}}^{\rm I, (u)}  =
0 +   \left (  {\alpha_s \over 4 \, \pi} \right )   \, \widetilde{\mathbb{T}}_{i, \,  {\rm pen}}^{\rm (1), (u)}
+ {\color{magenta} \left (  {\alpha_s \over 4 \, \pi} \right )^2    \,
\left [ C_{\rm FF}^{(1)}  \, \, \widetilde{\mathbb{T}}_{i, \,  {\rm pen}}^{\rm (1), (u)}
+ \widetilde{\mathbb{T}}_{i, \,  {\rm pen}}^{\rm (2), (u)}   \right ] }
+ {\cal O}(\alpha_s^3)  \qquad   (i=1, \, 2)  \,.
\label{master formula of the C-type functions}
\end{eqnarray}
The desired  expressions   for  the elementary  building blocks  at LO and NLO  accuracy  are  explicitly  given by
\begin{eqnarray}
\mathbb{T}_{1}^{(0)}  &=&  0 \,,   \qquad  \mathbb{T}_{2}^{(0)} =   1  \,,
\qquad   \widetilde{\mathbb{T}}_{1}^{(0)}  =  {C_F  \over N_c} \,,  \qquad
\widetilde{\mathbb{T}}_{2}^{(0)}  =  {1 \over N_c} \,,
\nonumber \\
\widetilde{\mathbb{T}}_{3}^{(0)}  &=&  {1  \over N_c} \,,  \qquad
\widetilde{\mathbb{T}}_{4}^{(0)}  =  {C_F \over N_c} \,, \qquad
\widetilde{\mathbb{T}}_{5}^{(0)}  =  {16 \over N_c} \,, \qquad
\widetilde{\mathbb{T}}_{6}^{(0)}  =  {16 \, C_F \over N_c} \,,
 \\
\nonumber \\
\widetilde{\mathbb{T}}_{1,  \, {\rm pen}}^{\rm (1),  (c)} &=& -  \left ( {C_F \over 2 \, N_c^2} \right ) \,
\mathbb{G}_{\rm pen} \left (z_c, \, \bar u \right )  \,,
\qquad
\widetilde{\mathbb{T}}_{2,  \, {\rm pen}}^{\rm (1),  (c)}
= (- 2 \, N_c)  \,\,  \widetilde{\mathbb{T}}_{1,  \,\, {\rm pen}}^{\rm (1),  (c)} \,,
\nonumber \\
C_{\rm FF}^{(1)} &=&  - {C_F \over 2} \, \left [ \ln^2 { \mu^2 \over m_b^2}  + 5 \, \ln { \mu^2 \over m_b^2}
+ { \pi^2 \over 6} + 12 \right ] \,,
\nonumber \\
\widetilde{\mathbb{T}}_{3}^{(1)}   &=& {C_F  \over N_c} \,
\left [ \mathbb{V}(u) + \mathbb{G}_{\rm pen} \left (0, \, \bar u \right )
+ \mathbb{G}_{\rm pen} \left (1, \, \bar u \right )  \right ]
+ \widetilde{\mathbb{T}}_{8}^{(1)}  \,,
\nonumber \\
\widetilde{\mathbb{T}}_{4}^{(1)}   &=& - {1 \over 2 \, N_c} \, \widetilde{\mathbb{T}}_{3}^{(1)}
+ {C_F  \over N_c} \,  \left [ (n_f -2) \,
\left (  \mathbb{G}_{\rm pen}  \,  \left (0, \, \bar u \right )  - {2 \over 3} \right )
+ \left (  \mathbb{G}_{\rm pen}  \,  \left ( z_c, \, \bar u \right )  - {2 \over 3} \right )
+ \left (  \mathbb{G}_{\rm pen}  \,  \left ( 1, \, \bar u \right )  - {2 \over 3} \right )   \right ] - C_F  \,,
\nonumber \\
\widetilde{\mathbb{T}}_{5}^{(1)}   &=& 16 \, \widetilde{\mathbb{T}}_{3}^{(1)}  + 4 \,\widetilde{\mathbb{T}}_{8}^{\rm (1)}
+ {16 \over 3} \, {C_F  \over N_c}    \,,
\nonumber \\
\widetilde{\mathbb{T}}_{6}^{(1)}   &=& -{3 \over N_c}  \, \widetilde{\mathbb{T}}_{3}^{(1)}
+ 10 \, \widetilde{\mathbb{T}}_{4}^{(1)} - {2 \over N_c}  \, \widetilde{\mathbb{T}}_{8}^{\rm (1)}
-  {4 \, C_F  \over N_c} \, \left ( {15 \over 2} \, N_c + {2 \over 3 \, N_c} - n_f  \right )    \,,
\nonumber \\
\widetilde{\mathbb{T}}_{8}^{\rm (1)} &=&  { C_F  \over N_c} \, \left ( - {2 \over \bar u} \right )\,,
\nonumber \\
\mathbb{T}_{1}^{(1)}  &=& { C_F  \over 2 \, N_c} \,  \mathbb{V}(u) \,,
\qquad   \mathbb{T}_{2}^{(1)}  = 0 \,,
\qquad  \widetilde{\mathbb{T}}_{1}^{(1)}  =  -{1 \over 3}  \, \mathbb{T}_{1}^{(1)}  - {4 \over 3} \,,
\qquad   \widetilde{\mathbb{T}}_{2}^{(1)}  =  2 \, \mathbb{T}_{1}^{(1)} \,,
\nonumber \\
\widetilde{\mathbb{T}}_{1, \,  {\rm pen}}^{\rm (1), (u)}  &=& - { C_F  \over 2 \, N_c^2} \,
\left [ \mathbb{G}_{\rm pen}(z_c, \bar u) - \mathbb{G}_{\rm pen}(0, \bar u) \right ]  \,,
\qquad \widetilde{\mathbb{T}}_{2, \,  {\rm pen}}^{\rm (1), (u)}
= (-2 \, N_c) \, \widetilde{\mathbb{T}}_{1, \,  {\rm pen}}^{\rm (1), (u)} \,,
\end{eqnarray}
where  $n_f=5$  stands for  the total number of quark flavours in the weak effective theory.
We have further defined  the dimensionless quantity  $z_q= {m_q^2 - i \, 0 \over m_b^2}$
and the  kernel  functions $\mathbb{G}_{\rm pen}(z, \bar u)$ and  $\mathbb{V}(u)$  for brevity
\begin{eqnarray}
\mathbb{G}_{\rm pen}(z, \bar u) &=& -{2 \over 3} \, \ln{ \mu^2  \over m_b^2} +  {2 \over 3}
- {2 \over 9 } \, \left ( 12 \, {z \over \bar u} + 5  - 3 \, \ln z \right )
+   {2 \over 3} \, s_z \,  \left ( 2 \, {z \over \bar u}  +  1 \right )  \,
\ln \left (   { s_z + 1 \over  s_z - 1 }  \right ) \,,
\nonumber \\
\mathbb{V}(u)   &=&  -6 \, \ln{ \mu^2  \over m_b^2}
+  \left [ 2 \, {\rm Li}_2(u) - \ln^2 u  - \left ( 1 + 2 \, i \, \pi - 2 \, { u \over \bar u} \right ) \, \ln u
- (u \leftrightarrow \bar u)  \right ] + 3 \, \left ( 1- {u \over \bar u} \right ) \, \ln u
- 3 \, i \, \pi -  22 \,,
\hspace{1.0 cm}
\end{eqnarray}
with the  kinematic variable  $s_z= \sqrt{ 1 - {4 \, z \over \bar u}}$.
The achieved  results for the various  NNLO  quantities  in (\ref{master formula of the C-type functions})  read
\begin{eqnarray}
\widetilde{\mathbb{T}}_{1,  \, {\rm pen}}^{\rm (2),  (c)}   &=& \widetilde{T}_{1c}^{(2)}  \,,
\qquad   \widetilde{\mathbb{T}}_{2,  \, {\rm pen}}^{\rm (2),  (c)}   = \widetilde{T}_{2c}^{(2)} \,,
\qquad \widetilde{\mathbb{T}}_{1, \,  {\rm pen}}^{\rm (2), (u)} =  \widetilde{T}_{1c}^{(2)}  - \widetilde{T}_{1u}^{(2)} \,,
\qquad  \widetilde{\mathbb{T}}_{2, \,  {\rm pen}}^{\rm (2), (u)} = \widetilde{T}_{2c}^{(2)}  - \widetilde{T}_{2u}^{(2)}  \,,
\nonumber \\
C_{\rm FF}^{(2)} &=&  \lim_{u \to 1} \, \left [ C_1^{(2)} + {u \over 2} \, C_2^{(2)} + C_3^{(2)} \right ] \,,
\qquad  C_{j}^{(2)} =  F_{j}^{(2)}  + F_{j}^{(1)} \, \left \{ \left [ Z_{\rm J}^{-1} \right ]^{(1)}
+ \xi_{45}^{(1)}  + Z_{\alpha}^{(1)}  \right \}   + \left [ Z_{\rm J}^{-1} \right ]^{(2)} \, F_{j}^{(0)}  \,,
\nonumber \\
C_{\rm  \bar q q}^{(2)}  &=&  {4  \over 3} \, C_F \, T_F \,
\bigg \{  \bigg [ \left ( \ln^2 {\mu^2 \over m_b^2}
+ \left ( 2 \, \ln{v \over u} - {10 \over 3}  \right ) \, \ln  {\mu^2 \over m_b^2} + {1 \over 2}  \, \ln^2 {v \over u}
- {5 \over 3} \, \ln {v \over u} + {28 \over 9} \right )  \, \mathbb{V}_{\rm ERBL}^{(0)}(u, v)
\nonumber \\
&& + \, \left ( 2 \,  \ln  {\mu^2 \over m_b^2} + \ln {v \over u} - {5 \over 3}  \right ) \,
{u \over v} \,\, \theta(v -u)  \bigg ]_{+}
+ \,  \left [ u \leftrightarrow \bar u, \, v \leftrightarrow \bar v  \right ]_{+}  \bigg \} \,,
\nonumber \\
\mathbb{T}_{1}^{(2)}  &=&  \left [ T_1^{\rm (2), \, re} + i \, \pi \, T_1^{\rm (2), \,  im} \right ]
+ n_l \, T_F \, \left [ T_{1, \, n_l}^{\rm (2), \, re} + i \, \pi \, T_{1, \, n_l}^{\rm (2), \,  im} \right ]
\nonumber \\
&&  + \,  T_F \, \left [ T_{1, \, T_F}^{\rm (2), \, re} + i \, \pi \, T_{1, \, T_F}^{\rm (2), \,  im} \right ]
+   T_F \, \left [ T_{1, \, \rm c}^{\rm (2), \, re} + i \, \pi \, T_{1, \, \rm c}^{\rm  (2), \,  im} \right ]    \,,
\nonumber \\
\mathbb{T}_{2}^{(2)}  &=&  \mathbb{T}_{2}^{\rm (2), \,  re}   + i \, \pi \, \mathbb{T}_{2}^{\rm (2), \, im}  \,,
\nonumber \\
\widetilde{\mathbb{T}}_{1}^{(2)}   &=& - {1 \over 3} \, \mathbb{T}_{1}^{(2)}  +  {4 \over 9}  \,  \mathbb{T}_{2}^{(2)}
- {20 \over 27} \, n_l \, T_F -  {20 \over 27} \,  T_F  - {49 \over 9}  \,,
\nonumber \\
\widetilde{\mathbb{T}}_{2}^{(2)}   &=& 2 \, \mathbb{T}_{1}^{(2)}  +  {1 \over 3}  \,  \mathbb{T}_{2}^{(2)}
+ 6  \,  \mathbb{T}_{1}^{(1)}    \,,
\label{master formulas for the 2-loop C-type functions}
\end{eqnarray}
where $T_F=1/2$ denotes the standard normalization for ${\rm SU}(N_c)$ gauge group
and $n_l =4$ represents  the number of  active quark flavours  in  ${\rm SCET}_{\rm I}$.
The one-loop RG  evolution function  $\mathbb{V}_{\rm ERBL}^{(0)}(u, v)$ for  the leading-twist distribution amplitude of the light meson
and the  ``+"  distribution  appearing in (\ref{master formulas for the 2-loop C-type functions}) are defined by
\begin{eqnarray}
\mathbb{V}_{\rm ERBL}^{(0)}(u, v) = \left ( {1 \over u -v} - 1 \right )  \, {u \over v} \,\,  \theta(v -u) \,,
\qquad
\left [f(u, v) \right ]_{+} \equiv   f(u, v) - \delta(u -v) \, \int_0^1 dt \, f (t, v)  \,.
\end{eqnarray}
The complete  expressions for the two-loop  coefficient  functions $\widetilde{T}_{i c }^{(2)}$, $\widetilde{T}_{i u}^{(2)}$,
$\widetilde{\mathbb{T}}_{3,..., 6}^{(2)}$  (with $i=1, \, 2$) and the additional one-loop function $\widetilde{\mathbb{T}}_{8}^{(2)}$
can be inferred  from \cite{Bell:2020qus}.
The renormalized  hard functions $C_{j}^{(2)}$  (with $j=1, \, 2, \, 3 $) from the ${\rm QCD} \to {\rm SCET}_{\rm I}$ matching
for the flavour-changing current $\bar q \, \gamma^{\mu} (1-\gamma_5) \, b$ have been previously determined in \cite{Beneke:2008ei}
(see \cite{Bonciani:2008wf,Asatrian:2008uk,Bell:2008ws}  for further  independent  analytical computations of these quantities).
The individual coefficients  in the NNLO hard  kernels $\mathbb{T}_{1, 2}^{(2)} $ and $\widetilde{\mathbb{T}}_{1, 2}^{(2)} $
relevant for the topological tree amplitudes of hadronic $B$-meson decays have been computed with the QCD factorization formalism
\cite{Bell:2007tv,Bell:2009nk,Beneke:2009ek}.
It still remains  to be  pointed out  that  the eventual  convolution integrals of  the  peculiar  terms
$C_{\rm  \bar q q}^{(2)} \, \mathbb{T}_{i}^{(0)}$ and $C_{\rm  \bar q q}^{(2)} \, \widetilde{\mathbb{T}}_{i}^{(0)}$
entering the master formulae (\ref{master formula of the C-type functions})  for the ${\rm C}$-type kernel  functions
with   the twist-two light-cone distribution amplitude $\phi_P$ must be understood in the following manner
\begin{eqnarray}
{\cal I}_{i,  \, \rm  \bar q q}^{(2)} \equiv  \int_0^1 d u \,  \int_0^1 d v \,\,
\left [ C_{\rm  \bar q q}^{(2)}(u, v) \, \mathbb{T}_{i}^{(0)}(u) \right ] \, \phi_P(v) \,,
\qquad
\widetilde{{\cal I}}_{i, \, \rm  \bar q q}^{(2)} \equiv  \int_0^1 d u \,  \int_0^1 d v \,\,
\left [ C_{\rm  \bar q q}^{(2)}(u, v) \, \widetilde{\mathbb{T}}_{i}^{(0)}(u) \right ] \, \phi_P(v) \,.
\label{two specical integrals}
\end{eqnarray}
Taking into account the very fact that the LO  hard functions $\mathbb{T}_{i}^{(0)}$
and $\widetilde{\mathbb{T}}_{i}^{(0)}$ are actually independent of the variable  $u$,
we can  readily carry out  the $u$ integration in  (\ref{two specical integrals})
by employing the  obtained result of the matching coefficient  $C_{\rm  \bar q q}^{(2)}$
\begin{eqnarray}
\int_0^1 d u \, C_{\rm  \bar q q}^{(2)}(u, v)  = 0 \,.
\end{eqnarray}
This interesting observation can be traced back to the conserved  {\it local} axial-vector current
in QCD with naive dimensional regularization (NDR) scheme of the Dirac matrix $\gamma_5$
(see \cite{Grozin:1998kf,Grozin:2006xm}  for more discussions on decoupling the heavy-quark loops for
flavour non-singlet light-light quark currents).
We can now proceed to deduce   the master formulae for the necessary  short-distance coefficients
in the ${\rm D}$-type hard functions  ${\cal C}_{\rm eff,\, D}^{(t, u)} $ up to the ${\cal O}(\alpha_s)$ accuracy
\begin{eqnarray}
\widetilde{\mathbb{H}}_{i,  \, {\rm pen}}^{\rm II,  (c)} &=&
0 + {\color{magenta}  \left ( { \alpha_s  \over 4 \, \pi} \right ) \,
\left [ \widetilde{\rm H}_{i,  \, {\rm pen}}^{\rm (1),  II,  (c)}
+ 2 \, \widetilde{\mathbb{T}}_{i,  \, {\rm pen}}^{\rm (1),  (c)}   \right ] } + {\cal O}(\alpha_s^2)
\qquad   (i=1, \, 2)  \,,
\nonumber \\
\widetilde{\mathbb{H}}_{i}^{\rm II}  &=&  \widetilde{\mathbb{H}}_{i}^{\rm (0),  II}
+ {\color{magenta}  \left ( { \alpha_s  \over 4 \, \pi} \right ) \,  \left [ \widetilde{\rm{H}}_{i}^{\rm (1),  II}
+ 2 \, \widetilde{\mathbb{T}}_{i}^{(1)} + (-2) \,\, C_{f_{+}}^{\rm (1), (B1)}  \,\, \widetilde{\mathbb{T}}_{i}^{(0)} \right ] }
+ {\cal O}(\alpha_s^2)  \qquad   (i=3, ..., 6) \,,
\nonumber \\
\widetilde{\mathbb{H}}_{8}^{\rm II}  &=&  0 + {\color{magenta}  \left ( { \alpha_s  \over 4 \, \pi} \right ) \,
\left [ \widetilde{\rm{H}}_{8}^{\rm (1), II}  + 2 \, \widetilde{\mathbb{T}}_{8}^{\rm (1)}  \right ] }
+ {\cal O}(\alpha_s^2)   \,,
\nonumber \\
\mathbb{H}_{i}^{\rm II,  (u)}  &=&  \mathbb{H}_{i}^{\rm (0),  II,  (u)}
+ {\color{magenta}  \left ( { \alpha_s  \over 4 \, \pi} \right ) \,
\left [ {\rm H}_{i}^{\rm (1),  II,  (u)} + 2 \, \mathbb{T}_{i}^{(1)}
+ (-2) \,\, C_{f_{+}}^{\rm (1), (B1)}  \,\,  \mathbb{T}_{i}^{(0)}   \right ] }
+ {\cal O}(\alpha_s^2) \qquad   (i=1, \, 2)  \,,
\nonumber \\
\widetilde{\mathbb{H}}_{i}^{\rm II, (u)} &=&  \widetilde{\mathbb{H}}_{i}^{\rm (0),  II,  (u)}
+ {\color{magenta}  \left ( { \alpha_s  \over 4 \, \pi} \right ) \, \left [ \widetilde{{\rm H}}_{i}^{\rm (1),  II,  (u)}
+ 2 \, \widetilde{\mathbb{T}}_{i}^{(1)}  + (-2) \,\, C_{f_{+}}^{\rm (1), (B1)}  \,\,  \widetilde{\mathbb{T}}_{i}^{(0)}   \right ] }
+ {\cal O}(\alpha_s^2)  \qquad   (i=1, \, 2)  \,,
\nonumber \\
\widetilde{\mathbb{H}}_{i, \, {\rm pen}}^{\rm II, (u)}  &=&
0 + {\color{magenta}  \left ( { \alpha_s  \over 4 \, \pi} \right ) \, \left [ \widetilde{\rm H}_{i,  \, {\rm pen}}^{\rm (1),  II,  (u)}
+ 2 \, \widetilde{\mathbb{T}}_{i,  \, {\rm pen}}^{\rm (1),  (u)}   \right ]}  + {\cal O}(\alpha_s^2)
\qquad   (i=1, \, 2)   \,.
\end{eqnarray}
The yielding  LO and NLO  results   for  these  fundamental  perturbative quantities can then be  cast in the  form
\begin{eqnarray}
\widetilde{\mathbb{H}}_{3}^{\rm (0),  II}  &=&  {2 \over N_c} \, \left (1 + {1 \over \bar u} \right ) \,,
\qquad
\widetilde{\mathbb{H}}_{4}^{\rm (0),  II}  = {1 \over N_c} \, \left ( 2 \, C_F - {1 \over N_c}  \,  {1 \over \bar u} \right ) \,,
\qquad
\widetilde{\mathbb{H}}_{5}^{\rm (0),  II}  =  16 \, \widetilde{\mathbb{H}}_{3}^{\rm (0),  II}  \,,
\qquad
\widetilde{\mathbb{H}}_{6}^{\rm (0),  II}  =  16 \, \widetilde{\mathbb{H}}_{4}^{\rm (0),  II}  \,,
\nonumber \\
\mathbb{H}_{1}^{\rm (0),  II,  (u)}  &=&  {1 \over N_c} \, {1 \over \bar u}\,,
\qquad \mathbb{H}_{2}^{\rm (0),  II,  (u)}  =  2 \,,
\qquad
\widetilde{\mathbb{H}}_{1}^{\rm (0),  II,  (u)} =  \widetilde{\mathbb{H}}_{4}^{\rm (0),  II}   \,,
\qquad  \widetilde{\mathbb{H}}_{2}^{\rm (0),  II,  (u)}  = \widetilde{\mathbb{H}}_{3}^{\rm (0),  II}   \,,
\\
\nonumber \\
\widetilde{\rm H}_{1,  \, {\rm pen}}^{\rm (1),  II,  (c)} &=&   {1 \over N_c}  \,
\left [ s_4(z_c) -  {1 \over N_c} \, s_3(z_c) \right ] \,,
\qquad
\widetilde{\rm H}_{2,  \, {\rm pen}}^{\rm (1),  II,  (c)} =   {2 \over N_c}  \,  s_3(z_c)  \,,
\nonumber \\
\widetilde{\rm{H}}_{3}^{\rm (1),  II} &=&  {2 \over 3} \, {2 \over  N_c} \, {1 \over \bar u}
+ {2 \over  N_c} \, \left \{ r_1(u, \bar \tau)  +  \left [ s_3(0) + s_3(1) \right ]
+ \left [ s_1(1) + s_1(z_c) +(n_f - 2) \, s_1(0) \right ]   \right \}
\nonumber \\
&& + \, {2 \over  N_c} \, \left \{ s_3^{\prime}(1)  +  \left [  s_1^{\prime}(1) + s_1^{\prime}(z_c) + (n_f - 2) \, s_1^{\prime}(0) \right ] \right \} \,,
\nonumber \\
\widetilde{\rm{H}}_{4}^{\rm (1),  II} &=&  -{11 \over 18} \, {2 \over  N_c} \, {1 \over \bar u}
+ {2 \over  N_c} \, \left ( - {1 \over 6} \right ) \, \left \{ r_1(u, \bar \tau)  +  \left [ s_3(0) + s_3(1) \right ]
+ \left [ s_1(1) + s_1(z_c) +(n_f - 2) \, s_1(0) \right ]   \right \}
\nonumber \\
&& + \, {2 \over  N_c} \, {1 \over 2} \, \left \{  r_2(u, \bar \tau)  +  \left [ s_4(0) + s_4(1) \right ]
+ \left [ s_2(1) + s_2(z_c) +(n_f - 2) \, s_2(0) \right ] \right \}
\nonumber \\
&& + \, {2 \over  N_c} \, \left ( - {1 \over 6} \right ) \,
\left \{ s_3^{\prime}(1)  +  \left [  s_1^{\prime}(1) + s_1^{\prime}(z_c) + (n_f - 2) \, s_1^{\prime}(0) \right ] \right \}
\nonumber \\
&& + \, {2 \over  N_c} \, {1 \over 2} \,
\left \{ s_4^{\prime}(1)  +  \left [  s_2^{\prime}(1) + s_2^{\prime}(z_c) + (n_f - 2) \, s_2^{\prime}(0) \right ] \right \}  \,,
\nonumber \\
\widetilde{\rm{H}}_{5}^{\rm (1),  II} &=&   {736 \over 9} \, {2 \over  N_c} \, {1 \over \bar u}
+ {2 \over  N_c} \, 16 \, \left \{ r_1(u, \bar \tau)  +  \left [ s_3(0) + s_3(1) \right ]
+ \left [ s_1(1) + s_1(z_c) +(n_f - 2) \, s_1(0) \right ]   \right \}
\nonumber \\
&& + \, {2 \over  N_c} \, 4 \,
\left \{ s_3^{\prime}(1)  +  \left [  s_1^{\prime}(1) + s_1^{\prime}(z_c) + (n_f - 2) \, s_1^{\prime}(0) \right ] \right \}  \,,
\nonumber \\
\widetilde{\rm{H}}_{6}^{\rm (1),  II} &=&  -{350 \over 27} \, {2 \over  N_c} \, {1 \over \bar u}
+ {2 \over  N_c} \, \left ( - {8 \over 3} \right ) \, \left \{ r_1(u, \bar \tau)  +  \left [ s_3(0) + s_3(1) \right ]
+ \left [ s_1(1) + s_1(z_c) +(n_f - 2) \, s_1(0) \right ]   \right \}
\nonumber \\
&& + \, {2 \over  N_c} \, 8 \, \left \{  r_2(u, \bar \tau)  +  \left [ s_4(0) + s_4(1) \right ]
+ \left [ s_2(1) + s_2(z_c) +(n_f - 2) \, s_2(0) \right ] \right \}
\nonumber \\
&& + \, {2 \over  N_c} \, \left ( - {2 \over 3} \right ) \,
\left \{ s_3^{\prime}(1)  +  \left [  s_1^{\prime}(1) + s_1^{\prime}(z_c) + (n_f - 2) \, s_1^{\prime}(0) \right ] \right \}
\nonumber \\
&& + \, {2 \over  N_c} \,  2 \,
\left \{ s_4^{\prime}(1)  +  \left [  s_2^{\prime}(1) + s_2^{\prime}(z_c) + (n_f - 2) \, s_2^{\prime}(0) \right ] \right \}  \,,
\nonumber \\
C_{f_{+}}^{\rm (1), (B1)}  &=&  C_F \, \left [ 2 \,  \ln^2 {\mu  \over m_b} +  \ln { \mu  \over m_b}
+ { 2 - \tau  \over {\bar \tau}^2} \, \ln \tau   + {1 \over \bar \tau} + { \pi^2 \over 12}  \right ]
\nonumber \\
&& + \, ( 2 \, C_F - C_A) \, {1 \over \tau} \, \left [ \left (-2 \, \ln { \mu  \over m_b}
+ \ln {\bar \tau}   + {1 \over \bar \tau} \, \ln \tau  - 1 \right ) \, \ln {\bar \tau}
+{\rm Li}_{2} (\tau) \right ]   \,,
\nonumber \\
\widetilde{\rm{H}}_{8}^{\rm (1), II}  &=&  0  \,,
\nonumber \\
{\rm H}_{1}^{\rm (1),  II,  (u)} &=&  {2 \over N_c} \, \left [ - {5 \over 6} \, {1 \over \bar u}
+ {1 \over 2} \, r_1(u, \bar \tau) -  {1 \over 6} \, r_2(u, \bar \tau)  \right ] \,,
\qquad
{\rm H}_{2}^{\rm (1),  II,  (u)} =  {2 \over N_c} \, \left [ - 2 \, {1 \over \bar u} +   r_2(u, \bar \tau)  \right ] \,,
\nonumber \\
\widetilde{{\rm H}}_{1}^{\rm (1),  II,  (u)} &=&   {2 \over N_c} \, \left [ - {11 \over 18} \, {1 \over \bar u}
- {1 \over 6} \, r_1(u, \bar \tau) +  {1 \over 2} \, r_2(u, \bar \tau)  \right ] \,,
\qquad
\widetilde{{\rm H}}_{2}^{\rm (1),  II,  (u)} =   {2 \over N_c} \, \left [  {2 \over 3} \, {1 \over \bar u}
+  \, r_1(u, \bar \tau)  \right ] \,,
\nonumber \\
\widetilde{\rm H}_{1,  \, {\rm pen}}^{\rm (1),  II,  (u)} &=&   {1 \over N_c}  \,
\left \{  \left [ s_4(z_c) - s_4(0)  \right ] - {1 \over N_c}  \, \left [ s_3(z_c) - s_3(0)  \right ]  \right \} \,,
\qquad
\widetilde{\rm H}_{2,  \, {\rm pen}}^{\rm (1),  II,  (u)} =   {2 \over N_c}  \,  \left [ s_3(z_c) - s_3(0)  \right ] \,.
\end{eqnarray}
The detailed expressions of  the  two  primitive  kernels $r_{1}(u, \bar \tau)$  and $r_{2}(u, \bar \tau)$
that  dictate  the hard  spectator-scattering corrections to the topological tree amplitudes of  exclusive non-leptonic $B$-meson decays
at one-loop  order have been  derived  from the ${\rm QCD} \to {\rm SCET}_{\rm I}$ matching of the current-current operators
${\cal Q}_{1, \, 2}^{u}$ in the effective weak Hamiltonian \cite{Beneke:2005vv}.
The remaining  kernel functions $s_{m}(z_q)$ and   $s_{m}^{\prime}(z_q)$ (with $m=1,..., 4$)
from  the QCD penguin contraction diagrams with insertions of the four-quark operators
${\cal Q}_{k}^{(c, u)}$ (with $k=1,..., 6$) have already  been determined analytically
for the purpose of evaluating  the NLO  spectator-scattering corrections  to the leading penguin amplitudes
$a_4^{(c, u)}$ in the heavy quark expansion  \cite{Beneke:2006mk}.
The resulting expression  of the  one-loop coefficient  function  $C_{f_{+}}^{\rm (1), (B1)}$ can be inferred from
perturbative  matching computations for the semileptonic $b \to u \ell \bar{\nu_{\ell}}$  transition
at  next-to-leading power accuracy \cite{Beneke:2004rc,Beneke:2005gs,Hill:2004if,Becher:2004kk},
by further  taking the maximal recoil limit $E \to m_b / 2$  for the energetic  light meson.

Subsequently, we can continue  to  present the desired   results   for the four distinct classes of jet functions
entering the soft-collinear  factorization formula (\ref{master factorization formula in SCET}) of
the exclusive  radiative   $B \to P \, \gamma^{\ast}$ decay form factors
\begin{eqnarray}
{\cal \overline J}_{\rm B} &=& \frac{n \cdot q \,\,  \omega}{2} \, \mathbb{\overline J}_{\rm B}
= (- 8 \, \pi^2)  \, \left (  {n \cdot q \over \bar n \cdot p^{\prime}} \, {1 \over \bar u} \right )  \,
{C_F \over N_c}  \,  \left [ 0 +    \left ( { \alpha_s  \over 4 \, \pi} \right )  \delta(\tau - \bar u)
+ {\color{magenta} \left ( { \alpha_s  \over 4 \, \pi} \right )^2 \,
j_{\parallel}(\tau;  \, u, \, \omega) } \right ]  + {\cal O}(\alpha_s^3) \,,
\nonumber \\
{\cal J}_{\rm C}^{(\rm 2P)} &=&   Q_q \,\, \omega \,\,  \mathbb{J}_{\rm C}^{(\rm 2P)}
=  {Q_q  \over (\bar n \cdot q  / \omega ) - 1  + i \, 0 } \,
\left [ 1 +  {\color{blue}  \left ( { \alpha_s  \over 4 \, \pi} \right )  \, \mathbb{J}_{\parallel, \, -}^{(1)} } \right ]
+ {\cal O}(\alpha_s^2) \,,
\nonumber \\
{\cal J}_{\rm C}^{(\rm 3P)} &=& Q_q \,\, \omega_1 \, \omega_2^2  \, \, \mathbb{J}_{\rm C}^{(\rm 3P)}
= Q_q \, \left \{  0 + {\color{magenta}  \left ({\alpha_s \over 4 \, \pi} \right )
\left  [   ( 2 \, C_F  - C_A) \,\,   \mathbb{K}_{\rm I} + C_A  \,\,  \mathbb{K}_{\rm II}  \right ] }  \right \}
+ {\cal O}(\alpha_s^2)   \,,
\nonumber \\
{\cal J}_{\rm D} &=&  Q_q \, \left (- {1 \over 2} \right )   \, { \omega  \over n \cdot q } \,  {m_B \over m_b} \,\,
\mathbb{J}_{\rm D}
=  Q_q \, \left (- {1 \over 2} \right )   \,  {m_B \over m_b}  \,
\left [  0 + {\color{blue}  \left ( { \alpha_s  \over 4 \, \pi} \right )  \, \mathbb{J}_{\parallel, \, +}^{(1)} } \right ]
+ {\cal O}(\alpha_s^2)   \,,
\end{eqnarray}
where $Q_q$ represents the electric charge of the spectator quark in the $B$-meson.
The lengthy  expression for  the NLO coefficient function $j_{\parallel}$  has been perturbatively  determined from   the second-step
${\rm SCET}_{\rm I} \to {\rm SCET}_{\rm II}$ matching  of  the ${\rm B}$-type effective  matrix element
$\Xi_P$  \cite{Beneke:2005gs,Becher:2004kk}, which constitutes an important prerequisite
for exploring   the  delicate  spectator-scattering mechanisms  in  heavy-to-light $B$-meson decay  form factors.
The one-loop valence  corrections  to the ${\rm C}$-type and   ${\rm D}$-type hard-collinear functions
$\mathbb{J}_{\parallel, \, \mp}^{(1)}$  can be directly taken from our previous SCET  calculations \cite{Huang:2024xii}
\begin{eqnarray}
\mathbb{J}_{\parallel, \, -}^{(1)} &=& C_F \, \left [  \ln^2 { {\hat \mu}^2  \over \omega  - \bar n \cdot q }
-  2 \, \ln { {\hat \mu}^2  \over \omega  - \bar n \cdot q } \, \ln \left (1 - {\omega  \over  \bar n \cdot q } \right )
- \ln^2  \left (1 - {\omega  \over  \bar n \cdot q } \right )   - \left (1 +{ 2 \,\,  \omega  \over  \bar n \cdot q} \right ) \,
\ln   \left (1 - {\omega  \over  \bar n \cdot q} \right )  -  {\pi^2 \over 6}  - 1 \right ],
\nonumber \\
\mathbb{J}_{\parallel, \, +}^{(1)} &=& 2 \, C_F \, \ln \left (1 - {\omega  \over  \bar n \cdot q } \right )  \,
(1-\tau) \,\,  \theta(\tau)  \,\,  \theta(1 - \tau).
\end{eqnarray}
The newly derived  perturbative kernels  $\mathbb{K}_{\rm I}$ and $\mathbb{K}_{\rm II}$ in the factorized expression
for the non-leading Fock state  contribution  of  the ${\rm C}$-type ${\rm SCET}_{\rm I}$ matrix element
are explicitly displayed in  (\ref{result of the 3P jet function}).

\subsection{Updated Theory Predictions for Exclusive  $B \to \left \{K, \pi \right \} \, \ell^{+} \ell^{-}$   Decay Observables}

We summarize explicitly  the yielding numerical   predictions of  the  CP-averaged  branching fractions, the isospin asymmetries,
and the  CP-violating  observables   for  the electroweak penguin $B \to P  \, \ell^{+} \ell^{-}$  decays
\begin{eqnarray}
{\cal BR}(B \to P  \, \ell^{+} \ell^{-}) [q_1^2, \, q_2^2] &\equiv& {1 \over 2}  \,
\int_{q_1^2}^{q_2^2} \, d q^2 \,\, \left [ \frac{d {\cal BR} (\bar  B \to \bar  P  \, \ell^{+} \ell^{-})}{d q^2}
+  \frac{d {\cal BR} (B \to P  \, \ell^{+} \ell^{-})}{d q^2}  \right ]  \,,
\nonumber  \\
{\cal A}_{\rm I}(B \to P  \, \ell^{+} \ell^{-}) [q_1^2, \, q_2^2] &\equiv&
\frac{\displaystyle \kappa \, {\cal BR}(B^{0} \to P^{0}  \, \ell^{+} \ell^{-}) [q_1^2, \, q_2^2]
- (\tau_{B^{0}} / \tau_{B^{+}}) \,\,  {\cal BR}(B^{+} \to P^{+}  \, \ell^{+} \ell^{-}) [q_1^2, \, q_2^2]}
{\displaystyle \kappa \, {\cal BR}(B^{0} \to P^{0}  \, \ell^{+} \ell^{-}) [q_1^2, \, q_2^2]
+ (\tau_{B^{0}} / \tau_{B^{+}}) \,\,  {\cal BR}(B^{+} \to P^{+}  \, \ell^{+} \ell^{-}) [q_1^2, \, q_2^2]} \,,
\nonumber  \\
{\cal A}_{\rm CP}(B \to P  \, \ell^{+} \ell^{-})  [q_1^2, \, q_2^2] &\equiv&  \dfrac{\displaystyle  \int_{q_1^2}^{q_2^2} \, d q^2 \,\,
\left [ \dfrac{d {\cal BR} (\bar  B \to \bar  P  \, \ell^{+} \ell^{-})}{d q^2}
-  \dfrac{d {\cal BR} (B \to P  \, \ell^{+} \ell^{-})}{d q^2}  \right ]}
{\displaystyle  \int_{q_1^2}^{q_2^2} \, d q^2 \,\,
\left [ \dfrac{d {\cal BR} (\bar  B \to \bar  P  \, \ell^{+} \ell^{-})}{d q^2}
+  \dfrac{d {\cal BR} (B \to P  \, \ell^{+} \ell^{-})}{d q^2}  \right ]}   \,.
\end{eqnarray}
Here  the constant factor $\kappa$ originates  from the quark flavour wavefunction
of the light pseudoscalar meson  \cite{Huang:2024xii,Hambrock:2015wka} and $\tau_{B^{0}} / \tau_{B^{+}}$  denotes
the measured   lifetime ratio for   the charged and neutral bottom mesons \cite{ParticleDataGroup:2024cfk}.

It can then be observed from Table \ref{tab:B-to-Kll-Observables}  that
the  updated numerical results for the partially integrated  branching fractions
of both  $B^{0} \to K^{0} \, \ell^{+} \ell^{-}$   and $B^{+} \to K^{+} \, \ell^{+} \ell^{-}$  decays
remain to be in sharp   tension with  the  available    Belle \cite{BELLE:2019xld},  CDF \cite{CDF:2011buy},
LHCb \cite{LHCb:2014cxe} and CMS \cite{CMS:2024syx}  measurements  in  the large hadronic recoil regime.
By contrast,  our  updated  theory  predictions for  the  CP-averaged  isospin-breaking  corrections to
the  exclusive   $B  \to K  \, \ell^{+} \ell^{-}$  decays coincide with  the complete set of
the measured  Belle \cite{BELLE:2019xld},  BaBar \cite{BaBar:2012mrf}, and LHCb data points \cite{LHCb:2014cxe},
which however suffer from   enormous  experimental  uncertainties due to the  presently  limited  statistical precision.
It also becomes evident that    our   numerical results   for the  direct  CP asymmetries
of  the charged  $B^{+} \to K^{+} \, \ell^{+} \ell^{-}$ decays  turn out to be   in excellent  agreement with
the  current   LHCb measurements  \cite{LHCb:2014mit}  from  their Run I dataset   merely.

Interestingly,  our improved  field-theoretic    predictions  for  the $q^2$-binned branching fractions
of the CKM-suppressed  $B^{+} \to \pi^{+} \, \ell^{+} \ell^{-}$ decays   in Table  \ref{tab:B-to-pill-Observables}
appear to   better   accommodate  the  first  LHCb  experimental   measurements  \cite{LHCb:2015hsa}
instead of the  most recent    LHCb  results  obtained  with the full  Run I and II  data   sample \cite{LHCb:2026huw}.
We can further   observe that  the state-of-the-art theory  predictions of   the  direct  CP asymmetries
for  the   $B^{+} \to \pi^{+} \, \ell^{+} \ell^{-}$ decays are compatible with  the encouraging  LHCb measurements \cite{LHCb:2026huw}
at the level of  approximately $(1.1\! - \!2.5) \, \sigma$ depending on the selected intervals of the  dilepton invariant mass.
Our  updated  results   for  the  CP-averaged  isospin asymmetries in these exclusive  $b  \to  s  \, \ell^{+} \ell^{-}$  decays
can be apparently  confronted  with the forthcoming  LHCb   measurements \cite{LHCb:2018roe}.


\begin{table*}[htp]
	\centering
	\renewcommand{\arraystretch}{1.0}
	\setlength{\tabcolsep}{2.5pt}
	\resizebox{1\textwidth}{!}{
		\begin{tabular}{|c|rrc|crc|crc|}
			\hline
			\multirow{2}{*}{$q^2~[\mathrm{GeV}^2]$}
			& \multicolumn{3}{c|}{$10^8\times\mathcal{BR}$}
			& \multicolumn{3}{c|}{$\mathcal A_{\rm CP}$}
			& \multicolumn{3}{c|}{$\mathcal A_{\rm I}$}\\
			\cline{2-10}
			& \multicolumn{1}{c}{\rule[-3pt]{0pt}{14pt}Theory}
            & \multicolumn{2}{c|}{Experiment}
            & Theory $(\times 10^2)$
            & \multicolumn{2}{c|}{Experiment $(\times 10)$} & Theory $(\times 10^2)$
            & \multicolumn{2}{c|}{Experiment}\\
			\hline
			 \multicolumn{10}{|c|}{\rule{0pt}{14pt}$\blue{B^0 \to K^0 \, \ell^+  \,  \ell^-}$} \\
			\hline
			$\rule[0pt]{0pt}{14pt} [1.0,\,\,2.0]$
			& $3.545^{+0.260}_{-0.240}$
			&  \multicolumn{2}{c|}{--}
			& $\phantom + 0.027^{+0.073}_{-0.111}$
			&  \multicolumn{2}{c|}{--}
			& $-0.949^{+0.126}_{-0.128}$
			&  \multicolumn{2}{c|}{--} \\
			$[2.0,\,\,3.0]$
			& $3.579^{+0.256}_{-0.235}$
			&  \multicolumn{2}{c|}{--}
			& $\phantom + 0.053^{+0.068}_{-0.103}$
			&  \multicolumn{2}{c|}{--}
			& $-0.533^{+0.069}_{-0.046}$
			&  \multicolumn{2}{c|}{--}\\
			$[3.0,\,\,4.0]$
			& $3.606^{+0.253}_{-0.230}$
			&  \multicolumn{2}{c|}{--}
			& $\phantom + 0.070^{+0.065}_{-0.098}$
			&  \multicolumn{2}{c|}{--}
			& $-0.338^{+0.071}_{-0.055}$
			&  \multicolumn{2}{c|}{--} \\
			$[4.0,\,\,5.0]$
			& $3.628^{+0.252}_{-0.227}$
			&  \multicolumn{2}{c|}{--}
			& $\phantom + 0.082^{+0.063}_{-0.094}$
			&  \multicolumn{2}{c|}{--}
			& $-0.262^{+0.041}_{-0.033}$
			&  \multicolumn{2}{c|}{--} \\
			$\rule[-6pt]{0pt}{14pt}[5.0,\,\,6.0]$
			& $3.648^{+0.253}_{-0.227}$
			&  \multicolumn{2}{c|}{--}
			& $\phantom + 0.092^{+0.061}_{-0.091}$
			&  \multicolumn{2}{c|}{--}
			& $-0.223^{+0.020}_{-0.016}$
			&  \multicolumn{2}{c|}{--}\\
			\hline
			\multirow{2}{*}{$[0.1,\,\,2.0]$}
			& \multirow{2}{*}{$6.736^{+0.500}_{-0.463}$}
			& \multirow{2}{*}{$2.32^{+1.13}_{-0.99}$}
            & \multirow{2}{*}{\cite{LHCb:2014cxe}}
			& \multirow{2}{*}{$\phantom + 0.001^{+0.076}_{-0.119}$}
			& \multicolumn{2}{c|}{\multirow{2}{*}{--}}
			& \multirow{2}{*}{$-0.763^{+0.287}_{-0.316}$}
			& $-0.37^{+0.18}_{-0.21}$
            & \cite{LHCb:2014cxe} \\
			& & & & & & & & $-0.51^{+0.49}_{-0.95}$
            & \cite{BaBar:2012mrf}\\
			\hline
			\rule[0pt]{0pt}{14pt}$[0.0,\,\,2.0]$
			& $7.105^{+0.528}_{-0.488}$
			& $3.12\pm3.73$
            & \cite{CDF:2011buy}
			& $-0.004^{+0.077}_{-0.121}$
			& \multicolumn{2}{c|}{--}
			& $-0.547^{+0.286}_{-0.297}$
			& \multicolumn{2}{c|}{--} \\
			$[2.0,\,\,4.0]$
			& $7.185^{+0.509}_{-0.464}$
			& $3.74^{+1.11}_{-1.00}$
            & \cite{LHCb:2014cxe}
			& $\phantom + 0.062^{+0.067}_{-0.100}$
			& \multicolumn{2}{c|}{--}
			& $-0.435^{+0.069}_{-0.050}$
			& $-0.15^{+0.13}_{-0.15}$
            & \cite{LHCb:2014cxe}\\
			$[4.0,\,\,6.0]$
			& $7.276^{+0.504}_{-0.453}$
			& $3.46^{+1.08}_{-0.98}$
            & \cite{LHCb:2014cxe}
			& $\phantom + 0.087^{+0.062}_{-0.092}$
			& \multicolumn{2}{c|}{--}
			& $-0.242^{+0.030}_{-0.024}$
			& $-0.10^{+0.13}_{-0.16}$
            & \cite{LHCb:2014cxe}\\
			$[2.0,\,\,4.3]$
			& $8.272^{+0.585}_{-0.533}$
			& $9.29\pm4.90$
            & \cite{CDF:2011buy}
			& $\phantom + 0.064^{+0.066}_{-0.099}$
			& \multicolumn{2}{c|}{--}
			& $-0.415^{+0.067}_{-0.049}$
			& $-0.73^{+0.48}_{-0.55}$
            & \cite{BaBar:2012mrf}\\
			\rule[-6pt]{0pt}{14pt}$[0.1,\,\,4.0]$
			& $13.922^{+1.006}_{-0.924}$
			& $12.40^{+6.01}_{-4.62}$
            & \cite{BELLE:2019xld}
			& $\phantom + 0.032^{+0.071}_{-0.109}$
			& \multicolumn{2}{c|}{--}
			& $-0.594^{+0.119}_{-0.122}$
			& $-0.11^{+0.20}_{-0.17}$
            & \cite{BELLE:2019xld}\\
			\hline
			\multirow{2}{*}{$[1.0,\,\,6.0]$}
			& \multirow{2}{*}{$18.006^{+1.268}_{-1.151}$}
			& $6.20^{+4.40}_{-3.21}$
            & \cite{BELLE:2019xld}
			& \multirow{2}{*}{$\phantom + 0.065^{+0.066}_{-0.099}$}
			& \multicolumn{2}{c|}{\multirow{2}{*}{--}}
			& \multirow{2}{*}{$-0.459^{+0.021}_{-0.020}$}
			& $-0.53^{+0.20}_{-0.17}$
            & \cite{BELLE:2019xld}\\
			& & $9.80\pm6.19$
            & \cite{CDF:2011buy} & & & &
            & $-0.41\pm 0.25$
            & \cite{BaBar:2012mrf}\\
			\hline
			\multicolumn{10}{|c|}{\rule[0pt]{0pt}{14pt}$\blue{B^{+} \to K^{+} \, \ell^+  \,  \ell^-}$}\\
			\hline
			\multirow{2}{*}{$[1.1,\,\,2.0]$}
			& \multirow{2}{*}{$3.506^{+0.256}_{-0.236}$}
			& $2.10\pm0.17$
            & \cite{LHCb:2014cxe}
			& \multirow{2}{*}{$\phantom + 0.720^{+0.168}_{-0.233}$}
			& \multirow{2}{*}{$-0.04\pm0.68$}
            & \multirow{2}{*}{\cite{LHCb:2014mit}}
			& \multirow{2}{*}{--}
			& \multicolumn{2}{c|}{\multirow{2}{*}{--}} \\
			& & $1.93\pm0.20$
            & \cite{CMS:2024syx} & & & & & & \\
			\hline
			\multirow{2}{*}{$[2.0,\,\,3.0]$}
			& \multirow{2}{*}{$3.901^{+0.278}_{-0.255}$}
			& $2.82\pm0.21$
            & \cite{LHCb:2014cxe}
			& \multirow{2}{*}{$\phantom + 0.373^{+0.115}_{-0.154}$}
			& \multirow{2}{*}{$0.42\pm0.59$}
            & \multirow{2}{*}{\cite{LHCb:2014mit}}
			& \multirow{2}{*}{--}
			& \multicolumn{2}{c|}{\multirow{2}{*}{--}} \\
			& & $3.06\pm0.25$
            & \cite{CMS:2024syx} & & & & & &\\
			\hline
			\multirow{2}{*}{$[3.0,\,\,4.0]$}
			& \multirow{2}{*}{$3.915^{+0.275}_{-0.249}$}
			& $2.54\pm0.20$
            &\cite{LHCb:2014cxe}
			& \multirow{2}{*}{$\phantom + 0.308^{+0.069}_{-0.103}$}
			& \multirow{2}{*}{$-0.34\pm0.63$}
            & \multirow{2}{*}{\cite{LHCb:2014mit}}
			& \multirow{2}{*}{--}
			& \multicolumn{2}{c|}{\multirow{2}{*}{--}}\\
			& & $2.54\pm0.23$
            & \cite{CMS:2024syx} & & & & & &\\
			\hline
			\multirow{2}{*}{$[4.0,\,\,5.0]$}
			& \multirow{2}{*}{$3.933^{+0.273}_{-0.246}$}
			& $2.21\pm0.18$
            & \cite{LHCb:2014cxe}
			& \multirow{2}{*}{$\phantom + 0.288^{+0.063}_{-0.094}$}
			& \multirow{2}{*}{$-0.21\pm0.64$}
            &  \multirow{2}{*}{\cite{LHCb:2014mit}}
			& \multirow{2}{*}{--}
			& \multicolumn{2}{c|}{\multirow{2}{*}{--}}\\
			& & $2.47\pm0.24$
            & \cite{CMS:2024syx} & & & & & &\\
			\hline
			\multirow{2}{*}{$[5.0,\,\,6.0]$}
			& \multirow{2}{*}{$3.951^{+0.274}_{-0.245}$}
			& $2.31\pm0.18$
            & \cite{LHCb:2014cxe}
			& \multirow{2}{*}{$\phantom + 0.265^{+0.062}_{-0.092}$}
			& \multirow{2}{*}{$0.31\pm0.62$}
            & \multirow{2}{*}{\cite{LHCb:2014mit}}
			& \multirow{2}{*}{--}
			& \multicolumn{2}{c|}{\multirow{2}{*}{--}}\\
			& & $2.53\pm0.27$
            & \cite{CMS:2024syx} & & & & & &\\
			\hline
			\rule[0pt]{0pt}{14pt}$[0.0,\,\,2.0]$
			& $7.746^{+0.578}_{-0.534}$
			& $3.60\pm1.14$
            & \cite{CDF:2011buy}
			& $\phantom + 1.674^{+0.104}_{-0.142}$
			& \multicolumn{2}{c|}{--}
			& --
			& \multicolumn{2}{c|}{--} \\
			$[2.0,\,\,4.3]$
			& $8.994^{+0.634}_{-0.577}$
			& $8.00\pm1.58$
            & \cite{CDF:2011buy}
			& $\phantom + 0.335^{+0.083}_{-0.118}$
			& \multicolumn{2}{c|}{--}
			& --
			& \multicolumn{2}{c|}{--} \\
			\rule[-6pt]{0pt}{14pt}$[0.1,\,\,4.0]$
			& $15.192^{+1.099}_{-1.008}$
			& $17.60^{+4.12}_{-3.72}$
            & \cite{BELLE:2019xld}
			& $\phantom + 0.930^{+0.074}_{-0.117}$
			& \multicolumn{2}{c|}{--}
			& --
			& \multicolumn{2}{c|}{--} \\
			\hline
			\multirow{2}{*}{$[1.0,\,\,6.0]$}
			& \multirow{2}{*}{$19.596^{+1.378}_{-1.250}$}
			& $23.00^{+4.13}_{-3.83}$
            & \cite{BELLE:2019xld}
			& \multirow{2}{*}{$\phantom + 0.400^{+0.081}_{-0.118}$}
			& \multicolumn{2}{c|}{\multirow{2}{*}{--}}
			& \multirow{2}{*}{--}
			& \multicolumn{2}{c|}{\multirow{2}{*}{--}}\\
			& & $14.10\pm2.24$
            & \cite{CDF:2011buy} & & & & & &\\
			\hline
		\end{tabular}
	}
\caption{Theory predictions for  the CP-averaged branching fractions ${\cal BR}$, the direct CP asymmetries ${\cal A}_{\rm CP}$
and  the isospin asymmetries ${\cal A}_{\rm I}$ of the electroweak penguin  $B \to K \, \ell^+ \, \ell^-$ decays obtained  by
combining  the unsuppressed  three-particle twist-three  contribution in the heavy quark expansion
and the various NNLO QCD corrections  to  the  leading two-particle valence   contributions
with  the previously computed leading-power contributions in the  QCD factorization framework \cite{Beneke:2001at,Beneke:2004dp,Huang:2024xii}.
The   available  experimental  measurements for these exclusive decay observables  from
the Belle \cite{BELLE:2019xld},   BaBar \cite{BaBar:2012mrf},  CDF \cite{CDF:2011buy}, LHCb \cite{LHCb:2014cxe,LHCb:2014mit}
and CMS \cite{CMS:2024syx}  Collaborations in  the  large  hadronic recoil region
are further displayed  for an instructive comparison.}
\label{tab:B-to-Kll-Observables}
\end{table*}
\renewcommand{\arraystretch}{1.0}

\begin{table*}[htp]
	\centering
	\renewcommand{\arraystretch}{1.0}
	\setlength{\tabcolsep}{6pt}
	\resizebox{1\textwidth}{!}{
		\begin{tabular}{|c|ccc|ccc|ccc|}
			\hline
			\multirow{2}{*}{$q^2~[\mathrm{GeV}^2]$}
			& \multicolumn{3}{c|}{$10^9\times\mathcal{BR}$}
			& \multicolumn{3}{c|}{$\mathcal A_{\rm CP}$}
			& \multicolumn{3}{c|}{$\mathcal A_{\rm I}$}\\
			\cline{2-10}
			& \rule[-3pt]{0pt}{14pt}Theory &  \multicolumn{2}{c|}{Experiment} & Theory &  \multicolumn{2}{c|}{Experiment} & Theory & \multicolumn{2}{c|}{Experiment}\\
			\hline
			\multicolumn{10}{|c|}{\rule[0pt]{0pt}{14pt}$\blue{B^0  \to  \pi^0  \, \ell^+  \, \ell^-}$}\\
			\hline
			\rule[0pt]{0pt}{14pt}$[1.0,\,\,2.0]$
			& $0.290^{+0.064}_{-0.059}$
			& \multicolumn{2}{c|}{--}
			& $-0.028^{+0.024}_{-0.016}$
			& \multicolumn{2}{c|}{--}
			& $-0.048^{+0.008}_{-0.009}$
			&  \multicolumn{2}{c|}{--} \\
			$[2.0,\,\,3.0]$
			& $0.303^{+0.062}_{-0.058}$
			& \multicolumn{2}{c|}{--}
			& $-0.025^{+0.022}_{-0.015}$
			& \multicolumn{2}{c|}{--}
			& $-0.019^{+0.005}_{-0.005}$
			& \multicolumn{2}{c|}{--} \\
			$[3.0,\,\,4.0]$
			& $0.315^{+0.060}_{-0.056}$
			& \multicolumn{2}{c|}{--}
			& $-0.024^{+0.022}_{-0.015}$
			& \multicolumn{2}{c|}{--}
			& $-0.011^{+0.003}_{-0.003}$
			& \multicolumn{2}{c|}{--} \\
			$[4.0,\,\,5.0]$
			& $0.327^{+0.059}_{-0.055}$
			& \multicolumn{2}{c|}{--}
			& $-0.025^{+0.021}_{-0.014}$
			& \multicolumn{2}{c|}{--}
			& $-0.008^{+0.002}_{-0.002}$
			& \multicolumn{2}{c|}{--} \\
            $[5.0,\,\,6.0]$
			& $0.339^{+0.058}_{-0.054}$
			& \multicolumn{2}{c|}{--}
			& $-0.026^{+0.020}_{-0.014}$
			& \multicolumn{2}{c|}{--}
			& $-0.007^{+0.001}_{-0.001}$
			& \multicolumn{2}{c|}{--} \\
            \rule[-6pt]{0pt}{14pt}$[1.0,\,\,6.0]$
            & $1.573^{+0.302}_{-0.281}$
            & \multicolumn{2}{c|}{--}
            & $-0.025^{+0.022}_{-0.015}$
            & \multicolumn{2}{c|}{--}
            & $-0.018^{+0.003}_{-0.003}$
            & \multicolumn{2}{c|}{--}\\
			\hline
			\multicolumn{10}{|c|}{$\blue{B^{+} \to \pi^{+} \, \ell^+  \, \ell^-}$}\\
			\hline
			\rule[0pt]{0pt}{14pt}$[1.0,\,\,2.0]$
			& $0.690^{+0.141}_{-0.130}$
			& \multicolumn{2}{c|}{--}
			& $-0.225^{+0.066}_{-0.050}$
			& \multicolumn{2}{c|}{--}
			& --
			& \multicolumn{2}{c|}{--}\\
			$[2.0,\,\,3.0]$
			& $0.678^{+0.136}_{-0.126}$
			& \multicolumn{2}{c|}{--}
			& $-0.111^{+0.042}_{-0.033}$
			& \multicolumn{2}{c|}{--}
			& --
			& \multicolumn{2}{c|}{--} \\
			$[3.0,\,\,4.0]$
			& $0.694^{+0.132}_{-0.122}$
			& \multicolumn{2}{c|}{--}
			& $-0.090^{+0.025}_{-0.017}$
			& \multicolumn{2}{c|}{--}
			& --
			& \multicolumn{2}{c|}{--} \\
			$[4.0,\,\,5.0]$
			& $0.716^{+0.128}_{-0.119}$
			& \multicolumn{2}{c|}{--}
			& $-0.082^{+0.022}_{-0.015}$
			& \multicolumn{2}{c|}{--}
			& --
			& \multicolumn{2}{c|}{--} \\
			\rule[-6pt]{0pt}{14pt}$[5.0,\,\,6.0]$
			& $0.741^{+0.125}_{-0.117}$
			& \multicolumn{2}{c|}{--}
			& $-0.073^{+0.021}_{-0.015}$
			& \multicolumn{2}{c|}{--}
			& --
			& \multicolumn{2}{c|}{--} \\
			\hline
			\multirow{2}{*}{$[0.1,\,\,2.0]$}
			& \multirow{2}{*}{$1.484^{+0.282}_{-0.257}$}
			& $3.59^{+0.90}_{-0.79}$ & ~\cite{LHCb:2015hsa}
			& \multirow{2}{*}{$-0.401^{+0.036}_{-0.030}$}
			& \multirow{2}{*}{$-0.09\pm0.12$} & \multirow{2}{*}{\cite{LHCb:2026huw}}
			& \multirow{2}{*}{--}
			& \multicolumn{2}{c|}{\multirow{2}{*}{--}} \\
			& & $3.71^{+0.49}_{-0.46}$ & \cite{LHCb:2026huw} & & &  & & &\\
			\hline
			\multirow{2}{*}{$[2.0,\,\,4.0]$}
			& \multirow{2}{*}{$1.371^{+0.267}_{-0.248}$}
			& $1.24^{+0.78}_{-0.66}$ & \cite{LHCb:2015hsa}
			& \multirow{2}{*}{$-0.100^{+0.032}_{-0.024}$}
			& \multirow{2}{*}{$-0.27\pm0.15$}
            & \multirow{2}{*}{\cite{LHCb:2026huw}}
			& \multirow{2}{*}{--}
			& \multicolumn{2}{c|}{\multirow{2}{*}{--}} \\
			& & $2.68^{+0.45}_{-0.41}$ & \cite{LHCb:2026huw} & & & & & &\\
			\hline
			\multirow{2}{*}{$[4.0,\,\,6.0]$}
			& \multirow{2}{*}{$1.457^{+0.253}_{-0.236}$}
			& $1.70^{+0.64}_{-0.54}$
            & \cite{LHCb:2015hsa}
			& \multirow{2}{*}{$-0.078^{+0.021}_{-0.015}$}
			& \multirow{2}{*}{$\phantom + 0.16\pm0.22$}
            &  \multirow{2}{*}{\cite{LHCb:2026huw}}
			& \multirow{2}{*}{--}
			& \multicolumn{2}{c|}{\multirow{2}{*}{--}} \\
			& & $1.26^{+0.29}_{-0.27}$
            & \cite{LHCb:2026huw} & & & & & &\\
			\hline
			\rule[-3pt]{0pt}{14pt}$[1.0,\,\,6.0]$
			& $3.518^{+0.659}_{-0.612}$
			& $4.55^{+1.06}_{-1.01}$
            & \cite{LHCb:2015hsa}
			& $-0.115^{+0.030}_{-0.022}$
			& \multicolumn{2}{c|}{--}
			& --
			& \multicolumn{2}{c|}{--} \\
			\hline
		\end{tabular}
	}
\caption{Theory predictions for  the CP-averaged branching fractions ${\cal BR}$, the direct CP asymmetries ${\cal A}_{\rm CP}$
and  the isospin asymmetries ${\cal A}_{\rm I}$ of the rare semileptonic  $B \to \pi \, \ell^+ \, \ell^-$ decays obtained from
adding  the unsuppressed  non-valence Fock state contribution in the  ${\Lambda_{\rm QCD} / m_b}$  expansion
and the various ${\cal O}(\alpha_s^2)$ radiative  corrections  to  the  leading  two-particle valence   contributions
on top of the previously determined    leading-power contributions with the QCD factorization method \cite{Beneke:2001at,Beneke:2004dp,Huang:2024xii}.
The currently available  experimental  measurements for these exclusive decay observables  from
the  LHCb \cite{LHCb:2015hsa,LHCb:2026huw}  Collaboration in  the kinematically enhanced large  hadronic recoil region
are further displayed  for an exploratory  comparison.}
\label{tab:B-to-pill-Observables}
\end{table*}
\renewcommand{\arraystretch}{1.0}

\end{widetext}

\bibliographystyle{apsrev4-1}

\FloatBarrier
\bibliography{References}

\begin{thebibliography}{143}%
\makeatletter
\providecommand \@ifxundefined [1]{%
 \@ifx{#1\undefined}
}%
\providecommand \@ifnum [1]{%
 \ifnum #1\expandafter \@firstoftwo
 \else \expandafter \@secondoftwo
 \fi
}%
\providecommand \@ifx [1]{%
 \ifx #1\expandafter \@firstoftwo
 \else \expandafter \@secondoftwo
 \fi
}%
\providecommand \natexlab [1]{#1}%
\providecommand \enquote  [1]{``#1''}%
\providecommand \bibnamefont  [1]{#1}%
\providecommand \bibfnamefont [1]{#1}%
\providecommand \citenamefont [1]{#1}%
\providecommand \href@noop [0]{\@secondoftwo}%
\providecommand \href [0]{\begingroup \@sanitize@url \@href}%
\providecommand \@href[1]{\@@startlink{#1}\@@href}%
\providecommand \@@href[1]{\endgroup#1\@@endlink}%
\providecommand \@sanitize@url [0]{\catcode `\\12\catcode `\$12\catcode
  `\&12\catcode `\#12\catcode `\^12\catcode `\_12\catcode `\%12\relax}%
\providecommand \@@startlink[1]{}%
\providecommand \@@endlink[0]{}%
\providecommand \url  [0]{\begingroup\@sanitize@url \@url }%
\providecommand \@url [1]{\endgroup\@href {#1}{\urlprefix }}%
\providecommand \urlprefix  [0]{URL }%
\providecommand \Eprint [0]{\href }%
\providecommand \doibase [0]{http://dx.doi.org/}%
\providecommand \selectlanguage [0]{\@gobble}%
\providecommand \bibinfo  [0]{\@secondoftwo}%
\providecommand \bibfield  [0]{\@secondoftwo}%
\providecommand \translation [1]{[#1]}%
\providecommand \BibitemOpen [0]{}%
\providecommand \bibitemStop [0]{}%
\providecommand \bibitemNoStop [0]{.\EOS\space}%
\providecommand \EOS [0]{\spacefactor3000\relax}%
\providecommand \BibitemShut  [1]{\csname bibitem#1\endcsname}%
\let\auto@bib@innerbib\@empty
\bibitem [{\citenamefont {Aaij}\ \emph {et~al.}(2020)\citenamefont {Aaij} \emph
  {et~al.}}]{LHCb:2020lmf}%
  \BibitemOpen
  \bibfield  {author} {\bibinfo {author} {\bibfnamefont {R.}~\bibnamefont
  {Aaij}} \emph {et~al.} (\bibinfo {collaboration} {LHCb}),\ }\href {\doibase
  10.1103/PhysRevLett.125.011802} {\bibfield  {journal} {\bibinfo  {journal}
  {Phys. Rev. Lett.}\ }\textbf {\bibinfo {volume} {125}},\ \bibinfo {pages}
  {011802} (\bibinfo {year} {2020})},\ \Eprint
  {http://arxiv.org/abs/2003.04831} {arXiv:2003.04831 [hep-ex]} \BibitemShut
  {NoStop}%
\bibitem [{\citenamefont {Aaij}\ \emph {et~al.}(2025)\citenamefont {Aaij} \emph
  {et~al.}}]{LHCb:2025mqb}%
  \BibitemOpen
  \bibfield  {author} {\bibinfo {author} {\bibfnamefont {R.}~\bibnamefont
  {Aaij}} \emph {et~al.} (\bibinfo {collaboration} {LHCb}),\ }\href@noop {} {\
  (\bibinfo {year} {2025})},\ \Eprint {http://arxiv.org/abs/2512.18053}
  {arXiv:2512.18053 [hep-ex]} \BibitemShut {NoStop}%
\bibitem [{\citenamefont {Hayrapetyan}\ \emph {et~al.}(2025)\citenamefont
  {Hayrapetyan} \emph {et~al.}}]{CMS:2024atz}%
  \BibitemOpen
  \bibfield  {author} {\bibinfo {author} {\bibfnamefont {A.}~\bibnamefont
  {Hayrapetyan}} \emph {et~al.} (\bibinfo {collaboration} {CMS}),\ }\href
  {\doibase 10.1016/j.physletb.2025.139406} {\bibfield  {journal} {\bibinfo
  {journal} {Phys. Lett. B}\ }\textbf {\bibinfo {volume} {864}},\ \bibinfo
  {pages} {139406} (\bibinfo {year} {2025})},\ \Eprint
  {http://arxiv.org/abs/2411.11820} {arXiv:2411.11820 [hep-ex]} \BibitemShut
  {NoStop}%
\bibitem [{\citenamefont {Aaij}\ \emph
  {et~al.}(2014{\natexlab{a}})\citenamefont {Aaij} \emph
  {et~al.}}]{LHCb:2014cxe}%
  \BibitemOpen
  \bibfield  {author} {\bibinfo {author} {\bibfnamefont {R.}~\bibnamefont
  {Aaij}} \emph {et~al.} (\bibinfo {collaboration} {LHCb}),\ }\href {\doibase
  10.1007/JHEP06(2014)133} {\bibfield  {journal} {\bibinfo  {journal} {JHEP}\
  }\textbf {\bibinfo {volume} {06}},\ \bibinfo {pages} {133} (\bibinfo {year}
  {2014}{\natexlab{a}})},\ \Eprint {http://arxiv.org/abs/1403.8044}
  {arXiv:1403.8044 [hep-ex]} \BibitemShut {NoStop}%
\bibitem [{\citenamefont {Aaij}\ \emph {et~al.}(2021)\citenamefont {Aaij} \emph
  {et~al.}}]{LHCb:2021zwz}%
  \BibitemOpen
  \bibfield  {author} {\bibinfo {author} {\bibfnamefont {R.}~\bibnamefont
  {Aaij}} \emph {et~al.} (\bibinfo {collaboration} {LHCb}),\ }\href {\doibase
  10.1103/PhysRevLett.127.151801} {\bibfield  {journal} {\bibinfo  {journal}
  {Phys. Rev. Lett.}\ }\textbf {\bibinfo {volume} {127}},\ \bibinfo {pages}
  {151801} (\bibinfo {year} {2021})},\ \Eprint
  {http://arxiv.org/abs/2105.14007} {arXiv:2105.14007 [hep-ex]} \BibitemShut
  {NoStop}%
\bibitem [{\citenamefont {Aaij}\ \emph {et~al.}(2015)\citenamefont {Aaij} \emph
  {et~al.}}]{LHCb:2015hsa}%
  \BibitemOpen
  \bibfield  {author} {\bibinfo {author} {\bibfnamefont {R.}~\bibnamefont
  {Aaij}} \emph {et~al.} (\bibinfo {collaboration} {LHCb}),\ }\href {\doibase
  10.1007/JHEP10(2015)034} {\bibfield  {journal} {\bibinfo  {journal} {JHEP}\
  }\textbf {\bibinfo {volume} {10}},\ \bibinfo {pages} {034} (\bibinfo {year}
  {2015})},\ \Eprint {http://arxiv.org/abs/1509.00414} {arXiv:1509.00414
  [hep-ex]} \BibitemShut {NoStop}%
\bibitem [{\citenamefont {Aaij}\ \emph
  {et~al.}(2026{\natexlab{a}})\citenamefont {Aaij} \emph
  {et~al.}}]{LHCb:2026xvw}%
  \BibitemOpen
  \bibfield  {author} {\bibinfo {author} {\bibfnamefont {R.}~\bibnamefont
  {Aaij}} \emph {et~al.} (\bibinfo {collaboration} {LHCb}),\ }\href@noop {} {\
  (\bibinfo {year} {2026}{\natexlab{a}})},\ \Eprint
  {http://arxiv.org/abs/2604.21987} {arXiv:2604.21987 [hep-ex]} \BibitemShut
  {NoStop}%
\bibitem [{\citenamefont {Aaij}\ \emph
  {et~al.}(2026{\natexlab{b}})\citenamefont {Aaij} \emph
  {et~al.}}]{LHCb:2026dfm}%
  \BibitemOpen
  \bibfield  {author} {\bibinfo {author} {\bibfnamefont {R.}~\bibnamefont
  {Aaij}} \emph {et~al.} (\bibinfo {collaboration} {LHCb}),\ }\href@noop {} {\
  (\bibinfo {year} {2026}{\natexlab{b}})},\ \Eprint
  {http://arxiv.org/abs/2604.26784} {arXiv:2604.26784 [hep-ex]} \BibitemShut
  {NoStop}%
\bibitem [{\citenamefont {Aaij}\ \emph
  {et~al.}(2018{\natexlab{a}})\citenamefont {Aaij} \emph
  {et~al.}}]{LHCb:2018rym}%
  \BibitemOpen
  \bibfield  {author} {\bibinfo {author} {\bibfnamefont {R.}~\bibnamefont
  {Aaij}} \emph {et~al.} (\bibinfo {collaboration} {LHCb}),\ }\href {\doibase
  10.1007/JHEP07(2018)020} {\bibfield  {journal} {\bibinfo  {journal} {JHEP}\
  }\textbf {\bibinfo {volume} {07}},\ \bibinfo {pages} {020} (\bibinfo {year}
  {2018}{\natexlab{a}})},\ \Eprint {http://arxiv.org/abs/1804.07167}
  {arXiv:1804.07167 [hep-ex]} \BibitemShut {NoStop}%
\bibitem [{\citenamefont {Beneke}\ \emph
  {et~al.}(2001{\natexlab{a}})\citenamefont {Beneke}, \citenamefont
  {Feldmann},\ and\ \citenamefont {Seidel}}]{Beneke:2001at}%
  \BibitemOpen
  \bibfield  {author} {\bibinfo {author} {\bibfnamefont {M.}~\bibnamefont
  {Beneke}}, \bibinfo {author} {\bibfnamefont {T.}~\bibnamefont {Feldmann}}, \
  and\ \bibinfo {author} {\bibfnamefont {D.}~\bibnamefont {Seidel}},\ }\href
  {\doibase 10.1016/S0550-3213(01)00366-2} {\bibfield  {journal} {\bibinfo
  {journal} {Nucl. Phys. B}\ }\textbf {\bibinfo {volume} {612}},\ \bibinfo
  {pages} {25} (\bibinfo {year} {2001}{\natexlab{a}})},\ \Eprint
  {http://arxiv.org/abs/hep-ph/0106067} {arXiv:hep-ph/0106067} \BibitemShut
  {NoStop}%
\bibitem [{\citenamefont {Beneke}\ \emph {et~al.}(2005)\citenamefont {Beneke},
  \citenamefont {Feldmann},\ and\ \citenamefont {Seidel}}]{Beneke:2004dp}%
  \BibitemOpen
  \bibfield  {author} {\bibinfo {author} {\bibfnamefont {M.}~\bibnamefont
  {Beneke}}, \bibinfo {author} {\bibfnamefont {T.}~\bibnamefont {Feldmann}}, \
  and\ \bibinfo {author} {\bibfnamefont {D.}~\bibnamefont {Seidel}},\ }\href
  {\doibase 10.1140/epjc/s2005-02181-5} {\bibfield  {journal} {\bibinfo
  {journal} {Eur. Phys. J. C}\ }\textbf {\bibinfo {volume} {41}},\ \bibinfo
  {pages} {173} (\bibinfo {year} {2005})},\ \Eprint
  {http://arxiv.org/abs/hep-ph/0412400} {arXiv:hep-ph/0412400} \BibitemShut
  {NoStop}%
\bibitem [{\citenamefont {Grinstein}\ and\ \citenamefont
  {Pirjol}(2004)}]{Grinstein:2004vb}%
  \BibitemOpen
  \bibfield  {author} {\bibinfo {author} {\bibfnamefont {B.}~\bibnamefont
  {Grinstein}}\ and\ \bibinfo {author} {\bibfnamefont {D.}~\bibnamefont
  {Pirjol}},\ }\href {\doibase 10.1103/PhysRevD.70.114005} {\bibfield
  {journal} {\bibinfo  {journal} {Phys. Rev. D}\ }\textbf {\bibinfo {volume}
  {70}},\ \bibinfo {pages} {114005} (\bibinfo {year} {2004})},\ \Eprint
  {http://arxiv.org/abs/hep-ph/0404250} {arXiv:hep-ph/0404250} \BibitemShut
  {NoStop}%
\bibitem [{\citenamefont {Beylich}\ \emph {et~al.}(2011)\citenamefont
  {Beylich}, \citenamefont {Buchalla},\ and\ \citenamefont
  {Feldmann}}]{Beylich:2011aq}%
  \BibitemOpen
  \bibfield  {author} {\bibinfo {author} {\bibfnamefont {M.}~\bibnamefont
  {Beylich}}, \bibinfo {author} {\bibfnamefont {G.}~\bibnamefont {Buchalla}}, \
  and\ \bibinfo {author} {\bibfnamefont {T.}~\bibnamefont {Feldmann}},\ }\href
  {\doibase 10.1140/epjc/s10052-011-1635-0} {\bibfield  {journal} {\bibinfo
  {journal} {Eur. Phys. J. C}\ }\textbf {\bibinfo {volume} {71}},\ \bibinfo
  {pages} {1635} (\bibinfo {year} {2011})},\ \Eprint
  {http://arxiv.org/abs/1101.5118} {arXiv:1101.5118 [hep-ph]} \BibitemShut
  {NoStop}%
\bibitem [{\citenamefont {Chay}\ and\ \citenamefont {Kim}(2003)}]{Chay:2003kb}%
  \BibitemOpen
  \bibfield  {author} {\bibinfo {author} {\bibfnamefont {J.-g.}\ \bibnamefont
  {Chay}}\ and\ \bibinfo {author} {\bibfnamefont {C.}~\bibnamefont {Kim}},\
  }\href {\doibase 10.1103/PhysRevD.68.034013} {\bibfield  {journal} {\bibinfo
  {journal} {Phys. Rev. D}\ }\textbf {\bibinfo {volume} {68}},\ \bibinfo
  {pages} {034013} (\bibinfo {year} {2003})},\ \Eprint
  {http://arxiv.org/abs/hep-ph/0305033} {arXiv:hep-ph/0305033} \BibitemShut
  {NoStop}%
\bibitem [{\citenamefont {Becher}\ \emph {et~al.}(2005)\citenamefont {Becher},
  \citenamefont {Hill},\ and\ \citenamefont {Neubert}}]{Becher:2005fg}%
  \BibitemOpen
  \bibfield  {author} {\bibinfo {author} {\bibfnamefont {T.}~\bibnamefont
  {Becher}}, \bibinfo {author} {\bibfnamefont {R.~J.}\ \bibnamefont {Hill}}, \
  and\ \bibinfo {author} {\bibfnamefont {M.}~\bibnamefont {Neubert}},\ }\href
  {\doibase 10.1103/PhysRevD.72.094017} {\bibfield  {journal} {\bibinfo
  {journal} {Phys. Rev. D}\ }\textbf {\bibinfo {volume} {72}},\ \bibinfo
  {pages} {094017} (\bibinfo {year} {2005})},\ \Eprint
  {http://arxiv.org/abs/hep-ph/0503263} {arXiv:hep-ph/0503263} \BibitemShut
  {NoStop}%
\bibitem [{\citenamefont {Ali}\ \emph {et~al.}(2006)\citenamefont {Ali},
  \citenamefont {Kramer},\ and\ \citenamefont {Zhu}}]{Ali:2006ew}%
  \BibitemOpen
  \bibfield  {author} {\bibinfo {author} {\bibfnamefont {A.}~\bibnamefont
  {Ali}}, \bibinfo {author} {\bibfnamefont {G.}~\bibnamefont {Kramer}}, \ and\
  \bibinfo {author} {\bibfnamefont {G.-H.}\ \bibnamefont {Zhu}},\ }\href
  {\doibase 10.1140/epjc/s2006-02596-4} {\bibfield  {journal} {\bibinfo
  {journal} {Eur. Phys. J. C}\ }\textbf {\bibinfo {volume} {47}},\ \bibinfo
  {pages} {625} (\bibinfo {year} {2006})},\ \Eprint
  {http://arxiv.org/abs/hep-ph/0601034} {arXiv:hep-ph/0601034} \BibitemShut
  {NoStop}%
\bibitem [{\citenamefont {Ali}\ \emph {et~al.}(2008)\citenamefont {Ali},
  \citenamefont {Pecjak},\ and\ \citenamefont {Greub}}]{Ali:2007sj}%
  \BibitemOpen
  \bibfield  {author} {\bibinfo {author} {\bibfnamefont {A.}~\bibnamefont
  {Ali}}, \bibinfo {author} {\bibfnamefont {B.~D.}\ \bibnamefont {Pecjak}}, \
  and\ \bibinfo {author} {\bibfnamefont {C.}~\bibnamefont {Greub}},\ }\href
  {\doibase 10.1140/epjc/s10052-008-0623-5} {\bibfield  {journal} {\bibinfo
  {journal} {Eur. Phys. J. C}\ }\textbf {\bibinfo {volume} {55}},\ \bibinfo
  {pages} {577} (\bibinfo {year} {2008})},\ \Eprint
  {http://arxiv.org/abs/0709.4422} {arXiv:0709.4422 [hep-ph]} \BibitemShut
  {NoStop}%
\bibitem [{\citenamefont {Beneke}\ \emph {et~al.}(1999)\citenamefont {Beneke},
  \citenamefont {Buchalla}, \citenamefont {Neubert},\ and\ \citenamefont
  {Sachrajda}}]{Beneke:1999br}%
  \BibitemOpen
  \bibfield  {author} {\bibinfo {author} {\bibfnamefont {M.}~\bibnamefont
  {Beneke}}, \bibinfo {author} {\bibfnamefont {G.}~\bibnamefont {Buchalla}},
  \bibinfo {author} {\bibfnamefont {M.}~\bibnamefont {Neubert}}, \ and\
  \bibinfo {author} {\bibfnamefont {C.~T.}\ \bibnamefont {Sachrajda}},\ }\href
  {\doibase 10.1103/PhysRevLett.83.1914} {\bibfield  {journal} {\bibinfo
  {journal} {Phys. Rev. Lett.}\ }\textbf {\bibinfo {volume} {83}},\ \bibinfo
  {pages} {1914} (\bibinfo {year} {1999})},\ \Eprint
  {http://arxiv.org/abs/hep-ph/9905312} {arXiv:hep-ph/9905312} \BibitemShut
  {NoStop}%
\bibitem [{\citenamefont {Beneke}\ \emph {et~al.}(2000)\citenamefont {Beneke},
  \citenamefont {Buchalla}, \citenamefont {Neubert},\ and\ \citenamefont
  {Sachrajda}}]{Beneke:2000ry}%
  \BibitemOpen
  \bibfield  {author} {\bibinfo {author} {\bibfnamefont {M.}~\bibnamefont
  {Beneke}}, \bibinfo {author} {\bibfnamefont {G.}~\bibnamefont {Buchalla}},
  \bibinfo {author} {\bibfnamefont {M.}~\bibnamefont {Neubert}}, \ and\
  \bibinfo {author} {\bibfnamefont {C.~T.}\ \bibnamefont {Sachrajda}},\ }\href
  {\doibase 10.1016/S0550-3213(00)00559-9} {\bibfield  {journal} {\bibinfo
  {journal} {Nucl. Phys. B}\ }\textbf {\bibinfo {volume} {591}},\ \bibinfo
  {pages} {313} (\bibinfo {year} {2000})},\ \Eprint
  {http://arxiv.org/abs/hep-ph/0006124} {arXiv:hep-ph/0006124} \BibitemShut
  {NoStop}%
\bibitem [{\citenamefont {Beneke}\ \emph
  {et~al.}(2001{\natexlab{b}})\citenamefont {Beneke}, \citenamefont {Buchalla},
  \citenamefont {Neubert},\ and\ \citenamefont {Sachrajda}}]{Beneke:2001ev}%
  \BibitemOpen
  \bibfield  {author} {\bibinfo {author} {\bibfnamefont {M.}~\bibnamefont
  {Beneke}}, \bibinfo {author} {\bibfnamefont {G.}~\bibnamefont {Buchalla}},
  \bibinfo {author} {\bibfnamefont {M.}~\bibnamefont {Neubert}}, \ and\
  \bibinfo {author} {\bibfnamefont {C.~T.}\ \bibnamefont {Sachrajda}},\ }\href
  {\doibase 10.1016/S0550-3213(01)00251-6} {\bibfield  {journal} {\bibinfo
  {journal} {Nucl. Phys. B}\ }\textbf {\bibinfo {volume} {606}},\ \bibinfo
  {pages} {245} (\bibinfo {year} {2001}{\natexlab{b}})},\ \Eprint
  {http://arxiv.org/abs/hep-ph/0104110} {arXiv:hep-ph/0104110} \BibitemShut
  {NoStop}%
\bibitem [{\citenamefont {Beneke}\ and\ \citenamefont
  {Neubert}(2003)}]{Beneke:2003zv}%
  \BibitemOpen
  \bibfield  {author} {\bibinfo {author} {\bibfnamefont {M.}~\bibnamefont
  {Beneke}}\ and\ \bibinfo {author} {\bibfnamefont {M.}~\bibnamefont
  {Neubert}},\ }\href {\doibase 10.1016/j.nuclphysb.2003.09.026} {\bibfield
  {journal} {\bibinfo  {journal} {Nucl. Phys. B}\ }\textbf {\bibinfo {volume}
  {675}},\ \bibinfo {pages} {333} (\bibinfo {year} {2003})},\ \Eprint
  {http://arxiv.org/abs/hep-ph/0308039} {arXiv:hep-ph/0308039} \BibitemShut
  {NoStop}%
\bibitem [{\citenamefont {L{\"u}}\ \emph {et~al.}(2023)\citenamefont {L{\"u}},
  \citenamefont {Shen}, \citenamefont {Wang},\ and\ \citenamefont
  {Wang}}]{Lu:2022kos}%
  \BibitemOpen
  \bibfield  {author} {\bibinfo {author} {\bibfnamefont {C.-D.}\ \bibnamefont
  {L{\"u}}}, \bibinfo {author} {\bibfnamefont {Y.-L.}\ \bibnamefont {Shen}},
  \bibinfo {author} {\bibfnamefont {C.}~\bibnamefont {Wang}}, \ and\ \bibinfo
  {author} {\bibfnamefont {Y.-M.}\ \bibnamefont {Wang}},\ }\href {\doibase
  10.1016/j.nuclphysb.2023.116175} {\bibfield  {journal} {\bibinfo  {journal}
  {Nucl. Phys. B}\ }\textbf {\bibinfo {volume} {990}},\ \bibinfo {pages}
  {116175} (\bibinfo {year} {2023})},\ \Eprint
  {http://arxiv.org/abs/2202.08073} {arXiv:2202.08073 [hep-ph]} \BibitemShut
  {NoStop}%
\bibitem [{\citenamefont {Huang}\ \emph {et~al.}(2025)\citenamefont {Huang},
  \citenamefont {Shen}, \citenamefont {Wang},\ and\ \citenamefont
  {Wang}}]{Huang:2024xii}%
  \BibitemOpen
  \bibfield  {author} {\bibinfo {author} {\bibfnamefont {Y.-K.}\ \bibnamefont
  {Huang}}, \bibinfo {author} {\bibfnamefont {Y.-L.}\ \bibnamefont {Shen}},
  \bibinfo {author} {\bibfnamefont {C.}~\bibnamefont {Wang}}, \ and\ \bibinfo
  {author} {\bibfnamefont {Y.-M.}\ \bibnamefont {Wang}},\ }\href {\doibase
  10.1103/PhysRevLett.134.091901} {\bibfield  {journal} {\bibinfo  {journal}
  {Phys. Rev. Lett.}\ }\textbf {\bibinfo {volume} {134}},\ \bibinfo {pages}
  {091901} (\bibinfo {year} {2025})},\ \Eprint
  {http://arxiv.org/abs/2403.11258} {arXiv:2403.11258 [hep-ph]} \BibitemShut
  {NoStop}%
\bibitem [{\citenamefont {Feldmann}\ and\ \citenamefont
  {Matias}(2003)}]{Feldmann:2002iw}%
  \BibitemOpen
  \bibfield  {author} {\bibinfo {author} {\bibfnamefont {T.}~\bibnamefont
  {Feldmann}}\ and\ \bibinfo {author} {\bibfnamefont {J.}~\bibnamefont
  {Matias}},\ }\href {\doibase 10.1088/1126-6708/2003/01/074} {\bibfield
  {journal} {\bibinfo  {journal} {JHEP}\ }\textbf {\bibinfo {volume} {01}},\
  \bibinfo {pages} {074} (\bibinfo {year} {2003})},\ \Eprint
  {http://arxiv.org/abs/hep-ph/0212158} {arXiv:hep-ph/0212158} \BibitemShut
  {NoStop}%
\bibitem [{\citenamefont {Khodjamirian}\ \emph {et~al.}(2010)\citenamefont
  {Khodjamirian}, \citenamefont {Mannel}, \citenamefont {Pivovarov},\ and\
  \citenamefont {Wang}}]{Khodjamirian:2010vf}%
  \BibitemOpen
  \bibfield  {author} {\bibinfo {author} {\bibfnamefont {A.}~\bibnamefont
  {Khodjamirian}}, \bibinfo {author} {\bibfnamefont {T.}~\bibnamefont
  {Mannel}}, \bibinfo {author} {\bibfnamefont {A.~A.}\ \bibnamefont
  {Pivovarov}}, \ and\ \bibinfo {author} {\bibfnamefont {Y.~M.}\ \bibnamefont
  {Wang}},\ }\href {\doibase 10.1007/JHEP09(2010)089} {\bibfield  {journal}
  {\bibinfo  {journal} {JHEP}\ }\textbf {\bibinfo {volume} {09}},\ \bibinfo
  {pages} {089} (\bibinfo {year} {2010})},\ \Eprint
  {http://arxiv.org/abs/1006.4945} {arXiv:1006.4945 [hep-ph]} \BibitemShut
  {NoStop}%
\bibitem [{\citenamefont {Khodjamirian}\ \emph {et~al.}(2013)\citenamefont
  {Khodjamirian}, \citenamefont {Mannel},\ and\ \citenamefont
  {Wang}}]{Khodjamirian:2012rm}%
  \BibitemOpen
  \bibfield  {author} {\bibinfo {author} {\bibfnamefont {A.}~\bibnamefont
  {Khodjamirian}}, \bibinfo {author} {\bibfnamefont {T.}~\bibnamefont
  {Mannel}}, \ and\ \bibinfo {author} {\bibfnamefont {Y.~M.}\ \bibnamefont
  {Wang}},\ }\href {\doibase 10.1007/JHEP02(2013)010} {\bibfield  {journal}
  {\bibinfo  {journal} {JHEP}\ }\textbf {\bibinfo {volume} {02}},\ \bibinfo
  {pages} {010} (\bibinfo {year} {2013})},\ \Eprint
  {http://arxiv.org/abs/1211.0234} {arXiv:1211.0234 [hep-ph]} \BibitemShut
  {NoStop}%
\bibitem [{\citenamefont {Gubernari}\ \emph {et~al.}(2021)\citenamefont
  {Gubernari}, \citenamefont {van Dyk},\ and\ \citenamefont
  {Virto}}]{Gubernari:2020eft}%
  \BibitemOpen
  \bibfield  {author} {\bibinfo {author} {\bibfnamefont {N.}~\bibnamefont
  {Gubernari}}, \bibinfo {author} {\bibfnamefont {D.}~\bibnamefont {van Dyk}},
  \ and\ \bibinfo {author} {\bibfnamefont {J.}~\bibnamefont {Virto}},\ }\href
  {\doibase 10.1007/JHEP02(2021)088} {\bibfield  {journal} {\bibinfo  {journal}
  {JHEP}\ }\textbf {\bibinfo {volume} {02}},\ \bibinfo {pages} {088} (\bibinfo
  {year} {2021})},\ \Eprint {http://arxiv.org/abs/2011.09813} {arXiv:2011.09813
  [hep-ph]} \BibitemShut {NoStop}%
\bibitem [{\citenamefont {Gubernari}\ \emph {et~al.}(2022)\citenamefont
  {Gubernari}, \citenamefont {Reboud}, \citenamefont {van Dyk},\ and\
  \citenamefont {Virto}}]{Gubernari:2022hxn}%
  \BibitemOpen
  \bibfield  {author} {\bibinfo {author} {\bibfnamefont {N.}~\bibnamefont
  {Gubernari}}, \bibinfo {author} {\bibfnamefont {M.}~\bibnamefont {Reboud}},
  \bibinfo {author} {\bibfnamefont {D.}~\bibnamefont {van Dyk}}, \ and\
  \bibinfo {author} {\bibfnamefont {J.}~\bibnamefont {Virto}},\ }\href
  {\doibase 10.1007/JHEP09(2022)133} {\bibfield  {journal} {\bibinfo  {journal}
  {JHEP}\ }\textbf {\bibinfo {volume} {09}},\ \bibinfo {pages} {133} (\bibinfo
  {year} {2022})},\ \Eprint {http://arxiv.org/abs/2206.03797} {arXiv:2206.03797
  [hep-ph]} \BibitemShut {NoStop}%
\bibitem [{\citenamefont {Gambino}\ \emph {et~al.}(2012)\citenamefont
  {Gambino}, \citenamefont {Mannel},\ and\ \citenamefont
  {Uraltsev}}]{Gambino:2012rd}%
  \BibitemOpen
  \bibfield  {author} {\bibinfo {author} {\bibfnamefont {P.}~\bibnamefont
  {Gambino}}, \bibinfo {author} {\bibfnamefont {T.}~\bibnamefont {Mannel}}, \
  and\ \bibinfo {author} {\bibfnamefont {N.}~\bibnamefont {Uraltsev}},\ }\href
  {\doibase 10.1007/JHEP10(2012)169} {\bibfield  {journal} {\bibinfo  {journal}
  {JHEP}\ }\textbf {\bibinfo {volume} {10}},\ \bibinfo {pages} {169} (\bibinfo
  {year} {2012})},\ \Eprint {http://arxiv.org/abs/1206.2296} {arXiv:1206.2296
  [hep-ph]} \BibitemShut {NoStop}%
\bibitem [{\citenamefont {Benzke}\ \emph {et~al.}(2010)\citenamefont {Benzke},
  \citenamefont {Lee}, \citenamefont {Neubert},\ and\ \citenamefont
  {Paz}}]{Benzke:2010js}%
  \BibitemOpen
  \bibfield  {author} {\bibinfo {author} {\bibfnamefont {M.}~\bibnamefont
  {Benzke}}, \bibinfo {author} {\bibfnamefont {S.~J.}\ \bibnamefont {Lee}},
  \bibinfo {author} {\bibfnamefont {M.}~\bibnamefont {Neubert}}, \ and\
  \bibinfo {author} {\bibfnamefont {G.}~\bibnamefont {Paz}},\ }\href {\doibase
  10.1007/JHEP08(2010)099} {\bibfield  {journal} {\bibinfo  {journal} {JHEP}\
  }\textbf {\bibinfo {volume} {08}},\ \bibinfo {pages} {099} (\bibinfo {year}
  {2010})},\ \Eprint {http://arxiv.org/abs/1003.5012} {arXiv:1003.5012
  [hep-ph]} \BibitemShut {NoStop}%
\bibitem [{\citenamefont {Kozachuk}\ and\ \citenamefont
  {Melikhov}(2018)}]{Kozachuk:2018yxf}%
  \BibitemOpen
  \bibfield  {author} {\bibinfo {author} {\bibfnamefont {A.}~\bibnamefont
  {Kozachuk}}\ and\ \bibinfo {author} {\bibfnamefont {D.}~\bibnamefont
  {Melikhov}},\ }\href {\doibase 10.1016/j.physletb.2018.10.026} {\bibfield
  {journal} {\bibinfo  {journal} {Phys. Lett. B}\ }\textbf {\bibinfo {volume}
  {786}},\ \bibinfo {pages} {378} (\bibinfo {year} {2018})},\ \Eprint
  {http://arxiv.org/abs/1805.05720} {arXiv:1805.05720 [hep-ph]} \BibitemShut
  {NoStop}%
\bibitem [{\citenamefont {Melikhov}(2022)}]{Melikhov:2022wct}%
  \BibitemOpen
  \bibfield  {author} {\bibinfo {author} {\bibfnamefont {D.}~\bibnamefont
  {Melikhov}},\ }\href {\doibase 10.1103/PhysRevD.106.054022} {\bibfield
  {journal} {\bibinfo  {journal} {Phys. Rev. D}\ }\textbf {\bibinfo {volume}
  {106}},\ \bibinfo {pages} {054022} (\bibinfo {year} {2022})},\ \Eprint
  {http://arxiv.org/abs/2208.04907} {arXiv:2208.04907 [hep-ph]} \BibitemShut
  {NoStop}%
\bibitem [{\citenamefont {Belov}\ \emph {et~al.}(2023)\citenamefont {Belov},
  \citenamefont {Berezhnoy},\ and\ \citenamefont {Melikhov}}]{Belov:2023xqk}%
  \BibitemOpen
  \bibfield  {author} {\bibinfo {author} {\bibfnamefont {I.}~\bibnamefont
  {Belov}}, \bibinfo {author} {\bibfnamefont {A.}~\bibnamefont {Berezhnoy}}, \
  and\ \bibinfo {author} {\bibfnamefont {D.}~\bibnamefont {Melikhov}},\ }\href
  {\doibase 10.1103/PhysRevD.108.094022} {\bibfield  {journal} {\bibinfo
  {journal} {Phys. Rev. D}\ }\textbf {\bibinfo {volume} {108}},\ \bibinfo
  {pages} {094022} (\bibinfo {year} {2023})},\ \Eprint
  {http://arxiv.org/abs/2309.00358} {arXiv:2309.00358 [hep-ph]} \BibitemShut
  {NoStop}%
\bibitem [{\citenamefont {Belov}\ \emph {et~al.}(2024)\citenamefont {Belov},
  \citenamefont {Berezhnoy},\ and\ \citenamefont {Melikhov}}]{Belov:2024vkv}%
  \BibitemOpen
  \bibfield  {author} {\bibinfo {author} {\bibfnamefont {I.}~\bibnamefont
  {Belov}}, \bibinfo {author} {\bibfnamefont {A.}~\bibnamefont {Berezhnoy}}, \
  and\ \bibinfo {author} {\bibfnamefont {D.}~\bibnamefont {Melikhov}},\ }\href
  {\doibase 10.1103/PhysRevD.109.114012} {\bibfield  {journal} {\bibinfo
  {journal} {Phys. Rev. D}\ }\textbf {\bibinfo {volume} {109}},\ \bibinfo
  {pages} {114012} (\bibinfo {year} {2024})},\ \Eprint
  {http://arxiv.org/abs/2404.01222} {arXiv:2404.01222 [hep-ph]} \BibitemShut
  {NoStop}%
\bibitem [{\citenamefont {Qin}\ \emph {et~al.}(2023)\citenamefont {Qin},
  \citenamefont {Shen}, \citenamefont {Wang},\ and\ \citenamefont
  {Wang}}]{Qin:2022rlk}%
  \BibitemOpen
  \bibfield  {author} {\bibinfo {author} {\bibfnamefont {Q.}~\bibnamefont
  {Qin}}, \bibinfo {author} {\bibfnamefont {Y.-L.}\ \bibnamefont {Shen}},
  \bibinfo {author} {\bibfnamefont {C.}~\bibnamefont {Wang}}, \ and\ \bibinfo
  {author} {\bibfnamefont {Y.-M.}\ \bibnamefont {Wang}},\ }\href {\doibase
  10.1103/PhysRevLett.131.091902} {\bibfield  {journal} {\bibinfo  {journal}
  {Phys. Rev. Lett.}\ }\textbf {\bibinfo {volume} {131}},\ \bibinfo {pages}
  {091902} (\bibinfo {year} {2023})},\ \Eprint
  {http://arxiv.org/abs/2207.02691} {arXiv:2207.02691 [hep-ph]} \BibitemShut
  {NoStop}%
\bibitem [{\citenamefont {Huang}\ \emph {et~al.}(2024)\citenamefont {Huang},
  \citenamefont {Ji}, \citenamefont {Shen}, \citenamefont {Wang}, \citenamefont
  {Wang},\ and\ \citenamefont {Zhao}}]{Huang:2023jdu}%
  \BibitemOpen
  \bibfield  {author} {\bibinfo {author} {\bibfnamefont {Y.-K.}\ \bibnamefont
  {Huang}}, \bibinfo {author} {\bibfnamefont {Y.}~\bibnamefont {Ji}}, \bibinfo
  {author} {\bibfnamefont {Y.-L.}\ \bibnamefont {Shen}}, \bibinfo {author}
  {\bibfnamefont {C.}~\bibnamefont {Wang}}, \bibinfo {author} {\bibfnamefont
  {Y.-M.}\ \bibnamefont {Wang}}, \ and\ \bibinfo {author} {\bibfnamefont
  {X.-C.}\ \bibnamefont {Zhao}},\ }\href {\doibase
  10.1103/PhysRevLett.133.171901} {\bibfield  {journal} {\bibinfo  {journal}
  {Phys. Rev. Lett.}\ }\textbf {\bibinfo {volume} {133}},\ \bibinfo {pages}
  {171901} (\bibinfo {year} {2024})},\ \Eprint
  {http://arxiv.org/abs/2312.15439} {arXiv:2312.15439 [hep-ph]} \BibitemShut
  {NoStop}%
\bibitem [{\citenamefont {Bartocci}\ \emph {et~al.}(2025)\citenamefont
  {Bartocci}, \citenamefont {B{\"o}er},\ and\ \citenamefont
  {Hurth}}]{Bartocci:2024bbf}%
  \BibitemOpen
  \bibfield  {author} {\bibinfo {author} {\bibfnamefont {R.}~\bibnamefont
  {Bartocci}}, \bibinfo {author} {\bibfnamefont {P.}~\bibnamefont {B{\"o}er}},
  \ and\ \bibinfo {author} {\bibfnamefont {T.}~\bibnamefont {Hurth}},\ }\href
  {\doibase 10.1007/JHEP04(2025)066} {\bibfield  {journal} {\bibinfo  {journal}
  {JHEP}\ }\textbf {\bibinfo {volume} {04}},\ \bibinfo {pages} {066} (\bibinfo
  {year} {2025})},\ \Eprint {http://arxiv.org/abs/2411.16634} {arXiv:2411.16634
  [hep-ph]} \BibitemShut {NoStop}%
\bibitem [{\citenamefont {Bartocci}\ \emph {et~al.}(2026)\citenamefont
  {Bartocci}, \citenamefont {B{\"o}er}, \citenamefont {Feldmann}, \citenamefont
  {Ferr{\'e}}, \citenamefont {Gubernari},\ and\ \citenamefont
  {Vladimirov}}]{Bartocci:2026lhe}%
  \BibitemOpen
  \bibfield  {author} {\bibinfo {author} {\bibfnamefont {R.}~\bibnamefont
  {Bartocci}}, \bibinfo {author} {\bibfnamefont {P.}~\bibnamefont {B{\"o}er}},
  \bibinfo {author} {\bibfnamefont {T.}~\bibnamefont {Feldmann}}, \bibinfo
  {author} {\bibfnamefont {M.}~\bibnamefont {Ferr{\'e}}}, \bibinfo {author}
  {\bibfnamefont {N.}~\bibnamefont {Gubernari}}, \ and\ \bibinfo {author}
  {\bibfnamefont {D.}~\bibnamefont {Vladimirov}},\ }\href@noop {} {\  (\bibinfo
  {year} {2026})},\ \Eprint {http://arxiv.org/abs/2606.20267} {arXiv:2606.20267
  [hep-ph]} \BibitemShut {NoStop}%
\bibitem [{\citenamefont {Khodjamirian}\ \emph {et~al.}(2024)\citenamefont
  {Khodjamirian}, \citenamefont {Meli{\'c}},\ and\ \citenamefont
  {Wang}}]{Khodjamirian:2023wol}%
  \BibitemOpen
  \bibfield  {author} {\bibinfo {author} {\bibfnamefont {A.}~\bibnamefont
  {Khodjamirian}}, \bibinfo {author} {\bibfnamefont {B.}~\bibnamefont
  {Meli{\'c}}}, \ and\ \bibinfo {author} {\bibfnamefont {Y.-M.}\ \bibnamefont
  {Wang}},\ }\href {\doibase 10.1140/epjs/s11734-023-01046-6} {\bibfield
  {journal} {\bibinfo  {journal} {Eur. Phys. J. ST}\ }\textbf {\bibinfo
  {volume} {233}},\ \bibinfo {pages} {271} (\bibinfo {year} {2024})},\ \Eprint
  {http://arxiv.org/abs/2311.08700} {arXiv:2311.08700 [hep-ph]} \BibitemShut
  {NoStop}%
\bibitem [{\citenamefont {Bosch}\ \emph {et~al.}(2003)\citenamefont {Bosch},
  \citenamefont {Hill}, \citenamefont {Lange},\ and\ \citenamefont
  {Neubert}}]{Bosch:2003fc}%
  \BibitemOpen
  \bibfield  {author} {\bibinfo {author} {\bibfnamefont {S.~W.}\ \bibnamefont
  {Bosch}}, \bibinfo {author} {\bibfnamefont {R.~J.}\ \bibnamefont {Hill}},
  \bibinfo {author} {\bibfnamefont {B.~O.}\ \bibnamefont {Lange}}, \ and\
  \bibinfo {author} {\bibfnamefont {M.}~\bibnamefont {Neubert}},\ }\href
  {\doibase 10.1103/PhysRevD.67.094014} {\bibfield  {journal} {\bibinfo
  {journal} {Phys. Rev. D}\ }\textbf {\bibinfo {volume} {67}},\ \bibinfo
  {pages} {094014} (\bibinfo {year} {2003})},\ \Eprint
  {http://arxiv.org/abs/hep-ph/0301123} {arXiv:hep-ph/0301123} \BibitemShut
  {NoStop}%
\bibitem [{\citenamefont {Beneke}\ and\ \citenamefont
  {Feldmann}(2004)}]{Beneke:2003pa}%
  \BibitemOpen
  \bibfield  {author} {\bibinfo {author} {\bibfnamefont {M.}~\bibnamefont
  {Beneke}}\ and\ \bibinfo {author} {\bibfnamefont {T.}~\bibnamefont
  {Feldmann}},\ }\href {\doibase 10.1016/j.nuclphysb.2004.02.033} {\bibfield
  {journal} {\bibinfo  {journal} {Nucl. Phys. B}\ }\textbf {\bibinfo {volume}
  {685}},\ \bibinfo {pages} {249} (\bibinfo {year} {2004})},\ \Eprint
  {http://arxiv.org/abs/hep-ph/0311335} {arXiv:hep-ph/0311335} \BibitemShut
  {NoStop}%
\bibitem [{\citenamefont {Hill}\ and\ \citenamefont
  {Neubert}(2003)}]{Hill:2002vw}%
  \BibitemOpen
  \bibfield  {author} {\bibinfo {author} {\bibfnamefont {R.~J.}\ \bibnamefont
  {Hill}}\ and\ \bibinfo {author} {\bibfnamefont {M.}~\bibnamefont {Neubert}},\
  }\href {\doibase 10.1016/S0550-3213(03)00116-0} {\bibfield  {journal}
  {\bibinfo  {journal} {Nucl. Phys. B}\ }\textbf {\bibinfo {volume} {657}},\
  \bibinfo {pages} {229} (\bibinfo {year} {2003})},\ \Eprint
  {http://arxiv.org/abs/hep-ph/0211018} {arXiv:hep-ph/0211018} \BibitemShut
  {NoStop}%
\bibitem [{\citenamefont {Ball}(2003)}]{Ball:2003bf}%
  \BibitemOpen
  \bibfield  {author} {\bibinfo {author} {\bibfnamefont {P.}~\bibnamefont
  {Ball}},\ }\href@noop {} {\  (\bibinfo {year} {2003})},\ \Eprint
  {http://arxiv.org/abs/hep-ph/0308249} {arXiv:hep-ph/0308249} \BibitemShut
  {NoStop}%
\bibitem [{\citenamefont {Lange}\ and\ \citenamefont
  {Neubert}(2004)}]{Lange:2003pk}%
  \BibitemOpen
  \bibfield  {author} {\bibinfo {author} {\bibfnamefont {B.~O.}\ \bibnamefont
  {Lange}}\ and\ \bibinfo {author} {\bibfnamefont {M.}~\bibnamefont
  {Neubert}},\ }\href {\doibase 10.1016/j.nuclphysb.2005.06.019} {\bibfield
  {journal} {\bibinfo  {journal} {Nucl. Phys. B}\ }\textbf {\bibinfo {volume}
  {690}},\ \bibinfo {pages} {249} (\bibinfo {year} {2004})},\ \bibinfo {note}
  {[Erratum: Nucl.Phys.B 723, 201--202 (2005)]},\ \Eprint
  {http://arxiv.org/abs/hep-ph/0311345} {arXiv:hep-ph/0311345} \BibitemShut
  {NoStop}%
\bibitem [{\citenamefont {Wang}\ and\ \citenamefont
  {Shen}(2015)}]{Wang:2015vgv}%
  \BibitemOpen
  \bibfield  {author} {\bibinfo {author} {\bibfnamefont {Y.-M.}\ \bibnamefont
  {Wang}}\ and\ \bibinfo {author} {\bibfnamefont {Y.-L.}\ \bibnamefont
  {Shen}},\ }\href {\doibase 10.1016/j.nuclphysb.2015.07.016} {\bibfield
  {journal} {\bibinfo  {journal} {Nucl. Phys. B}\ }\textbf {\bibinfo {volume}
  {898}},\ \bibinfo {pages} {563} (\bibinfo {year} {2015})},\ \Eprint
  {http://arxiv.org/abs/1506.00667} {arXiv:1506.00667 [hep-ph]} \BibitemShut
  {NoStop}%
\bibitem [{\citenamefont {Bauer}\ \emph
  {et~al.}(2002{\natexlab{a}})\citenamefont {Bauer}, \citenamefont {Pirjol},\
  and\ \citenamefont {Stewart}}]{Bauer:2001yt}%
  \BibitemOpen
  \bibfield  {author} {\bibinfo {author} {\bibfnamefont {C.~W.}\ \bibnamefont
  {Bauer}}, \bibinfo {author} {\bibfnamefont {D.}~\bibnamefont {Pirjol}}, \
  and\ \bibinfo {author} {\bibfnamefont {I.~W.}\ \bibnamefont {Stewart}},\
  }\href {\doibase 10.1103/PhysRevD.65.054022} {\bibfield  {journal} {\bibinfo
  {journal} {Phys. Rev. D}\ }\textbf {\bibinfo {volume} {65}},\ \bibinfo
  {pages} {054022} (\bibinfo {year} {2002}{\natexlab{a}})},\ \Eprint
  {http://arxiv.org/abs/hep-ph/0109045} {arXiv:hep-ph/0109045} \BibitemShut
  {NoStop}%
\bibitem [{\citenamefont {Bauer}\ \emph
  {et~al.}(2002{\natexlab{b}})\citenamefont {Bauer}, \citenamefont {Fleming},
  \citenamefont {Pirjol}, \citenamefont {Rothstein},\ and\ \citenamefont
  {Stewart}}]{Bauer:2002nz}%
  \BibitemOpen
  \bibfield  {author} {\bibinfo {author} {\bibfnamefont {C.~W.}\ \bibnamefont
  {Bauer}}, \bibinfo {author} {\bibfnamefont {S.}~\bibnamefont {Fleming}},
  \bibinfo {author} {\bibfnamefont {D.}~\bibnamefont {Pirjol}}, \bibinfo
  {author} {\bibfnamefont {I.~Z.}\ \bibnamefont {Rothstein}}, \ and\ \bibinfo
  {author} {\bibfnamefont {I.~W.}\ \bibnamefont {Stewart}},\ }\href {\doibase
  10.1103/PhysRevD.66.014017} {\bibfield  {journal} {\bibinfo  {journal} {Phys.
  Rev. D}\ }\textbf {\bibinfo {volume} {66}},\ \bibinfo {pages} {014017}
  (\bibinfo {year} {2002}{\natexlab{b}})},\ \Eprint
  {http://arxiv.org/abs/hep-ph/0202088} {arXiv:hep-ph/0202088} \BibitemShut
  {NoStop}%
\bibitem [{\citenamefont {Beneke}\ \emph {et~al.}(2002)\citenamefont {Beneke},
  \citenamefont {Chapovsky}, \citenamefont {Diehl},\ and\ \citenamefont
  {Feldmann}}]{Beneke:2002ph}%
  \BibitemOpen
  \bibfield  {author} {\bibinfo {author} {\bibfnamefont {M.}~\bibnamefont
  {Beneke}}, \bibinfo {author} {\bibfnamefont {A.~P.}\ \bibnamefont
  {Chapovsky}}, \bibinfo {author} {\bibfnamefont {M.}~\bibnamefont {Diehl}}, \
  and\ \bibinfo {author} {\bibfnamefont {T.}~\bibnamefont {Feldmann}},\ }\href
  {\doibase 10.1016/S0550-3213(02)00687-9} {\bibfield  {journal} {\bibinfo
  {journal} {Nucl. Phys. B}\ }\textbf {\bibinfo {volume} {643}},\ \bibinfo
  {pages} {431} (\bibinfo {year} {2002})},\ \Eprint
  {http://arxiv.org/abs/hep-ph/0206152} {arXiv:hep-ph/0206152} \BibitemShut
  {NoStop}%
\bibitem [{\citenamefont {Beneke}\ and\ \citenamefont
  {Feldmann}(2003)}]{Beneke:2002ni}%
  \BibitemOpen
  \bibfield  {author} {\bibinfo {author} {\bibfnamefont {M.}~\bibnamefont
  {Beneke}}\ and\ \bibinfo {author} {\bibfnamefont {T.}~\bibnamefont
  {Feldmann}},\ }\href {\doibase 10.1016/S0370-2693(02)03204-5} {\bibfield
  {journal} {\bibinfo  {journal} {Phys. Lett. B}\ }\textbf {\bibinfo {volume}
  {553}},\ \bibinfo {pages} {267} (\bibinfo {year} {2003})},\ \Eprint
  {http://arxiv.org/abs/hep-ph/0211358} {arXiv:hep-ph/0211358} \BibitemShut
  {NoStop}%
\bibitem [{\citenamefont {Becher}\ \emph {et~al.}(2015)\citenamefont {Becher},
  \citenamefont {Broggio},\ and\ \citenamefont {Ferroglia}}]{Becher:2014oda}%
  \BibitemOpen
  \bibfield  {author} {\bibinfo {author} {\bibfnamefont {T.}~\bibnamefont
  {Becher}}, \bibinfo {author} {\bibfnamefont {A.}~\bibnamefont {Broggio}}, \
  and\ \bibinfo {author} {\bibfnamefont {A.}~\bibnamefont {Ferroglia}},\ }\href
  {\doibase 10.1007/978-3-319-14848-9} {\emph {\bibinfo {title} {{Introduction
  to Soft-Collinear Effective Theory}}}},\ Vol.\ \bibinfo {volume} {896}\
  (\bibinfo  {publisher} {Springer},\ \bibinfo {year} {2015})\ \Eprint
  {http://arxiv.org/abs/1410.1892} {arXiv:1410.1892 [hep-ph]} \BibitemShut
  {NoStop}%
\bibitem [{\citenamefont {Beneke}(2015)}]{Beneke:2015wfa}%
  \BibitemOpen
  \bibfield  {author} {\bibinfo {author} {\bibfnamefont {M.}~\bibnamefont
  {Beneke}},\ }\href {\doibase 10.1016/j.nuclphysbps.2015.03.021} {\bibfield
  {journal} {\bibinfo  {journal} {Nucl. Part. Phys. Proc.}\ }\textbf {\bibinfo
  {volume} {261-262}},\ \bibinfo {pages} {311} (\bibinfo {year} {2015})},\
  \Eprint {http://arxiv.org/abs/1501.07374} {arXiv:1501.07374 [hep-ph]}
  \BibitemShut {NoStop}%
\bibitem [{\citenamefont {Chetyrkin}\ \emph {et~al.}(1997)\citenamefont
  {Chetyrkin}, \citenamefont {Misiak},\ and\ \citenamefont
  {Munz}}]{Chetyrkin:1996vx}%
  \BibitemOpen
  \bibfield  {author} {\bibinfo {author} {\bibfnamefont {K.~G.}\ \bibnamefont
  {Chetyrkin}}, \bibinfo {author} {\bibfnamefont {M.}~\bibnamefont {Misiak}}, \
  and\ \bibinfo {author} {\bibfnamefont {M.}~\bibnamefont {Munz}},\ }\href
  {\doibase 10.1016/S0370-2693(97)00324-9} {\bibfield  {journal} {\bibinfo
  {journal} {Phys. Lett. B}\ }\textbf {\bibinfo {volume} {400}},\ \bibinfo
  {pages} {206} (\bibinfo {year} {1997})},\ \bibinfo {note} {[Erratum:
  Phys.Lett.B 425, 414 (1998)]},\ \Eprint {http://arxiv.org/abs/hep-ph/9612313}
  {arXiv:hep-ph/9612313} \BibitemShut {NoStop}%
\bibitem [{\citenamefont {Chetyrkin}\ \emph {et~al.}(1998)\citenamefont
  {Chetyrkin}, \citenamefont {Misiak},\ and\ \citenamefont
  {Munz}}]{Chetyrkin:1997gb}%
  \BibitemOpen
  \bibfield  {author} {\bibinfo {author} {\bibfnamefont {K.~G.}\ \bibnamefont
  {Chetyrkin}}, \bibinfo {author} {\bibfnamefont {M.}~\bibnamefont {Misiak}}, \
  and\ \bibinfo {author} {\bibfnamefont {M.}~\bibnamefont {Munz}},\ }\href
  {\doibase 10.1016/S0550-3213(98)00131-X} {\bibfield  {journal} {\bibinfo
  {journal} {Nucl. Phys. B}\ }\textbf {\bibinfo {volume} {520}},\ \bibinfo
  {pages} {279} (\bibinfo {year} {1998})},\ \Eprint
  {http://arxiv.org/abs/hep-ph/9711280} {arXiv:hep-ph/9711280} \BibitemShut
  {NoStop}%
\bibitem [{\citenamefont {Beneke}(1998)}]{Beneke:1998rk}%
  \BibitemOpen
  \bibfield  {author} {\bibinfo {author} {\bibfnamefont {M.}~\bibnamefont
  {Beneke}},\ }\href {\doibase 10.1016/S0370-2693(98)00741-2} {\bibfield
  {journal} {\bibinfo  {journal} {Phys. Lett. B}\ }\textbf {\bibinfo {volume}
  {434}},\ \bibinfo {pages} {115} (\bibinfo {year} {1998})},\ \Eprint
  {http://arxiv.org/abs/hep-ph/9804241} {arXiv:hep-ph/9804241} \BibitemShut
  {NoStop}%
\bibitem [{\citenamefont {Beneke}\ and\ \citenamefont
  {Feldmann}(2001)}]{Beneke:2000wa}%
  \BibitemOpen
  \bibfield  {author} {\bibinfo {author} {\bibfnamefont {M.}~\bibnamefont
  {Beneke}}\ and\ \bibinfo {author} {\bibfnamefont {T.}~\bibnamefont
  {Feldmann}},\ }\href {\doibase 10.1016/S0550-3213(00)00585-X} {\bibfield
  {journal} {\bibinfo  {journal} {Nucl. Phys. B}\ }\textbf {\bibinfo {volume}
  {592}},\ \bibinfo {pages} {3} (\bibinfo {year} {2001})},\ \Eprint
  {http://arxiv.org/abs/hep-ph/0008255} {arXiv:hep-ph/0008255} \BibitemShut
  {NoStop}%
\bibitem [{\citenamefont {Khodjamirian}\ \emph {et~al.}(2011)\citenamefont
  {Khodjamirian}, \citenamefont {Mannel}, \citenamefont {Offen},\ and\
  \citenamefont {Wang}}]{Khodjamirian:2011ub}%
  \BibitemOpen
  \bibfield  {author} {\bibinfo {author} {\bibfnamefont {A.}~\bibnamefont
  {Khodjamirian}}, \bibinfo {author} {\bibfnamefont {T.}~\bibnamefont
  {Mannel}}, \bibinfo {author} {\bibfnamefont {N.}~\bibnamefont {Offen}}, \
  and\ \bibinfo {author} {\bibfnamefont {Y.~M.}\ \bibnamefont {Wang}},\ }\href
  {\doibase 10.1103/PhysRevD.83.094031} {\bibfield  {journal} {\bibinfo
  {journal} {Phys. Rev. D}\ }\textbf {\bibinfo {volume} {83}},\ \bibinfo
  {pages} {094031} (\bibinfo {year} {2011})},\ \Eprint
  {http://arxiv.org/abs/1103.2655} {arXiv:1103.2655 [hep-ph]} \BibitemShut
  {NoStop}%
\bibitem [{\citenamefont {L{\"u}}\ \emph {et~al.}(2019)\citenamefont {L{\"u}},
  \citenamefont {Shen}, \citenamefont {Wang},\ and\ \citenamefont
  {Wei}}]{Lu:2018cfc}%
  \BibitemOpen
  \bibfield  {author} {\bibinfo {author} {\bibfnamefont {C.-D.}\ \bibnamefont
  {L{\"u}}}, \bibinfo {author} {\bibfnamefont {Y.-L.}\ \bibnamefont {Shen}},
  \bibinfo {author} {\bibfnamefont {Y.-M.}\ \bibnamefont {Wang}}, \ and\
  \bibinfo {author} {\bibfnamefont {Y.-B.}\ \bibnamefont {Wei}},\ }\href
  {\doibase 10.1007/JHEP01(2019)024} {\bibfield  {journal} {\bibinfo  {journal}
  {JHEP}\ }\textbf {\bibinfo {volume} {01}},\ \bibinfo {pages} {024} (\bibinfo
  {year} {2019})},\ \Eprint {http://arxiv.org/abs/1810.00819} {arXiv:1810.00819
  [hep-ph]} \BibitemShut {NoStop}%
\bibitem [{\citenamefont {Cui}\ \emph {et~al.}(2023{\natexlab{a}})\citenamefont
  {Cui}, \citenamefont {Huang}, \citenamefont {Shen}, \citenamefont {Wang},\
  and\ \citenamefont {Wang}}]{Cui:2022zwm}%
  \BibitemOpen
  \bibfield  {author} {\bibinfo {author} {\bibfnamefont {B.-Y.}\ \bibnamefont
  {Cui}}, \bibinfo {author} {\bibfnamefont {Y.-K.}\ \bibnamefont {Huang}},
  \bibinfo {author} {\bibfnamefont {Y.-L.}\ \bibnamefont {Shen}}, \bibinfo
  {author} {\bibfnamefont {C.}~\bibnamefont {Wang}}, \ and\ \bibinfo {author}
  {\bibfnamefont {Y.-M.}\ \bibnamefont {Wang}},\ }\href {\doibase
  10.1007/JHEP03(2023)140} {\bibfield  {journal} {\bibinfo  {journal} {JHEP}\
  }\textbf {\bibinfo {volume} {03}},\ \bibinfo {pages} {140} (\bibinfo {year}
  {2023}{\natexlab{a}})},\ \Eprint {http://arxiv.org/abs/2212.11624}
  {arXiv:2212.11624 [hep-ph]} \BibitemShut {NoStop}%
\bibitem [{\citenamefont {Grozin}\ and\ \citenamefont
  {Neubert}(1997)}]{Grozin:1996pq}%
  \BibitemOpen
  \bibfield  {author} {\bibinfo {author} {\bibfnamefont {A.~G.}\ \bibnamefont
  {Grozin}}\ and\ \bibinfo {author} {\bibfnamefont {M.}~\bibnamefont
  {Neubert}},\ }\href {\doibase 10.1103/PhysRevD.55.272} {\bibfield  {journal}
  {\bibinfo  {journal} {Phys. Rev. D}\ }\textbf {\bibinfo {volume} {55}},\
  \bibinfo {pages} {272} (\bibinfo {year} {1997})},\ \Eprint
  {http://arxiv.org/abs/hep-ph/9607366} {arXiv:hep-ph/9607366} \BibitemShut
  {NoStop}%
\bibitem [{\citenamefont {Braun}\ \emph
  {et~al.}(2017{\natexlab{a}})\citenamefont {Braun}, \citenamefont {Ji},\ and\
  \citenamefont {Manashov}}]{Braun:2017liq}%
  \BibitemOpen
  \bibfield  {author} {\bibinfo {author} {\bibfnamefont {V.~M.}\ \bibnamefont
  {Braun}}, \bibinfo {author} {\bibfnamefont {Y.}~\bibnamefont {Ji}}, \ and\
  \bibinfo {author} {\bibfnamefont {A.~N.}\ \bibnamefont {Manashov}},\ }\href
  {\doibase 10.1007/JHEP05(2017)022} {\bibfield  {journal} {\bibinfo  {journal}
  {JHEP}\ }\textbf {\bibinfo {volume} {05}},\ \bibinfo {pages} {022} (\bibinfo
  {year} {2017}{\natexlab{a}})},\ \Eprint {http://arxiv.org/abs/1703.02446}
  {arXiv:1703.02446 [hep-ph]} \BibitemShut {NoStop}%
\bibitem [{\citenamefont {Lepage}\ and\ \citenamefont
  {Brodsky}(1980)}]{Lepage:1980fj}%
  \BibitemOpen
  \bibfield  {author} {\bibinfo {author} {\bibfnamefont {G.~P.}\ \bibnamefont
  {Lepage}}\ and\ \bibinfo {author} {\bibfnamefont {S.~J.}\ \bibnamefont
  {Brodsky}},\ }\href {\doibase 10.1103/PhysRevD.22.2157} {\bibfield  {journal}
  {\bibinfo  {journal} {Phys. Rev. D}\ }\textbf {\bibinfo {volume} {22}},\
  \bibinfo {pages} {2157} (\bibinfo {year} {1980})}\BibitemShut {NoStop}%
\bibitem [{\citenamefont {Efremov}\ and\ \citenamefont
  {Radyushkin}(1980)}]{Efremov:1979qk}%
  \BibitemOpen
  \bibfield  {author} {\bibinfo {author} {\bibfnamefont {A.~V.}\ \bibnamefont
  {Efremov}}\ and\ \bibinfo {author} {\bibfnamefont {A.~V.}\ \bibnamefont
  {Radyushkin}},\ }\href {\doibase 10.1016/0370-2693(80)90869-2} {\bibfield
  {journal} {\bibinfo  {journal} {Phys. Lett. B}\ }\textbf {\bibinfo {volume}
  {94}},\ \bibinfo {pages} {245} (\bibinfo {year} {1980})}\BibitemShut
  {NoStop}%
\bibitem [{\citenamefont {Beneke}\ \emph {et~al.}(2004)\citenamefont {Beneke},
  \citenamefont {Kiyo},\ and\ \citenamefont {Yang}}]{Beneke:2004rc}%
  \BibitemOpen
  \bibfield  {author} {\bibinfo {author} {\bibfnamefont {M.}~\bibnamefont
  {Beneke}}, \bibinfo {author} {\bibfnamefont {Y.}~\bibnamefont {Kiyo}}, \ and\
  \bibinfo {author} {\bibfnamefont {D.~S.}\ \bibnamefont {Yang}},\ }\href
  {\doibase 10.1016/j.nuclphysb.2004.05.018} {\bibfield  {journal} {\bibinfo
  {journal} {Nucl. Phys. B}\ }\textbf {\bibinfo {volume} {692}},\ \bibinfo
  {pages} {232} (\bibinfo {year} {2004})},\ \Eprint
  {http://arxiv.org/abs/hep-ph/0402241} {arXiv:hep-ph/0402241} \BibitemShut
  {NoStop}%
\bibitem [{\citenamefont {Beneke}\ and\ \citenamefont
  {Yang}(2006)}]{Beneke:2005gs}%
  \BibitemOpen
  \bibfield  {author} {\bibinfo {author} {\bibfnamefont {M.}~\bibnamefont
  {Beneke}}\ and\ \bibinfo {author} {\bibfnamefont {D.~S.}\ \bibnamefont
  {Yang}},\ }\href {\doibase 10.1016/j.nuclphysb.2005.11.027} {\bibfield
  {journal} {\bibinfo  {journal} {Nucl. Phys. B}\ }\textbf {\bibinfo {volume}
  {736}},\ \bibinfo {pages} {34} (\bibinfo {year} {2006})},\ \Eprint
  {http://arxiv.org/abs/hep-ph/0508250} {arXiv:hep-ph/0508250} \BibitemShut
  {NoStop}%
\bibitem [{\citenamefont {Beneke}\ and\ \citenamefont
  {Jager}(2006)}]{Beneke:2005vv}%
  \BibitemOpen
  \bibfield  {author} {\bibinfo {author} {\bibfnamefont {M.}~\bibnamefont
  {Beneke}}\ and\ \bibinfo {author} {\bibfnamefont {S.}~\bibnamefont {Jager}},\
  }\href {\doibase 10.1016/j.nuclphysb.2006.06.010} {\bibfield  {journal}
  {\bibinfo  {journal} {Nucl. Phys. B}\ }\textbf {\bibinfo {volume} {751}},\
  \bibinfo {pages} {160} (\bibinfo {year} {2006})},\ \Eprint
  {http://arxiv.org/abs/hep-ph/0512351} {arXiv:hep-ph/0512351} \BibitemShut
  {NoStop}%
\bibitem [{\citenamefont {Chay}\ and\ \citenamefont {Kim}(2004)}]{Chay:2003ju}%
  \BibitemOpen
  \bibfield  {author} {\bibinfo {author} {\bibfnamefont {J.}~\bibnamefont
  {Chay}}\ and\ \bibinfo {author} {\bibfnamefont {C.}~\bibnamefont {Kim}},\
  }\href {\doibase 10.1016/j.nuclphysb.2003.12.027} {\bibfield  {journal}
  {\bibinfo  {journal} {Nucl. Phys. B}\ }\textbf {\bibinfo {volume} {680}},\
  \bibinfo {pages} {302} (\bibinfo {year} {2004})},\ \Eprint
  {http://arxiv.org/abs/hep-ph/0301262} {arXiv:hep-ph/0301262} \BibitemShut
  {NoStop}%
\bibitem [{\citenamefont {Bauer}\ \emph {et~al.}(2004)\citenamefont {Bauer},
  \citenamefont {Pirjol}, \citenamefont {Rothstein},\ and\ \citenamefont
  {Stewart}}]{Bauer:2004tj}%
  \BibitemOpen
  \bibfield  {author} {\bibinfo {author} {\bibfnamefont {C.~W.}\ \bibnamefont
  {Bauer}}, \bibinfo {author} {\bibfnamefont {D.}~\bibnamefont {Pirjol}},
  \bibinfo {author} {\bibfnamefont {I.~Z.}\ \bibnamefont {Rothstein}}, \ and\
  \bibinfo {author} {\bibfnamefont {I.~W.}\ \bibnamefont {Stewart}},\ }\href
  {\doibase 10.1103/PhysRevD.70.054015} {\bibfield  {journal} {\bibinfo
  {journal} {Phys. Rev. D}\ }\textbf {\bibinfo {volume} {70}},\ \bibinfo
  {pages} {054015} (\bibinfo {year} {2004})},\ \Eprint
  {http://arxiv.org/abs/hep-ph/0401188} {arXiv:hep-ph/0401188} \BibitemShut
  {NoStop}%
\bibitem [{\citenamefont {Becher}\ and\ \citenamefont
  {Hill}(2004)}]{Becher:2004kk}%
  \BibitemOpen
  \bibfield  {author} {\bibinfo {author} {\bibfnamefont {T.}~\bibnamefont
  {Becher}}\ and\ \bibinfo {author} {\bibfnamefont {R.~J.}\ \bibnamefont
  {Hill}},\ }\href {\doibase 10.1088/1126-6708/2004/10/055} {\bibfield
  {journal} {\bibinfo  {journal} {JHEP}\ }\textbf {\bibinfo {volume} {10}},\
  \bibinfo {pages} {055} (\bibinfo {year} {2004})},\ \Eprint
  {http://arxiv.org/abs/hep-ph/0408344} {arXiv:hep-ph/0408344} \BibitemShut
  {NoStop}%
\bibitem [{\citenamefont {Hill}\ \emph {et~al.}(2004)\citenamefont {Hill},
  \citenamefont {Becher}, \citenamefont {Lee},\ and\ \citenamefont
  {Neubert}}]{Hill:2004if}%
  \BibitemOpen
  \bibfield  {author} {\bibinfo {author} {\bibfnamefont {R.~J.}\ \bibnamefont
  {Hill}}, \bibinfo {author} {\bibfnamefont {T.}~\bibnamefont {Becher}},
  \bibinfo {author} {\bibfnamefont {S.~J.}\ \bibnamefont {Lee}}, \ and\
  \bibinfo {author} {\bibfnamefont {M.}~\bibnamefont {Neubert}},\ }\href
  {\doibase 10.1088/1126-6708/2004/07/081} {\bibfield  {journal} {\bibinfo
  {journal} {JHEP}\ }\textbf {\bibinfo {volume} {07}},\ \bibinfo {pages} {081}
  (\bibinfo {year} {2004})},\ \Eprint {http://arxiv.org/abs/hep-ph/0404217}
  {arXiv:hep-ph/0404217} \BibitemShut {NoStop}%
\bibitem [{\citenamefont {Beneke}\ \emph
  {et~al.}(2018{\natexlab{a}})\citenamefont {Beneke}, \citenamefont {Garny},
  \citenamefont {Szafron},\ and\ \citenamefont {Wang}}]{Beneke:2018rbh}%
  \BibitemOpen
  \bibfield  {author} {\bibinfo {author} {\bibfnamefont {M.}~\bibnamefont
  {Beneke}}, \bibinfo {author} {\bibfnamefont {M.}~\bibnamefont {Garny}},
  \bibinfo {author} {\bibfnamefont {R.}~\bibnamefont {Szafron}}, \ and\
  \bibinfo {author} {\bibfnamefont {J.}~\bibnamefont {Wang}},\ }\href {\doibase
  10.1007/JHEP11(2018)112} {\bibfield  {journal} {\bibinfo  {journal} {JHEP}\
  }\textbf {\bibinfo {volume} {11}},\ \bibinfo {pages} {112} (\bibinfo {year}
  {2018}{\natexlab{a}})},\ \Eprint {http://arxiv.org/abs/1808.04742}
  {arXiv:1808.04742 [hep-ph]} \BibitemShut {NoStop}%
\bibitem [{\citenamefont {Beneke}\ and\ \citenamefont
  {Smirnov}(1998)}]{Beneke:1997zp}%
  \BibitemOpen
  \bibfield  {author} {\bibinfo {author} {\bibfnamefont {M.}~\bibnamefont
  {Beneke}}\ and\ \bibinfo {author} {\bibfnamefont {V.~A.}\ \bibnamefont
  {Smirnov}},\ }\href {\doibase 10.1016/S0550-3213(98)00138-2} {\bibfield
  {journal} {\bibinfo  {journal} {Nucl. Phys. B}\ }\textbf {\bibinfo {volume}
  {522}},\ \bibinfo {pages} {321} (\bibinfo {year} {1998})},\ \Eprint
  {http://arxiv.org/abs/hep-ph/9711391} {arXiv:hep-ph/9711391} \BibitemShut
  {NoStop}%
\bibitem [{\citenamefont {Smirnov}\ and\ \citenamefont
  {Rakhmetov}(1999)}]{Smirnov:1998vk}%
  \BibitemOpen
  \bibfield  {author} {\bibinfo {author} {\bibfnamefont {V.~A.}\ \bibnamefont
  {Smirnov}}\ and\ \bibinfo {author} {\bibfnamefont {E.~R.}\ \bibnamefont
  {Rakhmetov}},\ }\href {\doibase 10.1007/BF02557396} {\bibfield  {journal}
  {\bibinfo  {journal} {Theor. Math. Phys.}\ }\textbf {\bibinfo {volume}
  {120}},\ \bibinfo {pages} {870} (\bibinfo {year} {1999})},\ \Eprint
  {http://arxiv.org/abs/hep-ph/9812529} {arXiv:hep-ph/9812529} \BibitemShut
  {NoStop}%
\bibitem [{\citenamefont {Smirnov}(2002)}]{Smirnov:2002pj}%
  \BibitemOpen
  \bibfield  {author} {\bibinfo {author} {\bibfnamefont {V.~A.}\ \bibnamefont
  {Smirnov}},\ }\href@noop {} {\bibfield  {journal} {\bibinfo  {journal}
  {Springer Tracts Mod. Phys.}\ }\textbf {\bibinfo {volume} {177}},\ \bibinfo
  {pages} {1} (\bibinfo {year} {2002})}\BibitemShut {NoStop}%
\bibitem [{\citenamefont {Ma}(2025)}]{Ma:2025emu}%
  \BibitemOpen
  \bibfield  {author} {\bibinfo {author} {\bibfnamefont {Y.}~\bibnamefont
  {Ma}},\ }\href {\doibase 10.1142/S0217751X25300133} {\bibfield  {journal}
  {\bibinfo  {journal} {Int. J. Mod. Phys. A}\ }\textbf {\bibinfo {volume}
  {40}},\ \bibinfo {pages} {2530013} (\bibinfo {year} {2025})},\ \Eprint
  {http://arxiv.org/abs/2505.01368} {arXiv:2505.01368 [hep-ph]} \BibitemShut
  {NoStop}%
\bibitem [{\citenamefont {Braun}\ \emph {et~al.}(2015)\citenamefont {Braun},
  \citenamefont {Manashov},\ and\ \citenamefont {Offen}}]{Braun:2015pha}%
  \BibitemOpen
  \bibfield  {author} {\bibinfo {author} {\bibfnamefont {V.~M.}\ \bibnamefont
  {Braun}}, \bibinfo {author} {\bibfnamefont {A.~N.}\ \bibnamefont {Manashov}},
  \ and\ \bibinfo {author} {\bibfnamefont {N.}~\bibnamefont {Offen}},\ }\href
  {\doibase 10.1103/PhysRevD.92.074044} {\bibfield  {journal} {\bibinfo
  {journal} {Phys. Rev. D}\ }\textbf {\bibinfo {volume} {92}},\ \bibinfo
  {pages} {074044} (\bibinfo {year} {2015})},\ \Eprint
  {http://arxiv.org/abs/1507.03445} {arXiv:1507.03445 [hep-ph]} \BibitemShut
  {NoStop}%
\bibitem [{\citenamefont {Descotes-Genon}\ and\ \citenamefont
  {Offen}(2009)}]{Descotes-Genon:2009jif}%
  \BibitemOpen
  \bibfield  {author} {\bibinfo {author} {\bibfnamefont {S.}~\bibnamefont
  {Descotes-Genon}}\ and\ \bibinfo {author} {\bibfnamefont {N.}~\bibnamefont
  {Offen}},\ }\href {\doibase 10.1088/1126-6708/2009/05/091} {\bibfield
  {journal} {\bibinfo  {journal} {JHEP}\ }\textbf {\bibinfo {volume} {05}},\
  \bibinfo {pages} {091} (\bibinfo {year} {2009})},\ \Eprint
  {http://arxiv.org/abs/0903.0790} {arXiv:0903.0790 [hep-ph]} \BibitemShut
  {NoStop}%
\bibitem [{\citenamefont {Beneke}\ \emph
  {et~al.}(2018{\natexlab{b}})\citenamefont {Beneke}, \citenamefont {Braun},
  \citenamefont {Ji},\ and\ \citenamefont {Wei}}]{Beneke:2018wjp}%
  \BibitemOpen
  \bibfield  {author} {\bibinfo {author} {\bibfnamefont {M.}~\bibnamefont
  {Beneke}}, \bibinfo {author} {\bibfnamefont {V.~M.}\ \bibnamefont {Braun}},
  \bibinfo {author} {\bibfnamefont {Y.}~\bibnamefont {Ji}}, \ and\ \bibinfo
  {author} {\bibfnamefont {Y.-B.}\ \bibnamefont {Wei}},\ }\href {\doibase
  10.1007/JHEP07(2018)154} {\bibfield  {journal} {\bibinfo  {journal} {JHEP}\
  }\textbf {\bibinfo {volume} {07}},\ \bibinfo {pages} {154} (\bibinfo {year}
  {2018}{\natexlab{b}})},\ \Eprint {http://arxiv.org/abs/1804.04962}
  {arXiv:1804.04962 [hep-ph]} \BibitemShut {NoStop}%
\bibitem [{\citenamefont {Wang}\ \emph {et~al.}(2022)\citenamefont {Wang},
  \citenamefont {Wang},\ and\ \citenamefont {Wei}}]{Wang:2021yrr}%
  \BibitemOpen
  \bibfield  {author} {\bibinfo {author} {\bibfnamefont {C.}~\bibnamefont
  {Wang}}, \bibinfo {author} {\bibfnamefont {Y.-M.}\ \bibnamefont {Wang}}, \
  and\ \bibinfo {author} {\bibfnamefont {Y.-B.}\ \bibnamefont {Wei}},\ }\href
  {\doibase 10.1007/JHEP02(2022)141} {\bibfield  {journal} {\bibinfo  {journal}
  {JHEP}\ }\textbf {\bibinfo {volume} {02}},\ \bibinfo {pages} {141} (\bibinfo
  {year} {2022})},\ \Eprint {http://arxiv.org/abs/2111.11811} {arXiv:2111.11811
  [hep-ph]} \BibitemShut {NoStop}%
\bibitem [{\citenamefont {Wang}\ \emph {et~al.}(2020)\citenamefont {Wang},
  \citenamefont {Wang}, \citenamefont {Xu},\ and\ \citenamefont
  {Zhao}}]{Wang:2019msf}%
  \BibitemOpen
  \bibfield  {author} {\bibinfo {author} {\bibfnamefont {W.}~\bibnamefont
  {Wang}}, \bibinfo {author} {\bibfnamefont {Y.-M.}\ \bibnamefont {Wang}},
  \bibinfo {author} {\bibfnamefont {J.}~\bibnamefont {Xu}}, \ and\ \bibinfo
  {author} {\bibfnamefont {S.}~\bibnamefont {Zhao}},\ }\href {\doibase
  10.1103/PhysRevD.102.011502} {\bibfield  {journal} {\bibinfo  {journal}
  {Phys. Rev. D}\ }\textbf {\bibinfo {volume} {102}},\ \bibinfo {pages}
  {011502} (\bibinfo {year} {2020})},\ \Eprint
  {http://arxiv.org/abs/1908.09933} {arXiv:1908.09933 [hep-ph]} \BibitemShut
  {NoStop}%
\bibitem [{\citenamefont {Han}\ \emph {et~al.}(2025{\natexlab{a}})\citenamefont
  {Han}, \citenamefont {Hua}, \citenamefont {Ji}, \citenamefont {L{\"u}},
  \citenamefont {Wang}, \citenamefont {Xu}, \citenamefont {Zhang},\ and\
  \citenamefont {Zhao}}]{Han:2024fkr}%
  \BibitemOpen
  \bibfield  {author} {\bibinfo {author} {\bibfnamefont {X.-Y.}\ \bibnamefont
  {Han}}, \bibinfo {author} {\bibfnamefont {J.}~\bibnamefont {Hua}}, \bibinfo
  {author} {\bibfnamefont {X.}~\bibnamefont {Ji}}, \bibinfo {author}
  {\bibfnamefont {C.-D.}\ \bibnamefont {L{\"u}}}, \bibinfo {author}
  {\bibfnamefont {W.}~\bibnamefont {Wang}}, \bibinfo {author} {\bibfnamefont
  {J.}~\bibnamefont {Xu}}, \bibinfo {author} {\bibfnamefont {Q.-A.}\
  \bibnamefont {Zhang}}, \ and\ \bibinfo {author} {\bibfnamefont
  {S.}~\bibnamefont {Zhao}},\ }\href {\doibase 10.1103/2t8s-w8t6} {\bibfield
  {journal} {\bibinfo  {journal} {Phys. Rev. D}\ }\textbf {\bibinfo {volume}
  {111}},\ \bibinfo {pages} {L111503} (\bibinfo {year} {2025}{\natexlab{a}})},\
  \Eprint {http://arxiv.org/abs/2403.17492} {arXiv:2403.17492 [hep-ph]}
  \BibitemShut {NoStop}%
\bibitem [{\citenamefont {Han}\ \emph {et~al.}(2025{\natexlab{b}})\citenamefont
  {Han} \emph {et~al.}}]{LatticeParton:2024zko}%
  \BibitemOpen
  \bibfield  {author} {\bibinfo {author} {\bibfnamefont {X.-Y.}\ \bibnamefont
  {Han}} \emph {et~al.} (\bibinfo {collaboration} {Lattice Parton}),\ }\href
  {\doibase 10.1103/PhysRevD.111.034503} {\bibfield  {journal} {\bibinfo
  {journal} {Phys. Rev. D}\ }\textbf {\bibinfo {volume} {111}},\ \bibinfo
  {pages} {034503} (\bibinfo {year} {2025}{\natexlab{b}})},\ \Eprint
  {http://arxiv.org/abs/2410.18654} {arXiv:2410.18654 [hep-lat]} \BibitemShut
  {NoStop}%
\bibitem [{\citenamefont {Han}\ \emph {et~al.}(2026)\citenamefont {Han} \emph
  {et~al.}}]{LPC:2026ffe}%
  \BibitemOpen
  \bibfield  {author} {\bibinfo {author} {\bibfnamefont {X.-Y.}\ \bibnamefont
  {Han}} \emph {et~al.} (\bibinfo {collaboration} {LPC}),\ }\href@noop {} {\
  (\bibinfo {year} {2026})},\ \Eprint {http://arxiv.org/abs/2605.10946}
  {arXiv:2605.10946 [hep-lat]} \BibitemShut {NoStop}%
\bibitem [{\citenamefont {Gao}\ \emph {et~al.}(2026)\citenamefont {Gao} \emph
  {et~al.}}]{LPC:2026vyv}%
  \BibitemOpen
  \bibfield  {author} {\bibinfo {author} {\bibfnamefont {H.-F.}\ \bibnamefont
  {Gao}} \emph {et~al.} (\bibinfo {collaboration} {LPC}),\ }\href@noop {} {\
  (\bibinfo {year} {2026})},\ \Eprint {http://arxiv.org/abs/2604.25802}
  {arXiv:2604.25802 [hep-lat]} \BibitemShut {NoStop}%
\bibitem [{\citenamefont {Giusti}\ \emph {et~al.}(2023)\citenamefont {Giusti},
  \citenamefont {Kane}, \citenamefont {Lehner}, \citenamefont {Meinel},\ and\
  \citenamefont {Soni}}]{Giusti:2023pot}%
  \BibitemOpen
  \bibfield  {author} {\bibinfo {author} {\bibfnamefont {D.}~\bibnamefont
  {Giusti}}, \bibinfo {author} {\bibfnamefont {C.~F.}\ \bibnamefont {Kane}},
  \bibinfo {author} {\bibfnamefont {C.}~\bibnamefont {Lehner}}, \bibinfo
  {author} {\bibfnamefont {S.}~\bibnamefont {Meinel}}, \ and\ \bibinfo {author}
  {\bibfnamefont {A.}~\bibnamefont {Soni}},\ }\href {\doibase
  10.1103/PhysRevD.107.074507} {\bibfield  {journal} {\bibinfo  {journal}
  {Phys. Rev. D}\ }\textbf {\bibinfo {volume} {107}},\ \bibinfo {pages}
  {074507} (\bibinfo {year} {2023})},\ \Eprint
  {http://arxiv.org/abs/2302.01298} {arXiv:2302.01298 [hep-lat]} \BibitemShut
  {NoStop}%
\bibitem [{\citenamefont {Giusti}\ \emph {et~al.}(2025)\citenamefont {Giusti},
  \citenamefont {Kane}, \citenamefont {Lehner}, \citenamefont {Meinel},\ and\
  \citenamefont {Soni}}]{Giusti:2025ibe}%
  \BibitemOpen
  \bibfield  {author} {\bibinfo {author} {\bibfnamefont {D.}~\bibnamefont
  {Giusti}}, \bibinfo {author} {\bibfnamefont {C.~F.}\ \bibnamefont {Kane}},
  \bibinfo {author} {\bibfnamefont {C.}~\bibnamefont {Lehner}}, \bibinfo
  {author} {\bibfnamefont {S.}~\bibnamefont {Meinel}}, \ and\ \bibinfo {author}
  {\bibfnamefont {A.}~\bibnamefont {Soni}},\ }\href {\doibase
  10.1103/2pzm-v26v} {\bibfield  {journal} {\bibinfo  {journal} {Phys. Rev. D}\
  }\textbf {\bibinfo {volume} {112}},\ \bibinfo {pages} {054507} (\bibinfo
  {year} {2025})},\ \Eprint {http://arxiv.org/abs/2505.11757} {arXiv:2505.11757
  [hep-lat]} \BibitemShut {NoStop}%
\bibitem [{\citenamefont {Bali}\ \emph {et~al.}(2019)\citenamefont {Bali},
  \citenamefont {Braun}, \citenamefont {B{\"u}rger}, \citenamefont
  {G{\"o}ckeler}, \citenamefont {Gruber}, \citenamefont {Hutzler},
  \citenamefont {Korcyl}, \citenamefont {Sch{\"a}fer}, \citenamefont
  {Sternbeck},\ and\ \citenamefont {Wein}}]{RQCD:2019osh}%
  \BibitemOpen
  \bibfield  {author} {\bibinfo {author} {\bibfnamefont {G.~S.}\ \bibnamefont
  {Bali}}, \bibinfo {author} {\bibfnamefont {V.~M.}\ \bibnamefont {Braun}},
  \bibinfo {author} {\bibfnamefont {S.}~\bibnamefont {B{\"u}rger}}, \bibinfo
  {author} {\bibfnamefont {M.}~\bibnamefont {G{\"o}ckeler}}, \bibinfo {author}
  {\bibfnamefont {M.}~\bibnamefont {Gruber}}, \bibinfo {author} {\bibfnamefont
  {F.}~\bibnamefont {Hutzler}}, \bibinfo {author} {\bibfnamefont
  {P.}~\bibnamefont {Korcyl}}, \bibinfo {author} {\bibfnamefont
  {A.}~\bibnamefont {Sch{\"a}fer}}, \bibinfo {author} {\bibfnamefont
  {A.}~\bibnamefont {Sternbeck}}, \ and\ \bibinfo {author} {\bibfnamefont
  {P.}~\bibnamefont {Wein}} (\bibinfo {collaboration} {RQCD}),\ }\href
  {\doibase 10.1007/JHEP08(2019)065} {\bibfield  {journal} {\bibinfo  {journal}
  {JHEP}\ }\textbf {\bibinfo {volume} {08}},\ \bibinfo {pages} {065} (\bibinfo
  {year} {2019})},\ \bibinfo {note} {[Addendum: JHEP 11, 037 (2020)]},\ \Eprint
  {http://arxiv.org/abs/1903.08038} {arXiv:1903.08038 [hep-lat]} \BibitemShut
  {NoStop}%
\bibitem [{\citenamefont {Kniehl}\ and\ \citenamefont
  {Veretin}(2020{\natexlab{a}})}]{Kniehl:2020sgo}%
  \BibitemOpen
  \bibfield  {author} {\bibinfo {author} {\bibfnamefont {B.~A.}\ \bibnamefont
  {Kniehl}}\ and\ \bibinfo {author} {\bibfnamefont {O.~L.}\ \bibnamefont
  {Veretin}},\ }\href {\doibase 10.1016/j.physletb.2020.135398} {\bibfield
  {journal} {\bibinfo  {journal} {Phys. Lett. B}\ }\textbf {\bibinfo {volume}
  {804}},\ \bibinfo {pages} {135398} (\bibinfo {year} {2020}{\natexlab{a}})},\
  \Eprint {http://arxiv.org/abs/2002.10894} {arXiv:2002.10894 [hep-ph]}
  \BibitemShut {NoStop}%
\bibitem [{\citenamefont {Kniehl}\ and\ \citenamefont
  {Veretin}(2020{\natexlab{b}})}]{Kniehl:2020nhw}%
  \BibitemOpen
  \bibfield  {author} {\bibinfo {author} {\bibfnamefont {B.~A.}\ \bibnamefont
  {Kniehl}}\ and\ \bibinfo {author} {\bibfnamefont {O.~L.}\ \bibnamefont
  {Veretin}},\ }\href {\doibase 10.1016/j.nuclphysb.2020.115229} {\bibfield
  {journal} {\bibinfo  {journal} {Nucl. Phys. B}\ }\textbf {\bibinfo {volume}
  {961}},\ \bibinfo {pages} {115229} (\bibinfo {year} {2020}{\natexlab{b}})},\
  \Eprint {http://arxiv.org/abs/2009.11325} {arXiv:2009.11325 [hep-ph]}
  \BibitemShut {NoStop}%
\bibitem [{\citenamefont {Aoki}\ \emph {et~al.}(2026)\citenamefont {Aoki} \emph
  {et~al.}}]{FlavourLatticeAveragingGroupFLAG:2024oxs}%
  \BibitemOpen
  \bibfield  {author} {\bibinfo {author} {\bibfnamefont {Y.}~\bibnamefont
  {Aoki}} \emph {et~al.} (\bibinfo {collaboration} {Flavour Lattice Averaging
  Group (FLAG)}),\ }\href {\doibase 10.1103/nfzp-p5dn} {\bibfield  {journal}
  {\bibinfo  {journal} {Phys. Rev. D}\ }\textbf {\bibinfo {volume} {113}},\
  \bibinfo {pages} {014508} (\bibinfo {year} {2026})},\ \Eprint
  {http://arxiv.org/abs/2411.04268} {arXiv:2411.04268 [hep-lat]} \BibitemShut
  {NoStop}%
\bibitem [{\citenamefont {Gasser}\ and\ \citenamefont
  {Zarnauskas}(2010)}]{Gasser:2010wz}%
  \BibitemOpen
  \bibfield  {author} {\bibinfo {author} {\bibfnamefont {J.}~\bibnamefont
  {Gasser}}\ and\ \bibinfo {author} {\bibfnamefont {G.~R.~S.}\ \bibnamefont
  {Zarnauskas}},\ }\href {\doibase 10.1016/j.physletb.2010.08.021} {\bibfield
  {journal} {\bibinfo  {journal} {Phys. Lett. B}\ }\textbf {\bibinfo {volume}
  {693}},\ \bibinfo {pages} {122} (\bibinfo {year} {2010})},\ \Eprint
  {http://arxiv.org/abs/1008.3479} {arXiv:1008.3479 [hep-ph]} \BibitemShut
  {NoStop}%
\bibitem [{\citenamefont {Carrasco}\ \emph {et~al.}(2015)\citenamefont
  {Carrasco}, \citenamefont {Lubicz}, \citenamefont {Martinelli}, \citenamefont
  {Sachrajda}, \citenamefont {Tantalo}, \citenamefont {Tarantino},\ and\
  \citenamefont {Testa}}]{Carrasco:2015xwa}%
  \BibitemOpen
  \bibfield  {author} {\bibinfo {author} {\bibfnamefont {N.}~\bibnamefont
  {Carrasco}}, \bibinfo {author} {\bibfnamefont {V.}~\bibnamefont {Lubicz}},
  \bibinfo {author} {\bibfnamefont {G.}~\bibnamefont {Martinelli}}, \bibinfo
  {author} {\bibfnamefont {C.~T.}\ \bibnamefont {Sachrajda}}, \bibinfo {author}
  {\bibfnamefont {N.}~\bibnamefont {Tantalo}}, \bibinfo {author} {\bibfnamefont
  {C.}~\bibnamefont {Tarantino}}, \ and\ \bibinfo {author} {\bibfnamefont
  {M.}~\bibnamefont {Testa}},\ }\href {\doibase 10.1103/PhysRevD.91.074506}
  {\bibfield  {journal} {\bibinfo  {journal} {Phys. Rev. D}\ }\textbf {\bibinfo
  {volume} {91}},\ \bibinfo {pages} {074506} (\bibinfo {year} {2015})},\
  \Eprint {http://arxiv.org/abs/1502.00257} {arXiv:1502.00257 [hep-lat]}
  \BibitemShut {NoStop}%
\bibitem [{\citenamefont {Cornella}\ \emph {et~al.}(2026)\citenamefont
  {Cornella}, \citenamefont {Ferr{\'e}}, \citenamefont {K{\"o}nig},\ and\
  \citenamefont {Neubert}}]{Cornella:2026lkp}%
  \BibitemOpen
  \bibfield  {author} {\bibinfo {author} {\bibfnamefont {C.}~\bibnamefont
  {Cornella}}, \bibinfo {author} {\bibfnamefont {M.}~\bibnamefont {Ferr{\'e}}},
  \bibinfo {author} {\bibfnamefont {M.}~\bibnamefont {K{\"o}nig}}, \ and\
  \bibinfo {author} {\bibfnamefont {M.}~\bibnamefont {Neubert}},\ }\href
  {\doibase 10.1007/JHEP06(2026)027} {\bibfield  {journal} {\bibinfo  {journal}
  {JHEP}\ }\textbf {\bibinfo {volume} {06}},\ \bibinfo {pages} {027} (\bibinfo
  {year} {2026})},\ \Eprint {http://arxiv.org/abs/2601.14361} {arXiv:2601.14361
  [hep-ph]} \BibitemShut {NoStop}%
\bibitem [{\citenamefont {Braun}\ \emph {et~al.}(2019)\citenamefont {Braun},
  \citenamefont {Ji},\ and\ \citenamefont {Manashov}}]{Braun:2019wyx}%
  \BibitemOpen
  \bibfield  {author} {\bibinfo {author} {\bibfnamefont {V.~M.}\ \bibnamefont
  {Braun}}, \bibinfo {author} {\bibfnamefont {Y.}~\bibnamefont {Ji}}, \ and\
  \bibinfo {author} {\bibfnamefont {A.~N.}\ \bibnamefont {Manashov}},\ }\href
  {\doibase 10.3204/PUBDB-2019-02451} {\bibfield  {journal} {\bibinfo
  {journal} {Phys. Rev. D}\ }\textbf {\bibinfo {volume} {100}},\ \bibinfo
  {pages} {014023} (\bibinfo {year} {2019})},\ \Eprint
  {http://arxiv.org/abs/1905.04498} {arXiv:1905.04498 [hep-ph]} \BibitemShut
  {NoStop}%
\bibitem [{\citenamefont {Liu}\ and\ \citenamefont
  {Neubert}(2020)}]{Liu:2020ydl}%
  \BibitemOpen
  \bibfield  {author} {\bibinfo {author} {\bibfnamefont {Z.~L.}\ \bibnamefont
  {Liu}}\ and\ \bibinfo {author} {\bibfnamefont {M.}~\bibnamefont {Neubert}},\
  }\href {\doibase 10.1007/JHEP06(2020)060} {\bibfield  {journal} {\bibinfo
  {journal} {JHEP}\ }\textbf {\bibinfo {volume} {06}},\ \bibinfo {pages} {060}
  (\bibinfo {year} {2020})},\ \Eprint {http://arxiv.org/abs/2003.03393}
  {arXiv:2003.03393 [hep-ph]} \BibitemShut {NoStop}%
\bibitem [{\citenamefont {Galda}\ and\ \citenamefont
  {Neubert}(2020)}]{Galda:2020epp}%
  \BibitemOpen
  \bibfield  {author} {\bibinfo {author} {\bibfnamefont {A.~M.}\ \bibnamefont
  {Galda}}\ and\ \bibinfo {author} {\bibfnamefont {M.}~\bibnamefont
  {Neubert}},\ }\href {\doibase 10.1103/PhysRevD.102.071501} {\bibfield
  {journal} {\bibinfo  {journal} {Phys. Rev. D}\ }\textbf {\bibinfo {volume}
  {102}},\ \bibinfo {pages} {071501} (\bibinfo {year} {2020})},\ \Eprint
  {http://arxiv.org/abs/2006.05428} {arXiv:2006.05428 [hep-ph]} \BibitemShut
  {NoStop}%
\bibitem [{\citenamefont {Braun}\ \emph
  {et~al.}(2017{\natexlab{b}})\citenamefont {Braun}, \citenamefont {Manashov},
  \citenamefont {Moch},\ and\ \citenamefont {Strohmaier}}]{Braun:2017cih}%
  \BibitemOpen
  \bibfield  {author} {\bibinfo {author} {\bibfnamefont {V.~M.}\ \bibnamefont
  {Braun}}, \bibinfo {author} {\bibfnamefont {A.~N.}\ \bibnamefont {Manashov}},
  \bibinfo {author} {\bibfnamefont {S.}~\bibnamefont {Moch}}, \ and\ \bibinfo
  {author} {\bibfnamefont {M.}~\bibnamefont {Strohmaier}},\ }\href {\doibase
  10.1007/JHEP06(2017)037} {\bibfield  {journal} {\bibinfo  {journal} {JHEP}\
  }\textbf {\bibinfo {volume} {06}},\ \bibinfo {pages} {037} (\bibinfo {year}
  {2017}{\natexlab{b}})},\ \Eprint {http://arxiv.org/abs/1703.09532}
  {arXiv:1703.09532 [hep-ph]} \BibitemShut {NoStop}%
\bibitem [{\citenamefont {Strohmaier}(2018)}]{Strohmaier:2018tjo}%
  \BibitemOpen
  \bibfield  {author} {\bibinfo {author} {\bibfnamefont {M.}~\bibnamefont
  {Strohmaier}},\ }\emph {\bibinfo {title} {{Conformal symmetry breaking and
  evolution equations in Quantum Chromodynamics}}},\ \href {\doibase
  10.5283/epub.37432} {Ph.D. thesis},\ \bibinfo  {school} {Regensburg U.}
  (\bibinfo {year} {2018})\BibitemShut {NoStop}%
\bibitem [{\citenamefont {Bourrely}\ \emph {et~al.}(2009)\citenamefont
  {Bourrely}, \citenamefont {Caprini},\ and\ \citenamefont
  {Lellouch}}]{Bourrely:2008za}%
  \BibitemOpen
  \bibfield  {author} {\bibinfo {author} {\bibfnamefont {C.}~\bibnamefont
  {Bourrely}}, \bibinfo {author} {\bibfnamefont {I.}~\bibnamefont {Caprini}}, \
  and\ \bibinfo {author} {\bibfnamefont {L.}~\bibnamefont {Lellouch}},\ }\href
  {\doibase 10.1103/PhysRevD.82.099902} {\bibfield  {journal} {\bibinfo
  {journal} {Phys. Rev. D}\ }\textbf {\bibinfo {volume} {79}},\ \bibinfo
  {pages} {013008} (\bibinfo {year} {2009})},\ \bibinfo {note} {[Erratum:
  Phys.Rev.D 82, 099902 (2010)]},\ \Eprint {http://arxiv.org/abs/0807.2722}
  {arXiv:0807.2722 [hep-ph]} \BibitemShut {NoStop}%
\bibitem [{\citenamefont {De~Fazio}\ \emph {et~al.}(2006)\citenamefont
  {De~Fazio}, \citenamefont {Feldmann},\ and\ \citenamefont
  {Hurth}}]{DeFazio:2005dx}%
  \BibitemOpen
  \bibfield  {author} {\bibinfo {author} {\bibfnamefont {F.}~\bibnamefont
  {De~Fazio}}, \bibinfo {author} {\bibfnamefont {T.}~\bibnamefont {Feldmann}},
  \ and\ \bibinfo {author} {\bibfnamefont {T.}~\bibnamefont {Hurth}},\ }\href
  {\doibase 10.1016/j.nuclphysb.2008.03.022} {\bibfield  {journal} {\bibinfo
  {journal} {Nucl. Phys. B}\ }\textbf {\bibinfo {volume} {733}},\ \bibinfo
  {pages} {1} (\bibinfo {year} {2006})},\ \bibinfo {note} {[Erratum:
  Nucl.Phys.B 800, 405 (2008)]},\ \Eprint {http://arxiv.org/abs/hep-ph/0504088}
  {arXiv:hep-ph/0504088} \BibitemShut {NoStop}%
\bibitem [{\citenamefont {De~Fazio}\ \emph {et~al.}(2008)\citenamefont
  {De~Fazio}, \citenamefont {Feldmann},\ and\ \citenamefont
  {Hurth}}]{DeFazio:2007hw}%
  \BibitemOpen
  \bibfield  {author} {\bibinfo {author} {\bibfnamefont {F.}~\bibnamefont
  {De~Fazio}}, \bibinfo {author} {\bibfnamefont {T.}~\bibnamefont {Feldmann}},
  \ and\ \bibinfo {author} {\bibfnamefont {T.}~\bibnamefont {Hurth}},\ }\href
  {\doibase 10.1088/1126-6708/2008/02/031} {\bibfield  {journal} {\bibinfo
  {journal} {JHEP}\ }\textbf {\bibinfo {volume} {02}},\ \bibinfo {pages} {031}
  (\bibinfo {year} {2008})},\ \Eprint {http://arxiv.org/abs/0711.3999}
  {arXiv:0711.3999 [hep-ph]} \BibitemShut {NoStop}%
\bibitem [{\citenamefont {Gao}\ \emph {et~al.}(2020)\citenamefont {Gao},
  \citenamefont {L{\"u}}, \citenamefont {Shen}, \citenamefont {Wang},\ and\
  \citenamefont {Wei}}]{Gao:2019lta}%
  \BibitemOpen
  \bibfield  {author} {\bibinfo {author} {\bibfnamefont {J.}~\bibnamefont
  {Gao}}, \bibinfo {author} {\bibfnamefont {C.-D.}\ \bibnamefont {L{\"u}}},
  \bibinfo {author} {\bibfnamefont {Y.-L.}\ \bibnamefont {Shen}}, \bibinfo
  {author} {\bibfnamefont {Y.-M.}\ \bibnamefont {Wang}}, \ and\ \bibinfo
  {author} {\bibfnamefont {Y.-B.}\ \bibnamefont {Wei}},\ }\href {\doibase
  10.1103/PhysRevD.101.074035} {\bibfield  {journal} {\bibinfo  {journal}
  {Phys. Rev. D}\ }\textbf {\bibinfo {volume} {101}},\ \bibinfo {pages}
  {074035} (\bibinfo {year} {2020})},\ \Eprint
  {http://arxiv.org/abs/1907.11092} {arXiv:1907.11092 [hep-ph]} \BibitemShut
  {NoStop}%
\bibitem [{\citenamefont {Gao}\ \emph {et~al.}(2022)\citenamefont {Gao},
  \citenamefont {Huber}, \citenamefont {Ji}, \citenamefont {Wang},
  \citenamefont {Wang},\ and\ \citenamefont {Wei}}]{Gao:2021sav}%
  \BibitemOpen
  \bibfield  {author} {\bibinfo {author} {\bibfnamefont {J.}~\bibnamefont
  {Gao}}, \bibinfo {author} {\bibfnamefont {T.}~\bibnamefont {Huber}}, \bibinfo
  {author} {\bibfnamefont {Y.}~\bibnamefont {Ji}}, \bibinfo {author}
  {\bibfnamefont {C.}~\bibnamefont {Wang}}, \bibinfo {author} {\bibfnamefont
  {Y.-M.}\ \bibnamefont {Wang}}, \ and\ \bibinfo {author} {\bibfnamefont
  {Y.-B.}\ \bibnamefont {Wei}},\ }\href {\doibase 10.1007/JHEP05(2022)024}
  {\bibfield  {journal} {\bibinfo  {journal} {JHEP}\ }\textbf {\bibinfo
  {volume} {05}},\ \bibinfo {pages} {024} (\bibinfo {year} {2022})},\ \Eprint
  {http://arxiv.org/abs/2112.12674} {arXiv:2112.12674 [hep-ph]} \BibitemShut
  {NoStop}%
\bibitem [{\citenamefont {Cui}\ \emph {et~al.}(2023{\natexlab{b}})\citenamefont
  {Cui}, \citenamefont {Huang}, \citenamefont {Wang},\ and\ \citenamefont
  {Zhao}}]{Cui:2023jiw}%
  \BibitemOpen
  \bibfield  {author} {\bibinfo {author} {\bibfnamefont {B.-Y.}\ \bibnamefont
  {Cui}}, \bibinfo {author} {\bibfnamefont {Y.-K.}\ \bibnamefont {Huang}},
  \bibinfo {author} {\bibfnamefont {Y.-M.}\ \bibnamefont {Wang}}, \ and\
  \bibinfo {author} {\bibfnamefont {X.-C.}\ \bibnamefont {Zhao}},\ }\href
  {\doibase 10.1103/PhysRevD.108.L071504} {\bibfield  {journal} {\bibinfo
  {journal} {Phys. Rev. D}\ }\textbf {\bibinfo {volume} {108}},\ \bibinfo
  {pages} {L071504} (\bibinfo {year} {2023}{\natexlab{b}})},\ \Eprint
  {http://arxiv.org/abs/2301.12391} {arXiv:2301.12391 [hep-ph]} \BibitemShut
  {NoStop}%
\bibitem [{\citenamefont {Wang}\ \emph {et~al.}(2017)\citenamefont {Wang},
  \citenamefont {Wei}, \citenamefont {Shen},\ and\ \citenamefont
  {L{\"u}}}]{Wang:2017jow}%
  \BibitemOpen
  \bibfield  {author} {\bibinfo {author} {\bibfnamefont {Y.-M.}\ \bibnamefont
  {Wang}}, \bibinfo {author} {\bibfnamefont {Y.-B.}\ \bibnamefont {Wei}},
  \bibinfo {author} {\bibfnamefont {Y.-L.}\ \bibnamefont {Shen}}, \ and\
  \bibinfo {author} {\bibfnamefont {C.-D.}\ \bibnamefont {L{\"u}}},\ }\href
  {\doibase 10.1007/JHEP06(2017)062} {\bibfield  {journal} {\bibinfo  {journal}
  {JHEP}\ }\textbf {\bibinfo {volume} {06}},\ \bibinfo {pages} {062} (\bibinfo
  {year} {2017})},\ \Eprint {http://arxiv.org/abs/1701.06810} {arXiv:1701.06810
  [hep-ph]} \BibitemShut {NoStop}%
\bibitem [{\citenamefont {Wang}(2016)}]{Wang:2016qii}%
  \BibitemOpen
  \bibfield  {author} {\bibinfo {author} {\bibfnamefont {Y.-M.}\ \bibnamefont
  {Wang}},\ }\href {\doibase 10.1007/JHEP09(2016)159} {\bibfield  {journal}
  {\bibinfo  {journal} {JHEP}\ }\textbf {\bibinfo {volume} {09}},\ \bibinfo
  {pages} {159} (\bibinfo {year} {2016})},\ \Eprint
  {http://arxiv.org/abs/1606.03080} {arXiv:1606.03080 [hep-ph]} \BibitemShut
  {NoStop}%
\bibitem [{\citenamefont {Bailey}\ \emph
  {et~al.}(2015{\natexlab{a}})\citenamefont {Bailey} \emph
  {et~al.}}]{FermilabLattice:2015mwy}%
  \BibitemOpen
  \bibfield  {author} {\bibinfo {author} {\bibfnamefont {J.~A.}\ \bibnamefont
  {Bailey}} \emph {et~al.} (\bibinfo {collaboration} {Fermilab Lattice,
  MILC}),\ }\href {\doibase 10.1103/PhysRevD.92.014024} {\bibfield  {journal}
  {\bibinfo  {journal} {Phys. Rev. D}\ }\textbf {\bibinfo {volume} {92}},\
  \bibinfo {pages} {014024} (\bibinfo {year} {2015}{\natexlab{a}})},\ \Eprint
  {http://arxiv.org/abs/1503.07839} {arXiv:1503.07839 [hep-lat]} \BibitemShut
  {NoStop}%
\bibitem [{\citenamefont {Flynn}\ \emph {et~al.}(2015)\citenamefont {Flynn},
  \citenamefont {Izubuchi}, \citenamefont {Kawanai}, \citenamefont {Lehner},
  \citenamefont {Soni}, \citenamefont {Van~de Water},\ and\ \citenamefont
  {Witzel}}]{Flynn:2015mha}%
  \BibitemOpen
  \bibfield  {author} {\bibinfo {author} {\bibfnamefont {J.~M.}\ \bibnamefont
  {Flynn}}, \bibinfo {author} {\bibfnamefont {T.}~\bibnamefont {Izubuchi}},
  \bibinfo {author} {\bibfnamefont {T.}~\bibnamefont {Kawanai}}, \bibinfo
  {author} {\bibfnamefont {C.}~\bibnamefont {Lehner}}, \bibinfo {author}
  {\bibfnamefont {A.}~\bibnamefont {Soni}}, \bibinfo {author} {\bibfnamefont
  {R.~S.}\ \bibnamefont {Van~de Water}}, \ and\ \bibinfo {author}
  {\bibfnamefont {O.}~\bibnamefont {Witzel}},\ }\href {\doibase
  10.1103/PhysRevD.91.074510} {\bibfield  {journal} {\bibinfo  {journal} {Phys.
  Rev. D}\ }\textbf {\bibinfo {volume} {91}},\ \bibinfo {pages} {074510}
  (\bibinfo {year} {2015})},\ \Eprint {http://arxiv.org/abs/1501.05373}
  {arXiv:1501.05373 [hep-lat]} \BibitemShut {NoStop}%
\bibitem [{\citenamefont {Bailey}\ \emph
  {et~al.}(2015{\natexlab{b}})\citenamefont {Bailey} \emph
  {et~al.}}]{FermilabLattice:2015cdh}%
  \BibitemOpen
  \bibfield  {author} {\bibinfo {author} {\bibfnamefont {J.~A.}\ \bibnamefont
  {Bailey}} \emph {et~al.} (\bibinfo {collaboration} {Fermilab Lattice,
  MILC}),\ }\href {\doibase 10.1103/PhysRevLett.115.152002} {\bibfield
  {journal} {\bibinfo  {journal} {Phys. Rev. Lett.}\ }\textbf {\bibinfo
  {volume} {115}},\ \bibinfo {pages} {152002} (\bibinfo {year}
  {2015}{\natexlab{b}})},\ \Eprint {http://arxiv.org/abs/1507.01618}
  {arXiv:1507.01618 [hep-ph]} \BibitemShut {NoStop}%
\bibitem [{\citenamefont {Bailey}\ \emph {et~al.}(2016)\citenamefont {Bailey}
  \emph {et~al.}}]{Bailey:2015dka}%
  \BibitemOpen
  \bibfield  {author} {\bibinfo {author} {\bibfnamefont {J.~A.}\ \bibnamefont
  {Bailey}} \emph {et~al.},\ }\href {\doibase 10.1103/PhysRevD.93.025026}
  {\bibfield  {journal} {\bibinfo  {journal} {Phys. Rev. D}\ }\textbf {\bibinfo
  {volume} {93}},\ \bibinfo {pages} {025026} (\bibinfo {year} {2016})},\
  \Eprint {http://arxiv.org/abs/1509.06235} {arXiv:1509.06235 [hep-lat]}
  \BibitemShut {NoStop}%
\bibitem [{\citenamefont {Shen}\ \emph {et~al.}(2020)\citenamefont {Shen},
  \citenamefont {Wang},\ and\ \citenamefont {Wei}}]{Shen:2020hfq}%
  \BibitemOpen
  \bibfield  {author} {\bibinfo {author} {\bibfnamefont {Y.-L.}\ \bibnamefont
  {Shen}}, \bibinfo {author} {\bibfnamefont {Y.-M.}\ \bibnamefont {Wang}}, \
  and\ \bibinfo {author} {\bibfnamefont {Y.-B.}\ \bibnamefont {Wei}},\ }\href
  {\doibase 10.1007/JHEP12(2020)169} {\bibfield  {journal} {\bibinfo  {journal}
  {JHEP}\ }\textbf {\bibinfo {volume} {12}},\ \bibinfo {pages} {169} (\bibinfo
  {year} {2020})},\ \Eprint {http://arxiv.org/abs/2009.02723} {arXiv:2009.02723
  [hep-ph]} \BibitemShut {NoStop}%
\bibitem [{Inp()}]{InputParameter:2026}%
  \BibitemOpen
  \href@noop {} {}\bibinfo {note} {As an exception, we prefer to employ the
  numerical interval of the combination $2 \, \lambda_E^2 + \lambda_H^2$ for
  the three-particle light-cone distribution amplitude $\phi_{B, \, 3}$ from
  the classical equations-of-motion constraints \cite{Braun:2017liq} and merely
  take the leading-order estimate of the ratio $\lambda_E^2 / \lambda_H^2$ from
  the method of two-point QCD sum rules \cite{Rahimi:2020zzo}.}\BibitemShut
  {Stop}%
\bibitem [{NLL()}]{NLLprime:2026}%
  \BibitemOpen
  \href@noop {} {}\bibinfo {note} {Here we have introduced the shorthand
  notation ${\rm NLL}^{\prime}$ to emphasize that the unsuppressed higher Fock
  state corrections to the non-local $B \to P \, \gamma^{\ast}$ form factors
  from the three-particle twist-three bottom-meson distribution amplitude at
  order ${\cal O}(\alpha_s)$ are, however, not included in this
  approximation.}\BibitemShut {Stop}%
\bibitem [{\citenamefont {Beneke}\ \emph {et~al.}(2010)\citenamefont {Beneke},
  \citenamefont {Huber},\ and\ \citenamefont {Li}}]{Beneke:2009ek}%
  \BibitemOpen
  \bibfield  {author} {\bibinfo {author} {\bibfnamefont {M.}~\bibnamefont
  {Beneke}}, \bibinfo {author} {\bibfnamefont {T.}~\bibnamefont {Huber}}, \
  and\ \bibinfo {author} {\bibfnamefont {X.-Q.}\ \bibnamefont {Li}},\ }\href
  {\doibase 10.1016/j.nuclphysb.2010.02.002} {\bibfield  {journal} {\bibinfo
  {journal} {Nucl. Phys. B}\ }\textbf {\bibinfo {volume} {832}},\ \bibinfo
  {pages} {109} (\bibinfo {year} {2010})},\ \Eprint
  {http://arxiv.org/abs/0911.3655} {arXiv:0911.3655 [hep-ph]} \BibitemShut
  {NoStop}%
\bibitem [{\citenamefont {Bell}\ \emph {et~al.}(2020)\citenamefont {Bell},
  \citenamefont {Beneke}, \citenamefont {Huber},\ and\ \citenamefont
  {Li}}]{Bell:2020qus}%
  \BibitemOpen
  \bibfield  {author} {\bibinfo {author} {\bibfnamefont {G.}~\bibnamefont
  {Bell}}, \bibinfo {author} {\bibfnamefont {M.}~\bibnamefont {Beneke}},
  \bibinfo {author} {\bibfnamefont {T.}~\bibnamefont {Huber}}, \ and\ \bibinfo
  {author} {\bibfnamefont {X.-Q.}\ \bibnamefont {Li}},\ }\href {\doibase
  10.1007/JHEP04(2020)055} {\bibfield  {journal} {\bibinfo  {journal} {JHEP}\
  }\textbf {\bibinfo {volume} {04}},\ \bibinfo {pages} {055} (\bibinfo {year}
  {2020})},\ \Eprint {http://arxiv.org/abs/2002.03262} {arXiv:2002.03262
  [hep-ph]} \BibitemShut {NoStop}%
\bibitem [{\citenamefont {Buras}\ \emph {et~al.}(2003)\citenamefont {Buras},
  \citenamefont {Fleischer}, \citenamefont {Recksiegel},\ and\ \citenamefont
  {Schwab}}]{Buras:2003yc}%
  \BibitemOpen
  \bibfield  {author} {\bibinfo {author} {\bibfnamefont {A.~J.}\ \bibnamefont
  {Buras}}, \bibinfo {author} {\bibfnamefont {R.}~\bibnamefont {Fleischer}},
  \bibinfo {author} {\bibfnamefont {S.}~\bibnamefont {Recksiegel}}, \ and\
  \bibinfo {author} {\bibfnamefont {F.}~\bibnamefont {Schwab}},\ }\href
  {\doibase 10.1140/epjc/s2003-01379-9} {\bibfield  {journal} {\bibinfo
  {journal} {Eur. Phys. J. C}\ }\textbf {\bibinfo {volume} {32}},\ \bibinfo
  {pages} {45} (\bibinfo {year} {2003})},\ \Eprint
  {http://arxiv.org/abs/hep-ph/0309012} {arXiv:hep-ph/0309012} \BibitemShut
  {NoStop}%
\bibitem [{\citenamefont {Mishima}\ and\ \citenamefont
  {Yoshikawa}(2004)}]{Mishima:2004um}%
  \BibitemOpen
  \bibfield  {author} {\bibinfo {author} {\bibfnamefont {S.}~\bibnamefont
  {Mishima}}\ and\ \bibinfo {author} {\bibfnamefont {T.}~\bibnamefont
  {Yoshikawa}},\ }\href {\doibase 10.1103/PhysRevD.70.094024} {\bibfield
  {journal} {\bibinfo  {journal} {Phys. Rev. D}\ }\textbf {\bibinfo {volume}
  {70}},\ \bibinfo {pages} {094024} (\bibinfo {year} {2004})},\ \Eprint
  {http://arxiv.org/abs/hep-ph/0408090} {arXiv:hep-ph/0408090} \BibitemShut
  {NoStop}%
\bibitem [{\citenamefont {Fleischer}\ \emph {et~al.}(2007)\citenamefont
  {Fleischer}, \citenamefont {Recksiegel},\ and\ \citenamefont
  {Schwab}}]{Fleischer:2007mq}%
  \BibitemOpen
  \bibfield  {author} {\bibinfo {author} {\bibfnamefont {R.}~\bibnamefont
  {Fleischer}}, \bibinfo {author} {\bibfnamefont {S.}~\bibnamefont
  {Recksiegel}}, \ and\ \bibinfo {author} {\bibfnamefont {F.}~\bibnamefont
  {Schwab}},\ }\href {\doibase 10.1140/epjc/s10052-007-0277-8} {\bibfield
  {journal} {\bibinfo  {journal} {Eur. Phys. J. C}\ }\textbf {\bibinfo {volume}
  {51}},\ \bibinfo {pages} {55} (\bibinfo {year} {2007})},\ \Eprint
  {http://arxiv.org/abs/hep-ph/0702275} {arXiv:hep-ph/0702275} \BibitemShut
  {NoStop}%
\bibitem [{\citenamefont {Fleischer}\ \emph {et~al.}(2018)\citenamefont
  {Fleischer}, \citenamefont {Jaarsma}, \citenamefont {Malami},\ and\
  \citenamefont {Vos}}]{Fleischer:2018bld}%
  \BibitemOpen
  \bibfield  {author} {\bibinfo {author} {\bibfnamefont {R.}~\bibnamefont
  {Fleischer}}, \bibinfo {author} {\bibfnamefont {R.}~\bibnamefont {Jaarsma}},
  \bibinfo {author} {\bibfnamefont {E.}~\bibnamefont {Malami}}, \ and\ \bibinfo
  {author} {\bibfnamefont {K.~K.}\ \bibnamefont {Vos}},\ }\href {\doibase
  10.1140/epjc/s10052-018-6397-5} {\bibfield  {journal} {\bibinfo  {journal}
  {Eur. Phys. J. C}\ }\textbf {\bibinfo {volume} {78}},\ \bibinfo {pages} {943}
  (\bibinfo {year} {2018})},\ \Eprint {http://arxiv.org/abs/1806.08783}
  {arXiv:1806.08783 [hep-ph]} \BibitemShut {NoStop}%
\bibitem [{\citenamefont {Fang}\ \emph {et~al.}(2026)\citenamefont {Fang},
  \citenamefont {Huber}, \citenamefont {Li}, \citenamefont {Malami},\ and\
  \citenamefont {Tetlalmatzi-Xolocotzi}}]{Fang:2026fhl}%
  \BibitemOpen
  \bibfield  {author} {\bibinfo {author} {\bibfnamefont {W.-S.}\ \bibnamefont
  {Fang}}, \bibinfo {author} {\bibfnamefont {T.}~\bibnamefont {Huber}},
  \bibinfo {author} {\bibfnamefont {X.-Q.}\ \bibnamefont {Li}}, \bibinfo
  {author} {\bibfnamefont {E.}~\bibnamefont {Malami}}, \ and\ \bibinfo {author}
  {\bibfnamefont {G.}~\bibnamefont {Tetlalmatzi-Xolocotzi}},\ }\href@noop {} {\
   (\bibinfo {year} {2026})},\ \Eprint {http://arxiv.org/abs/2604.19612}
  {arXiv:2604.19612 [hep-ph]} \BibitemShut {NoStop}%
\bibitem [{\citenamefont {Wei}\ \emph {et~al.}(2009)\citenamefont {Wei} \emph
  {et~al.}}]{Belle:2009zue}%
  \BibitemOpen
  \bibfield  {author} {\bibinfo {author} {\bibfnamefont {J.~T.}\ \bibnamefont
  {Wei}} \emph {et~al.} (\bibinfo {collaboration} {Belle}),\ }\href {\doibase
  10.1103/PhysRevLett.103.171801} {\bibfield  {journal} {\bibinfo  {journal}
  {Phys. Rev. Lett.}\ }\textbf {\bibinfo {volume} {103}},\ \bibinfo {pages}
  {171801} (\bibinfo {year} {2009})},\ \Eprint {http://arxiv.org/abs/0904.0770}
  {arXiv:0904.0770 [hep-ex]} \BibitemShut {NoStop}%
\bibitem [{\citenamefont {Choudhury}\ \emph {et~al.}(2021)\citenamefont
  {Choudhury} \emph {et~al.}}]{BELLE:2019xld}%
  \BibitemOpen
  \bibfield  {author} {\bibinfo {author} {\bibfnamefont {S.}~\bibnamefont
  {Choudhury}} \emph {et~al.} (\bibinfo {collaboration} {BELLE}),\ }\href
  {\doibase 10.1007/JHEP03(2021)105} {\bibfield  {journal} {\bibinfo  {journal}
  {JHEP}\ }\textbf {\bibinfo {volume} {03}},\ \bibinfo {pages} {105} (\bibinfo
  {year} {2021})},\ \Eprint {http://arxiv.org/abs/1908.01848} {arXiv:1908.01848
  [hep-ex]} \BibitemShut {NoStop}%
\bibitem [{\citenamefont {Lees}\ \emph {et~al.}(2012)\citenamefont {Lees} \emph
  {et~al.}}]{BaBar:2012mrf}%
  \BibitemOpen
  \bibfield  {author} {\bibinfo {author} {\bibfnamefont {J.~P.}\ \bibnamefont
  {Lees}} \emph {et~al.} (\bibinfo {collaboration} {BaBar}),\ }\href {\doibase
  10.1103/PhysRevD.86.032012} {\bibfield  {journal} {\bibinfo  {journal} {Phys.
  Rev. D}\ }\textbf {\bibinfo {volume} {86}},\ \bibinfo {pages} {032012}
  (\bibinfo {year} {2012})},\ \Eprint {http://arxiv.org/abs/1204.3933}
  {arXiv:1204.3933 [hep-ex]} \BibitemShut {NoStop}%
\bibitem [{\citenamefont {Aaltonen}\ \emph
  {et~al.}(2011{\natexlab{a}})\citenamefont {Aaltonen} \emph
  {et~al.}}]{CDF:2011grz}%
  \BibitemOpen
  \bibfield  {author} {\bibinfo {author} {\bibfnamefont {T.}~\bibnamefont
  {Aaltonen}} \emph {et~al.} (\bibinfo {collaboration} {CDF}),\ }\href
  {\doibase 10.1103/PhysRevLett.106.161801} {\bibfield  {journal} {\bibinfo
  {journal} {Phys. Rev. Lett.}\ }\textbf {\bibinfo {volume} {106}},\ \bibinfo
  {pages} {161801} (\bibinfo {year} {2011}{\natexlab{a}})},\ \Eprint
  {http://arxiv.org/abs/1101.1028} {arXiv:1101.1028 [hep-ex]} \BibitemShut
  {NoStop}%
\bibitem [{\citenamefont {Aaltonen}\ \emph
  {et~al.}(2011{\natexlab{b}})\citenamefont {Aaltonen} \emph
  {et~al.}}]{CDF:2011buy}%
  \BibitemOpen
  \bibfield  {author} {\bibinfo {author} {\bibfnamefont {T.}~\bibnamefont
  {Aaltonen}} \emph {et~al.} (\bibinfo {collaboration} {CDF}),\ }\href
  {\doibase 10.1103/PhysRevLett.107.201802} {\bibfield  {journal} {\bibinfo
  {journal} {Phys. Rev. Lett.}\ }\textbf {\bibinfo {volume} {107}},\ \bibinfo
  {pages} {201802} (\bibinfo {year} {2011}{\natexlab{b}})},\ \Eprint
  {http://arxiv.org/abs/1107.3753} {arXiv:1107.3753 [hep-ex]} \BibitemShut
  {NoStop}%
\bibitem [{\citenamefont {Aaij}\ \emph
  {et~al.}(2014{\natexlab{b}})\citenamefont {Aaij} \emph
  {et~al.}}]{LHCb:2014mit}%
  \BibitemOpen
  \bibfield  {author} {\bibinfo {author} {\bibfnamefont {R.}~\bibnamefont
  {Aaij}} \emph {et~al.} (\bibinfo {collaboration} {LHCb}),\ }\href {\doibase
  10.1007/JHEP09(2014)177} {\bibfield  {journal} {\bibinfo  {journal} {JHEP}\
  }\textbf {\bibinfo {volume} {09}},\ \bibinfo {pages} {177} (\bibinfo {year}
  {2014}{\natexlab{b}})},\ \Eprint {http://arxiv.org/abs/1408.0978}
  {arXiv:1408.0978 [hep-ex]} \BibitemShut {NoStop}%
\bibitem [{\citenamefont {Aaij}\ \emph
  {et~al.}(2026{\natexlab{c}})\citenamefont {Aaij} \emph
  {et~al.}}]{LHCb:2026suh}%
  \BibitemOpen
  \bibfield  {author} {\bibinfo {author} {\bibfnamefont {R.}~\bibnamefont
  {Aaij}} \emph {et~al.} (\bibinfo {collaboration} {LHCb}),\ }\href@noop {} {\
  (\bibinfo {year} {2026}{\natexlab{c}})},\ \Eprint
  {http://arxiv.org/abs/2603.12477} {arXiv:2603.12477 [hep-ex]} \BibitemShut
  {NoStop}%
\bibitem [{\citenamefont {Aaij}\ \emph
  {et~al.}(2026{\natexlab{d}})\citenamefont {Aaij} \emph
  {et~al.}}]{LHCb:2026huw}%
  \BibitemOpen
  \bibfield  {author} {\bibinfo {author} {\bibfnamefont {R.}~\bibnamefont
  {Aaij}} \emph {et~al.} (\bibinfo {collaboration} {LHCb}),\ }\href@noop {} {\
  (\bibinfo {year} {2026}{\natexlab{d}})},\ \Eprint
  {http://arxiv.org/abs/2606.23646} {arXiv:2606.23646 [hep-ex]} \BibitemShut
  {NoStop}%
\bibitem [{\citenamefont {Hayrapetyan}\ \emph {et~al.}(2024)\citenamefont
  {Hayrapetyan} \emph {et~al.}}]{CMS:2024syx}%
  \BibitemOpen
  \bibfield  {author} {\bibinfo {author} {\bibfnamefont {A.}~\bibnamefont
  {Hayrapetyan}} \emph {et~al.} (\bibinfo {collaboration} {CMS}),\ }\href
  {\doibase 10.1088/1361-6633/ad4e65} {\bibfield  {journal} {\bibinfo
  {journal} {Rept. Prog. Phys.}\ }\textbf {\bibinfo {volume} {87}},\ \bibinfo
  {pages} {077802} (\bibinfo {year} {2024})},\ \Eprint
  {http://arxiv.org/abs/2401.07090} {arXiv:2401.07090 [hep-ex]} \BibitemShut
  {NoStop}%
\bibitem [{\citenamefont {Seidel}(2004)}]{Seidel:2004jh}%
  \BibitemOpen
  \bibfield  {author} {\bibinfo {author} {\bibfnamefont {D.}~\bibnamefont
  {Seidel}},\ }\href {\doibase 10.1103/PhysRevD.70.094038} {\bibfield
  {journal} {\bibinfo  {journal} {Phys. Rev. D}\ }\textbf {\bibinfo {volume}
  {70}},\ \bibinfo {pages} {094038} (\bibinfo {year} {2004})},\ \Eprint
  {http://arxiv.org/abs/hep-ph/0403185} {arXiv:hep-ph/0403185} \BibitemShut
  {NoStop}%
\bibitem [{\citenamefont {Asatryan}\ \emph {et~al.}(2002)\citenamefont
  {Asatryan}, \citenamefont {Asatrian}, \citenamefont {Greub},\ and\
  \citenamefont {Walker}}]{Asatryan:2001zw}%
  \BibitemOpen
  \bibfield  {author} {\bibinfo {author} {\bibfnamefont {H.~H.}\ \bibnamefont
  {Asatryan}}, \bibinfo {author} {\bibfnamefont {H.~M.}\ \bibnamefont
  {Asatrian}}, \bibinfo {author} {\bibfnamefont {C.}~\bibnamefont {Greub}}, \
  and\ \bibinfo {author} {\bibfnamefont {M.}~\bibnamefont {Walker}},\ }\href
  {\doibase 10.1103/PhysRevD.65.074004} {\bibfield  {journal} {\bibinfo
  {journal} {Phys. Rev. D}\ }\textbf {\bibinfo {volume} {65}},\ \bibinfo
  {pages} {074004} (\bibinfo {year} {2002})},\ \Eprint
  {http://arxiv.org/abs/hep-ph/0109140} {arXiv:hep-ph/0109140} \BibitemShut
  {NoStop}%
\bibitem [{\citenamefont {Asatrian}\ \emph {et~al.}(2020)\citenamefont
  {Asatrian}, \citenamefont {Greub},\ and\ \citenamefont
  {Virto}}]{Asatrian:2019kbk}%
  \BibitemOpen
  \bibfield  {author} {\bibinfo {author} {\bibfnamefont {H.~M.}\ \bibnamefont
  {Asatrian}}, \bibinfo {author} {\bibfnamefont {C.}~\bibnamefont {Greub}}, \
  and\ \bibinfo {author} {\bibfnamefont {J.}~\bibnamefont {Virto}},\ }\href
  {\doibase 10.1007/JHEP04(2020)012} {\bibfield  {journal} {\bibinfo  {journal}
  {JHEP}\ }\textbf {\bibinfo {volume} {04}},\ \bibinfo {pages} {012} (\bibinfo
  {year} {2020})},\ \Eprint {http://arxiv.org/abs/1912.09099} {arXiv:1912.09099
  [hep-ph]} \BibitemShut {NoStop}%
\bibitem [{\citenamefont {Beneke}\ \emph {et~al.}(2009)\citenamefont {Beneke},
  \citenamefont {Huber},\ and\ \citenamefont {Li}}]{Beneke:2008ei}%
  \BibitemOpen
  \bibfield  {author} {\bibinfo {author} {\bibfnamefont {M.}~\bibnamefont
  {Beneke}}, \bibinfo {author} {\bibfnamefont {T.}~\bibnamefont {Huber}}, \
  and\ \bibinfo {author} {\bibfnamefont {X.~Q.}\ \bibnamefont {Li}},\ }\href
  {\doibase 10.1016/j.nuclphysb.2008.11.019} {\bibfield  {journal} {\bibinfo
  {journal} {Nucl. Phys. B}\ }\textbf {\bibinfo {volume} {811}},\ \bibinfo
  {pages} {77} (\bibinfo {year} {2009})},\ \Eprint
  {http://arxiv.org/abs/0810.1230} {arXiv:0810.1230 [hep-ph]} \BibitemShut
  {NoStop}%
\bibitem [{\citenamefont {Bonciani}\ and\ \citenamefont
  {Ferroglia}(2008)}]{Bonciani:2008wf}%
  \BibitemOpen
  \bibfield  {author} {\bibinfo {author} {\bibfnamefont {R.}~\bibnamefont
  {Bonciani}}\ and\ \bibinfo {author} {\bibfnamefont {A.}~\bibnamefont
  {Ferroglia}},\ }\href {\doibase 10.1088/1126-6708/2008/11/065} {\bibfield
  {journal} {\bibinfo  {journal} {JHEP}\ }\textbf {\bibinfo {volume} {11}},\
  \bibinfo {pages} {065} (\bibinfo {year} {2008})},\ \Eprint
  {http://arxiv.org/abs/0809.4687} {arXiv:0809.4687 [hep-ph]} \BibitemShut
  {NoStop}%
\bibitem [{\citenamefont {Asatrian}\ \emph {et~al.}(2008)\citenamefont
  {Asatrian}, \citenamefont {Greub},\ and\ \citenamefont
  {Pecjak}}]{Asatrian:2008uk}%
  \BibitemOpen
  \bibfield  {author} {\bibinfo {author} {\bibfnamefont {H.~M.}\ \bibnamefont
  {Asatrian}}, \bibinfo {author} {\bibfnamefont {C.}~\bibnamefont {Greub}}, \
  and\ \bibinfo {author} {\bibfnamefont {B.~D.}\ \bibnamefont {Pecjak}},\
  }\href {\doibase 10.1103/PhysRevD.78.114028} {\bibfield  {journal} {\bibinfo
  {journal} {Phys. Rev. D}\ }\textbf {\bibinfo {volume} {78}},\ \bibinfo
  {pages} {114028} (\bibinfo {year} {2008})},\ \Eprint
  {http://arxiv.org/abs/0810.0987} {arXiv:0810.0987 [hep-ph]} \BibitemShut
  {NoStop}%
\bibitem [{\citenamefont {Bell}(2009{\natexlab{a}})}]{Bell:2008ws}%
  \BibitemOpen
  \bibfield  {author} {\bibinfo {author} {\bibfnamefont {G.}~\bibnamefont
  {Bell}},\ }\href {\doibase 10.1016/j.nuclphysb.2008.12.018} {\bibfield
  {journal} {\bibinfo  {journal} {Nucl. Phys. B}\ }\textbf {\bibinfo {volume}
  {812}},\ \bibinfo {pages} {264} (\bibinfo {year} {2009}{\natexlab{a}})},\
  \Eprint {http://arxiv.org/abs/0810.5695} {arXiv:0810.5695 [hep-ph]}
  \BibitemShut {NoStop}%
\bibitem [{\citenamefont {Bell}(2008)}]{Bell:2007tv}%
  \BibitemOpen
  \bibfield  {author} {\bibinfo {author} {\bibfnamefont {G.}~\bibnamefont
  {Bell}},\ }\href {\doibase 10.1016/j.nuclphysb.2007.09.006} {\bibfield
  {journal} {\bibinfo  {journal} {Nucl. Phys. B}\ }\textbf {\bibinfo {volume}
  {795}},\ \bibinfo {pages} {1} (\bibinfo {year} {2008})},\ \Eprint
  {http://arxiv.org/abs/0705.3127} {arXiv:0705.3127 [hep-ph]} \BibitemShut
  {NoStop}%
\bibitem [{\citenamefont {Bell}(2009{\natexlab{b}})}]{Bell:2009nk}%
  \BibitemOpen
  \bibfield  {author} {\bibinfo {author} {\bibfnamefont {G.}~\bibnamefont
  {Bell}},\ }\href {\doibase 10.1016/j.nuclphysb.2009.07.012} {\bibfield
  {journal} {\bibinfo  {journal} {Nucl. Phys. B}\ }\textbf {\bibinfo {volume}
  {822}},\ \bibinfo {pages} {172} (\bibinfo {year} {2009}{\natexlab{b}})},\
  \Eprint {http://arxiv.org/abs/0902.1915} {arXiv:0902.1915 [hep-ph]}
  \BibitemShut {NoStop}%
\bibitem [{\citenamefont {Grozin}(1998)}]{Grozin:1998kf}%
  \BibitemOpen
  \bibfield  {author} {\bibinfo {author} {\bibfnamefont {A.~G.}\ \bibnamefont
  {Grozin}},\ }\href {\doibase 10.1016/S0370-2693(98)01439-7} {\bibfield
  {journal} {\bibinfo  {journal} {Phys. Lett. B}\ }\textbf {\bibinfo {volume}
  {445}},\ \bibinfo {pages} {165} (\bibinfo {year} {1998})},\ \Eprint
  {http://arxiv.org/abs/hep-ph/9810358} {arXiv:hep-ph/9810358} \BibitemShut
  {NoStop}%
\bibitem [{\citenamefont {Grozin}\ \emph {et~al.}(2006)\citenamefont {Grozin},
  \citenamefont {Smirnov},\ and\ \citenamefont {Smirnov}}]{Grozin:2006xm}%
  \BibitemOpen
  \bibfield  {author} {\bibinfo {author} {\bibfnamefont {A.~G.}\ \bibnamefont
  {Grozin}}, \bibinfo {author} {\bibfnamefont {A.~V.}\ \bibnamefont {Smirnov}},
  \ and\ \bibinfo {author} {\bibfnamefont {V.~A.}\ \bibnamefont {Smirnov}},\
  }\href {\doibase 10.1088/1126-6708/2006/11/022} {\bibfield  {journal}
  {\bibinfo  {journal} {JHEP}\ }\textbf {\bibinfo {volume} {11}},\ \bibinfo
  {pages} {022} (\bibinfo {year} {2006})},\ \Eprint
  {http://arxiv.org/abs/hep-ph/0609280} {arXiv:hep-ph/0609280} \BibitemShut
  {NoStop}%
\bibitem [{\citenamefont {Beneke}\ and\ \citenamefont
  {Jager}(2007)}]{Beneke:2006mk}%
  \BibitemOpen
  \bibfield  {author} {\bibinfo {author} {\bibfnamefont {M.}~\bibnamefont
  {Beneke}}\ and\ \bibinfo {author} {\bibfnamefont {S.}~\bibnamefont {Jager}},\
  }\href {\doibase 10.1016/j.nuclphysb.2007.01.016} {\bibfield  {journal}
  {\bibinfo  {journal} {Nucl. Phys. B}\ }\textbf {\bibinfo {volume} {768}},\
  \bibinfo {pages} {51} (\bibinfo {year} {2007})},\ \Eprint
  {http://arxiv.org/abs/hep-ph/0610322} {arXiv:hep-ph/0610322} \BibitemShut
  {NoStop}%
\bibitem [{\citenamefont {Hambrock}\ \emph {et~al.}(2015)\citenamefont
  {Hambrock}, \citenamefont {Khodjamirian},\ and\ \citenamefont
  {Rusov}}]{Hambrock:2015wka}%
  \BibitemOpen
  \bibfield  {author} {\bibinfo {author} {\bibfnamefont {C.}~\bibnamefont
  {Hambrock}}, \bibinfo {author} {\bibfnamefont {A.}~\bibnamefont
  {Khodjamirian}}, \ and\ \bibinfo {author} {\bibfnamefont {A.}~\bibnamefont
  {Rusov}},\ }\href {\doibase 10.1103/PhysRevD.92.074020} {\bibfield  {journal}
  {\bibinfo  {journal} {Phys. Rev. D}\ }\textbf {\bibinfo {volume} {92}},\
  \bibinfo {pages} {074020} (\bibinfo {year} {2015})},\ \Eprint
  {http://arxiv.org/abs/1506.07760} {arXiv:1506.07760 [hep-ph]} \BibitemShut
  {NoStop}%
\bibitem [{\citenamefont {Navas}\ \emph {et~al.}(2024)\citenamefont {Navas}
  \emph {et~al.}}]{ParticleDataGroup:2024cfk}%
  \BibitemOpen
  \bibfield  {author} {\bibinfo {author} {\bibfnamefont {S.}~\bibnamefont
  {Navas}} \emph {et~al.} (\bibinfo {collaboration} {Particle Data Group}),\
  }\href {\doibase 10.1103/PhysRevD.110.030001} {\bibfield  {journal} {\bibinfo
   {journal} {Phys. Rev. D}\ }\textbf {\bibinfo {volume} {110}},\ \bibinfo
  {pages} {030001} (\bibinfo {year} {2024})}\BibitemShut {NoStop}%
\bibitem [{\citenamefont {Aaij}\ \emph
  {et~al.}(2018{\natexlab{b}})\citenamefont {Aaij} \emph
  {et~al.}}]{LHCb:2018roe}%
  \BibitemOpen
  \bibfield  {author} {\bibinfo {author} {\bibfnamefont {R.}~\bibnamefont
  {Aaij}} \emph {et~al.} (\bibinfo {collaboration} {LHCb}),\ }\href@noop {} {\
  (\bibinfo {year} {2018}{\natexlab{b}})},\ \Eprint
  {http://arxiv.org/abs/1808.08865} {arXiv:1808.08865 [hep-ex]} \BibitemShut
  {NoStop}%
\end{thebibliography}%

\end{document}